\documentclass[%
 aip,
 amsmath,amssymb,
preprint,
]{revtex4-1}

\usepackage{graphicx}
\usepackage{dcolumn}
\usepackage{bm}

\usepackage{CJK}
\usepackage[utf8]{inputenc}
\usepackage[T1]{fontenc}
\usepackage{mathptmx}
\usepackage{gensymb}
\usepackage{graphicx}
\usepackage{epstopdf}
\usepackage{dcolumn}
\usepackage{bm}
\usepackage{amsmath}
\usepackage{mathtools}
\usepackage{relsize}

\usepackage[colorlinks]{hyperref}
\begin{document}


\title{How Carbon Vacancies Can Affect the Properties of Group IV Color Centers in Diamond: A Study of Thermodynamics and Kinetics}

\author{Rodrick Kuate Defo \footnote{E-mail: kuate@physics.harvard.edu}}
\affiliation{Department of Physics, Harvard University, Cambridge, MA 02138, USA}
\author{Efthimios Kaxiras} 
\affiliation{Department of Physics, Harvard University, Cambridge, MA 02138, USA}
\affiliation{John A. Paulson School of Engineering and Applied Sciences, Harvard University, Cambridge, MA 02138, USA}
\author{Steven L. Richardson}  
\affiliation{John A. Paulson School of Engineering and Applied Sciences, Harvard University, Cambridge, MA 02138, USA}
\affiliation{Department of Electrical and Computer Engineering, Howard University, Washington, DC 20059, USA}

\date{\today}
\begin{abstract}
Recently there has been much interest in using Group IV elements from the Periodic Table to fabricate and study X$V$ color centers in diamond where X = Si, Ge, Sn, or Pb and $V$ is a carbon vacancy. These Group IV color centers have a number of interesting spin and optical properties which could potentially make them better candidates than N$V^-$ centers for important applications in quantum computing and quantum information processing. Unfortunately, the very same ion implantation process that is required to create these X$V$ color centers in diamond necessarily  also produces many carbon vacancies ($V_{\rm C}$) which can form complexes with these color centers ($V_{\rm C}-$X$V$) that can dramatically affect the properties of the isolated X$V$ color centers. The main focus of this work is to use density-functional theory (DFT) to study the thermodynamics and kinetics of the formation of these $V_{\rm C}-$X$V$ complexes and to suggest experimental ways to impede this process such as varying the Fermi level of the host diamond material through chemical doping or applying an external electrical bias. We also include a discussion of how the simple presence of many $V_{\rm C}$ can negatively impact the spin coherence times ($T_2$) of Group IV color centers through the presence of acoustic phonons.

\end{abstract}
\maketitle


\section{\label{intro}INTRODUCTION}
Solid state single photon emitters (SPEs) have become important systems in both basic and applied research because of their important applications in quantum computing, quantum metrology, and quantum information processing.~\cite{Doherty2013nitrogen,Deak,Gali2,awschalom2018quant,DEFO2019,Kuate} The most studied SPE to date is the negatively charged nitrogen vacancy (N$V^-$) in diamond which can be made by replacing one carbon atom with a nitrogen atom, removing a neighboring carbon atom to form a vacancy, and then adding an electron to the system (see Fig. \ref{fig:structure}(a)).  The N$V^-$ color center has an electron spin with an excellent spin coherence time ($T_2$) even at room temperature~\cite{Walsworth} and this spin state can be prepared, manipulated, and read out using light and radio-frequency fields. While the spin coherence time of N$V ^-$ can be negatively affected by coupling to the $^{13}$C nuclear spins in the local environment of the diamond host material,~\cite{Balasubramanian} the question of how a N$V^-$ center can couple to or decouple from its neighboring nuclear spins has been studied as a means of implementing multi-qubit registers.~\cite{Abobeih} Regarding the optical properties of the N$V^-$ center, unfortunately only about 4\% of the fluorescence is found in the zero-phonon line (ZPL).~\cite{Jelezko} Another problem with using the N$V^-$ color center as a viable SPE is that it has an electric dipole moment thus making it susceptible to external noise and local fields thus causing broadening in the transition level inhomogeneities. One way to solve this problem is to consider solid state SPEs which have inversion symmetry and therefore lack a dipole moment making them insensitive to external fields.  

In a quest to identify a suitable solid state SPE with inversion symmetry, there has been a considerable amount of both experimental and theoretical work in the literature on Si$V$ color centers.~\cite{zaitsev1981cathodoluminescence, goss1996twelve,clark1995silicon, gali2013ab, neu2011single,hepp2014electronic, rogers2014all, sipahigil2014indistinguishable, Pingault, pingault2014all, dietrich2014isotopically, becker2016ultrafast,Jahnke, sukachev2017silicon, becker2018all, nguyen2018all,d2011optical,rose2018observation, iakoubovskii2002characterization, edmonds2008electron,thiering2019the,metsch2019init} The Si$V$ centers in diamond can be formed by removing two adjacent carbon atoms along the <111> lattice direction and inserting an interstitial Si atom midway along this direction between the two vacant sites  (see Fig. \ref{fig:structure}(b) where X = Si). While the fluorescence of the  charge state Si$V^-$ color center is quite an improvement from the N$V^-$, with 60\% of its luminescence now seen in the ZPL, it has a much shorter spin coherence time than the N$V^-$. It has been demonstrated that by applying strain  on the Si$V^-$ one can enhance its spin coherence time.~\cite{sohn2018cont} The neutral charge state color center Si$V^0$ has also been extensively studied since it is expected to have an intrinsically longer spin coherence time than the Si$V^-$.~\cite{thiering2019the,rose2018observation}

One can explore other color centers beyond Si$V$ and its various charge states by considering the general class of X$V$ color centers, where X is a Group IV element (X = Ge, Sn, or Pb) and $V$ is a carbon vacancy in the diamond structure. Group IV color centers are particularly interesting choices for solid state SPEs because the selection of heavier elements for X results in a larger energy split in the ground state for the singly negatively charged color center which will increase their spin coherence time even at higher temperatures.~\cite{Iwasaki2}   
These X$V$ color centers would also possess inversion symmetry and thus potentially be better candidates for SPEs than the N$V^-$ in diamond. 
Recently, a new Group IV color center in diamond, the Ge$V^-$, has been produced by ion implantation and chemical vapor deposition techniques \cite{Iwasaki} and also by using a microwave plasma chemical reactor.~\cite{Ralchenko} The Ge$V^-$ color center has a structure very similar to Si$V^-$ and it has a sharp and strong ZPL at 602~nm at room temperature. Subsequent experiments have spectroscopically confirmed that Ge$V^-$ is a promising candidate for a SPE.~\cite{Ekimov,Palyanov2015germanium,Palyanov, Ekimov2,Bhaskar,bray2018single,ekimov2019effect} The Sn$V^-$ color center has been created and identified at room temperature and it has a geometry similar to Ge$V^-$ and a photoluminesce spectrum that shows a sharp ZPL at 619 nm.~\cite{Iwasaki2, tchernij2017single} Sn$V^-$ centers have also been created under high pressure~\cite{Ekimov2018tin, palyanov2019high}  and can be used in luminescent thermometry.~\cite{Alkahtani2018tin} As demonstrated for N$V^-$ centers, it is possible to structurally confine Sn$V^-$ color centers to reside within diamond nanopillars and thus enhance their optical properties.~\cite{rugar2019char}  Finally, the Pb$V^-$ color center has been fabricated and characterized as a SPE with a structure similar to that of the Si$V^-$, Ge$V^-$ and Sn$V^-$ centers, though there is disagreement in the literature as to the value of its ZPL with measured values of 520~nm~\cite{Trusheim} or over 550~nm~\cite{tchernij2018single} and a predicted value of 517~nm.~\cite{Trusheim,Gali3} 

In this paper we will explore how a single carbon vacancy ($V_{\rm C}$) can affect the properties of X$V$ color centers in diamond. Let us first, however, review how single carbon vacancies can affect N$V^-$ color centers in diamond. Indeed, to generate these N$V^-$ centers, atomic and molecular nitrogen ions (e.g. $^{15}$N$^{+}$ and  $^{15}$N$^{+}_{2}$) are introduced at low to high energies (2 keV - 2 MeV) in a host diamond material where they substitutionally replace carbon atoms.~\cite{meijer2005generation, pezzagna2010creat,rabeau2006implantation,weis2008single,Luhmann2018screen,schwartz2011situ,Naydenov} 
 This process produces a large number of single carbon vacancies throughout the crystal especially along the pathway of the implanted nitrogen ions. Subsequent annealing of the host diamond material at high temperature eliminates many of these newly created carbon vacancies, while others can diffuse towards the implanted nitrogen atoms to form the desired N$V^-$ color centers. Unfortunately, the remaining single carbon vacancies in the host material can cause several problems for these newly created N$V^-$ color centers.  One issue is that a $V_{\rm C}$ can form a complex with a N$V^-$ color center (e.g. $V_{\rm C}-$N$V^-$), thus affecting its optical properties by quenching its fluorescence (an issue that becomes more prevalent the heavier the color center).~\cite{Luhmann2018screen} Another problem occurs when single carbon vacancies combine to form paramagnetic clusters of various shapes and sizes ($V_{\rm C}$)$_{n = 2,3, ...}$ which can interfere with the spin of the N$V^-$ centers thus decreasing their spin coherence times.~\cite{Yamamoto,hounsome2005optical, iakoubovskii2004vacancy}  
 A clever way of eliminating the formation of vacancy clusters when creating N$V^-$ centers in diamond has been recently proposed.~\cite{Oliveira} By charging the vacancies in the space charge layer of free carriers generated by a boron-doped diamond structure, the formation of thermally stable paramagnetic di-vacancy complexes ($V_2$) was suppressed resulting in a tenfold-improved spin coherence time for the color centers and a twofold-improved formation yield of nitrogen vacancy centers in diamond. We believe that all of these concerns would also apply for the carbon vacancies which were generated during the formation of X$V$ color centers.

As a first step in understanding how carbon vacancy clusters can affect the properties of X$V$ color centers in diamond, we consider in this paper the simplest example of a carbon vacancy cluster, the single isolated $V_{\rm C} $ in diamond, and study the thermodynamics of the formation of a $V_{\rm C}-$X$V$ complex using density-functional theory (DFT). We compute the diffusion barrier height of an isolated $V_{\rm C} $ in diamond and study how it is altered in the presence of a nearby X$V$ color center. As motivated by the work on carbon vacancy clusters on the N$V^-$ center,~\cite{Oliveira} we then model the formation of the $V_{\rm C}-$X$V$ complex as a function its charge state and the Fermi level of the host diamond material. These results might propose ways to mitigate the formation of $V_{\rm C}-$X$V$ clusters by simple electromagnetic and chemical considerations under certain doping conditions.   Finally, we address the problem that while acoustic phonons can have a detrimental effect on the spin coherence times of X$V$ color centers, and the N$V^-$ color center as well, this effect can be mitigated by varying the density of carbon vacancies in the host diamond material. 

In this paper we first review and discuss methods we will use in our computations (Section \ref{methods}). In Section \ref{barriers} we use DFT to compute the optimized geometries (Section \ref{sec:opt_geom}) and formation energies and charge transition levels (Section \ref{sec:form_en}) for the isolated $V_{\rm C}$, the N$V$ defect, and the X$V$ defects as well, where X is a Group IV element (X = Si, Ge, Sn, or Pb). Our calculated results in Section \ref{barriers} are also compared with available experimental and theoretical results. In Section \ref{sec:VC_iso} we compute the lowest energy diffusion pathway for an isolated $V_{\rm C}$ and determine its diffusion barrier height and demonstrate that it is a function of the charge state of $V_{\rm C}$ and thus the Fermi level of the host material. In an effort to evaluate the ease with which the $V_{\rm C}-$X$V$ complex can form kinetically, in Section \ref{sec:vc_xv} we then study the effect of a neighboring X$V$ color center on the diffusion barrier height of a $V_{\rm C}$ as a function of the charge state of the $V_{\rm C}$ and the Fermi level of the host material. Since we show in this paper that the formation energies for the isolated $V_{\rm C}$ and the isolated X$V$ color center depend on both their individual defect charge states and the Fermi level of the host material, we develop in Section \ref{sec:therm_comp} a detailed model which will enable us to compute the formation energy of a $V_{\rm C}-$X$V$ complex while ensuring charge conservation for the system. In Section \ref{sec:phons} we propose a model to explore the effect of carbon vacancies on the acoustic phonons of the host diamond material and how this could affect the spin coherence time of X$V$ color centers and then summarize the conclusions of this work.

\section{\label{methods}COMPUTATIONAL METHODS}

We performed first-principles DFT calculations for the various defect structures using the VASP code.~\cite{Kresse1,Kresse2,Kresse3}
For the exchange-correlation energy of electrons we use the generalized gradient approximation (GGA), as parametrized by Perdew, Burke and Erzenhof (PBE).~\cite{Perdew}
The atomic positions were relaxed until the magnitude of the Hellmann-Feynman forces was smaller than 0.01 eV$\cdot$\AA$^{-1}$~on each atom and the lattice parameters were concurrently relaxed.
The wavefunctions were expanded in a plane wave basis with a cutoff energy of 600 eV and a Monkhorst-Pack grid of $18\times18\times18$ k-points was used for integrations in reciprocal space for the stoichiometric conventional unit cell.
The relaxed lattice parameters of the stoichiometric conventional unit cell were then used for all other structures.
As a test of the level of convergence of energies and structural features, we checked that the increase in grid size from $12\times12\times12$ to $18\times18\times18$ and in cutoff size from 400 eV to 600 eV caused a change in the total energy of less than 0.02~eV and a change in the lattice constants of less than 0.01~\AA.
Formation energies and transition states were calculated using a supercell with 216 atoms ($3\times3\times3$ multiple of the conventional unit cell) with appropriately scaled k-point grids and a cutoff energy of 400 eV. Supercells that were a $4\times4\times4$ multiple of the conventional unit cell were also investigated to check the convergence of the results.  

The formation energies of X$V^{(q)}$ (X = N, Si, Ge, Sn, Pb) in various charge states were calculated according to the formula,~\cite{zhang1991chemical, RevModPhys.86.253}
\begin{equation}
\label{eq:form_eq}
E_f(q) = E_{\text{def}}(q) - E_0 - \sum_i\mu_in_i + q(E_{\text{VBM}} +E_{\text{F}}) + E_{\text{corr}}(q)
\end{equation}
where $q$ denotes the charge state, with $q \in [-3,+2]$, $E_{\text{def}}(q)$ is the total energy for the defect supercell with charge state $q$, $E_0$ is the total energy for the stoichiometric neutral supercell, $\mu_i$ is the chemical potential of atom $i$, $n_i$ is a positive (negative) integer representing the number of atoms added (removed) from the system relative to the stoichiometric cell, $E_{\text{VBM}}$ is the absolute position of the valence band maximum, $E_{\text{F}}$ is the position of the Fermi level with respect to the valence band maximum (generally treated as a parameter), and $E_{\text{corr}}(q)$ is a correction term to account for the finite size of the supercell when performing calculations for charged defects.~\cite{Vinichenko} This correction term does not simply treat the charged defect as a point charge, but rather considers the extended charge distribution. The chemical potentials of all the reference elements used in our calculations are listed as follows as a function of their crystal structure and total energy per atom: N ($\beta$ hexagonal close-packed structure, $-8.29$ eV/atom); C (diamond structure, $-9.10$ eV/atom);  Si (diamond structure, $-5.42$ eV/atom); Ge (diamond structure, $-4.49$ eV/atom); Sn (body-centered tetragonal structure, $-3.80$ eV/atom); and Pb (face-centered cubic structure, $-3.57$ eV/atom).  For diffusion studies, all defect atoms were located at the same position in the crystal lattice relative to the $V_{\rm C}$ with the exception of N, which was considered at two positions such that its average position in the crystal lattice relative to the $V_{\rm C}$ was the same as that of the other defect atoms.

For calculating the barriers for diffusion, we used the NEB method.~\cite{Henkelman,Jansson} The atomic positions were first relaxed in the initial and final $3\times3\times3$ supercell configurations until the threshold of 0.01~eV$\cdot$\AA$^{-1}$ was reached. Three images between the endpoints were then constructed by linearly interpolating between the endpoints and each image was relaxed to the force threshold of less than 0.01~eV$\cdot$\AA$^{-1}$ on each atom. A spring force was set up between neighboring images such that the relaxation would occur predominantly in the direction perpendicular to the hypertangent between images thus ensuring the preservation of equal distances between images.

To fully characterize diffusion, the diffusivity, $D$, is given by
\begin{equation}
D = \nu_0d^2e^{-\varepsilon_b/k_BT}
\label{eq:diffusivity}
\end{equation}
where $\nu_0 = \nu_0(T)$ is the attempt frequency, $d$ is the migration distance, in the case of $V_{\rm C}$ equal to the nearest neighbor hop between C sites, and $\varepsilon_b$ is the activation energy barrier.
From harmonic transition-state theory, we calculate the attempt frequency as \cite{Voter},
\begin{equation}
\nu_0 = \frac{k_BT}{h}\prod_{j=1}^{m'}\frac{e^{-h\nu_j'/2k_BT}}{(1-e^{-h\nu_j'/k_BT})}\left(\prod_{j=1}^{m}\frac{e^{-h\nu_j/2k_BT}}{(1-e^{-h\nu_j/k_BT})}\right)^{-1},
\end{equation}
where $m$, $m'$ and $\nu_j$, $\nu_j'$  are the corresponding number of normal modes and phonon frequencies, respectively, at the initial ($I$) and saddle-point ($S$) configurations (there is one fewer normal mode at the saddle point than at the equilibrium configuration, $m = m'+1$).

The charge at the $V_{\rm C}$ was obtained using the DDEC6 method.~\cite{Manz2016} These density derived electrostatic and chemical (DDEC) methods for calculating the charge assign to each atom a charge density. Such charge densities are optimized with respect to distance functions designed to respect ionic and covalent bonding and with the constraint that the sum of the atomic charge densities equals the total charge density.~\cite{Manz2010} The DDEC6 method is preferable to Bader's quantum chemical topology (QCT) which can lead to non-nuclear attractors and thereby undefined net atomic charges (NAC).~\cite{Cao1987} The DDEC6 method is an improvement from the DDEC3 approach which does not always converge to a unique solution or one that respects the symmetries of the system.~\cite{Manz2016} Finally, the DDEC6 method is also preferable to such methods as the Mulliken population analysis which is not independent of the chosen basis.~\cite{Reed1985}  

As input to our first-principles phonon calculations which were performed using Phonopy,~\cite{Togo} a supercell was constructed containing 64 atoms ($2\times2\times2$ multiple of the conventional unit cell) with appropriately scaled k-point grids and a cutoff energy of 500 eV. In constructing the input supercells with defects for Phonopy, the atomic positions were relaxed until the magnitude of the Hellmann-Feynman forces was smaller than $10^{-4}$ eV$\cdot$\AA$^{-1}$. 



\section{\label{barriers} RESULTS AND DISCUSSION}

\subsection{\label{sec:opt_geom}Optimized geometries}

\begin{figure}[ht!] 
\centering
\includegraphics[width=0.75\textwidth]{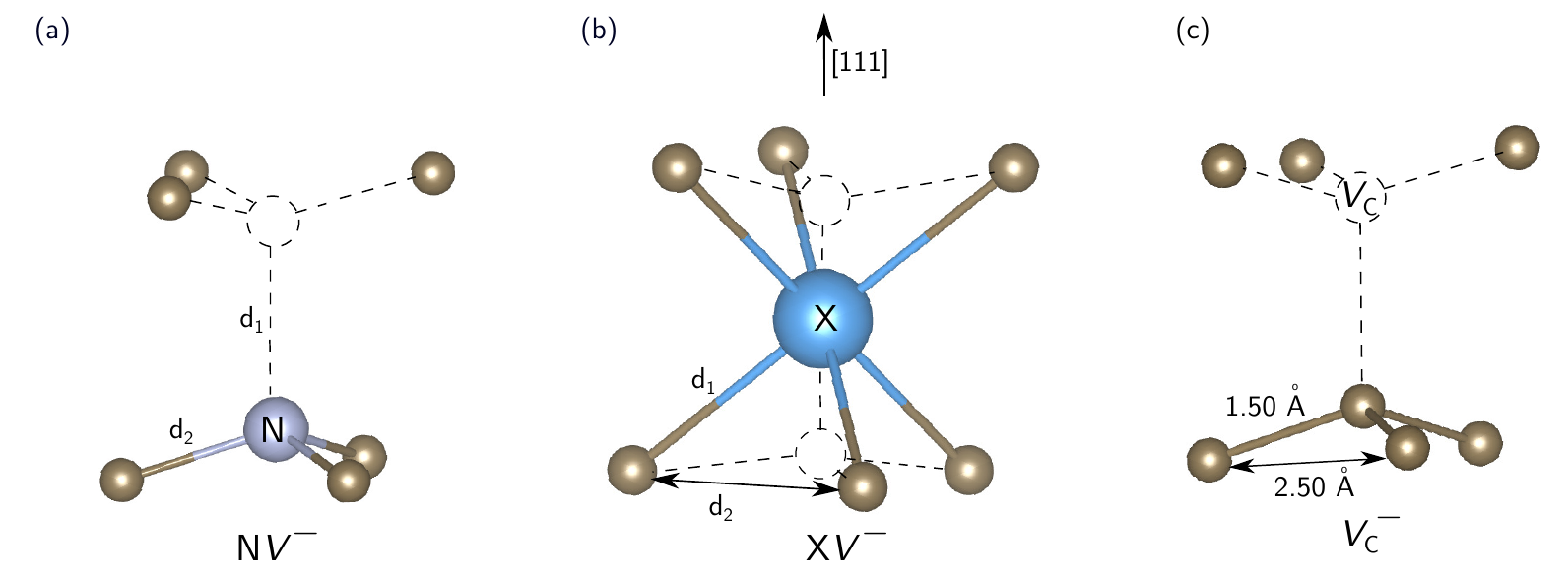}
\caption{Structure of (a) N$V^-$, (b) X$V^-$ and (c) $V_{\rm C}^-$ defects, with the same orientation, with distances defined corresponding to the values in Table \ref{tab:parameters}. Carbon atoms are indicated in brown.} 
\label{fig:structure}
\end{figure}

For the structural features of diamond, we obtain $a=3.572$ \AA \  for the lattice constant.
Our value agrees well with the experimental value of $a=3.567$ \AA \  \cite{Levinshtein}  and with a theoretical value of $a=3.570$~\AA .~\cite{Deak} Various structural constants are defined in Fig. \ref{fig:structure} and the values for the different color centers are presented in Table \ref{tab:parameters}, as obtained from our calculations and from previous theoretical investigations.~\cite{Deak,Gali3} 

\begin{table}[ht!]
\caption{ Structural constants $d_1$ and $d_2$ (in \AA, see Fig. \ref{fig:structure}) for X$V^-$ centers and corresponding values from previous work in brackets. }
\centering
\vspace{1 mm}
\begin{tabular}{|l|c|c|}
\hline\hline 
X$V$ &$d_1$ & $d_2$ \\
\hline 
N$V$ & 1.80 [1.87$^{\rm a}$] & 1.48 [1.42$^{\rm a}$]  \\
Si$V$ &  1.98 [1.96$^{\rm b}$] & 2.69 [2.67$^{\rm b}$]\\
Ge$V$ &  2.03 [2.01$^{\rm b}$] & 2.76 [2.73$^{\rm b}$] \\
Sn$V$ &   2.10 [2.08$^{\rm b}$] & 2.86 [2.83$^{\rm b}$]\\
Pb$V$ &  2.15 [2.12$^{\rm b}$] &  2.92 [2.88$^{\rm b}$] \\
\hline
\end{tabular}
\begin{flushleft}
a. From Ref. \cite{Deak}

b. From Ref. \cite{Gali3}
\end{flushleft}
\label{tab:parameters}
\end{table}

\subsection{Formation energies}
\label{sec:form_en}
\begin{figure}[ht!] 
\centering
\includegraphics[width=0.9\textwidth]{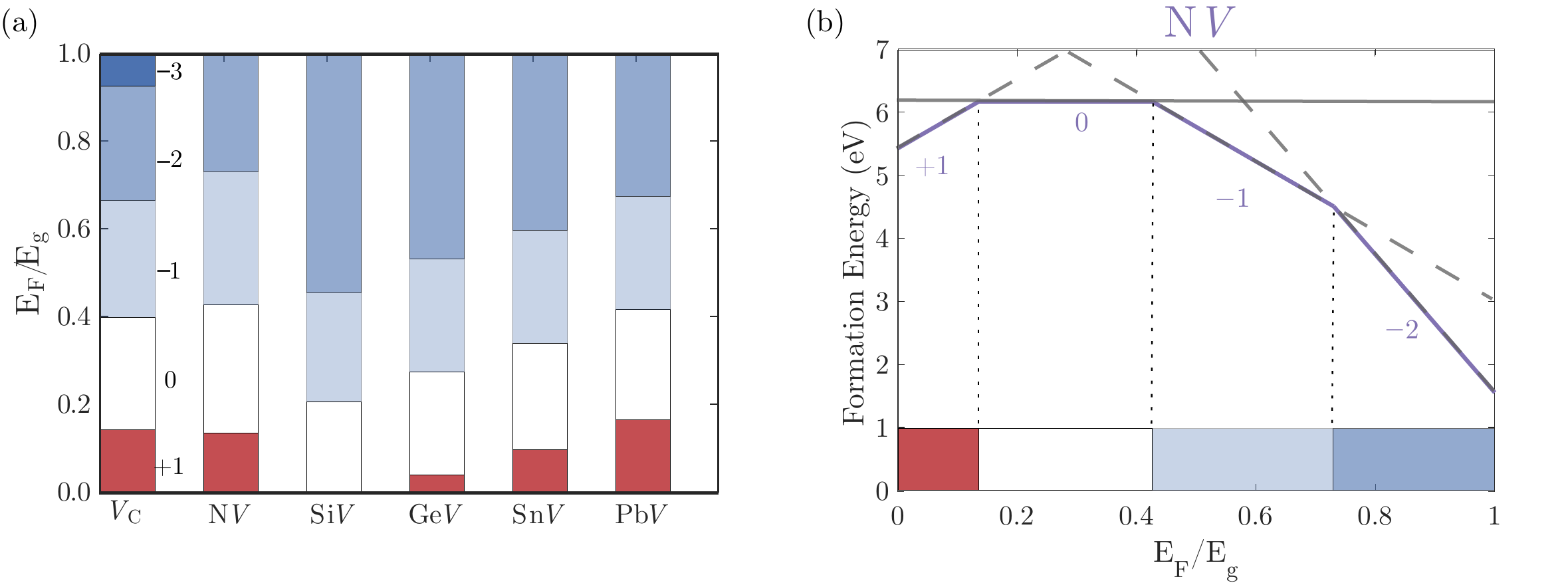}
\caption{(a) Charge transition levels shown relative to the experimental gap E$_{\rm g} = 5.47$ eV,~\cite{Isberg} for the unit cell with 216 atoms for $V_{\rm C}$, N$V$, Si$V$, Ge$V$, Sn$V$ and Pb$V$. The calculated gap was E$^{\rm DFT}_{\rm g} = 4.10$ eV.} (b) Sample formation energy plot for the N$V$ defect, from which charge transition levels are obtained. 
\label{fig:form_en_all}
\end{figure}

In order to relate the barrier height as a function of charge of $V_{\rm C}$ to the Fermi level we computed formation energies and charge transition levels for the defects, which are shown in Fig. \ref{fig:form_en_all}(a). Formation energies for a representative case, the N$V$ defect, are shown in Fig. \ref{fig:form_en_all}(b). Our results for the formation energy of the N$V$ defect in diamond are in good agreement with previous results~\cite{Deak} and our results agree with the charge state of Ge$V$ found experimentally for intrinsic diamond.~\cite{Bhaskar} 

\begin{figure}[ht!] 
\centering
\includegraphics[width=0.4\textwidth]{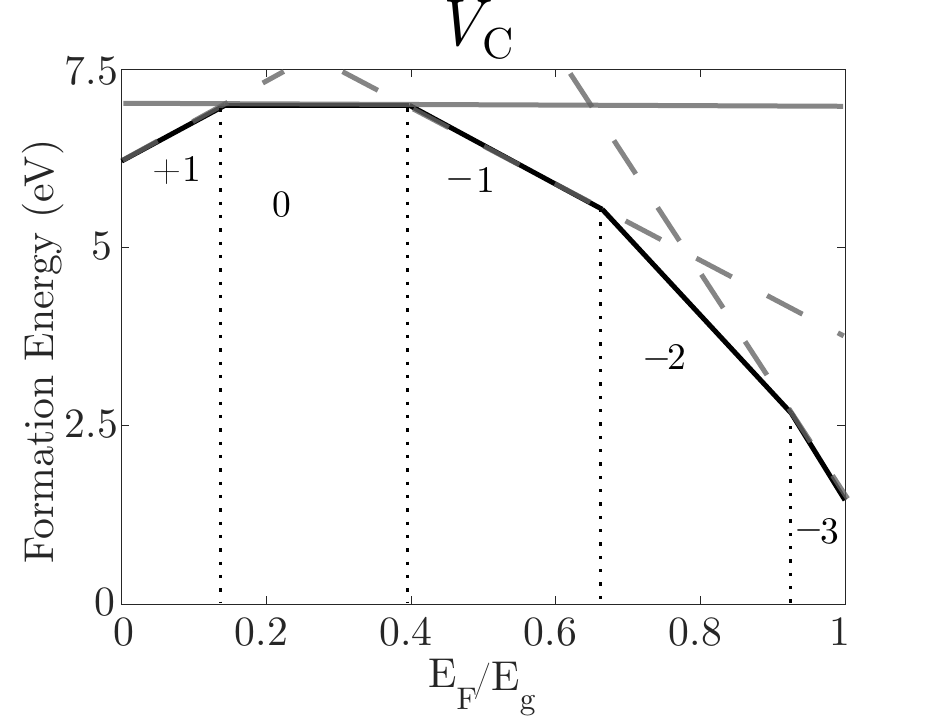}
\caption{Formation energy for the most stable charge state of the $V_{\rm C}$ as a function of the normalized Fermi level. Vertical dotted lines indicate transitions between charge states.} 
\label{fig:formVC}
\end{figure}

Our formation energy for the neutral $V_{\rm C}$, shown in Fig. \ref{fig:formVC}, of 7.00~eV is in good agreement with values from the literature of 6.74~eV and 6.86~eV, using a 128-atom supercell with the B3LYP functional and using a 456-atom hydrogen-terminated cluster through the CRYSTAL code, respectively,~\cite{Salustro2017} or 7.01~eV and 6.99~eV using 32- and 64-atom supercells, respectively, with the HSE06 functional~\cite{Zelferino2016} or 6.78~eV using Vanderbilt ultra-soft pseudopotentials with a 64-atom supercell.~\cite{Miyazaki2002} We note that the calculated formation energies for the $V_{\rm C}$ are consistent with experimental observations of the dependence of the GR1 color center (which has been assigned to the neutral defect) on the Fermi level in diamond.~\cite{Collins_2002} We also note that our transition levels for the Si$V$ are in good agreement with Ref. \cite{gali2013ab} where the charge correction scheme of Lany and Zunger was used~\cite{Lany} as well as the HSE06 functional with a 512-atom supercell. Given the lower order of approximation possible with the PBE functional, we miss the $+1$ charge state as and our transition levels are shifted slightly. Using the Kr\"oger-Vink notation to describe a transition level,~\cite{Kroger,RevModPhys.86.253} the $(-|0)$ level is at a reduced Fermi level of 0.21 compared to 0.26\cite{gali2013ab} and the $(2-|-)$ level is at a reduced Fermi level of 0.45 compared to 0.39\cite{gali2013ab}. Thus, we assign an error to our transition levels of roughly 6\% of E$_{\rm g}$.   

\subsection{Diffusivity of isolated $V_{\rm C}$ in diamond}
\label{sec:VC_iso}

\begin{figure}[ht!] 
\centering
\includegraphics[width=0.4\textwidth]{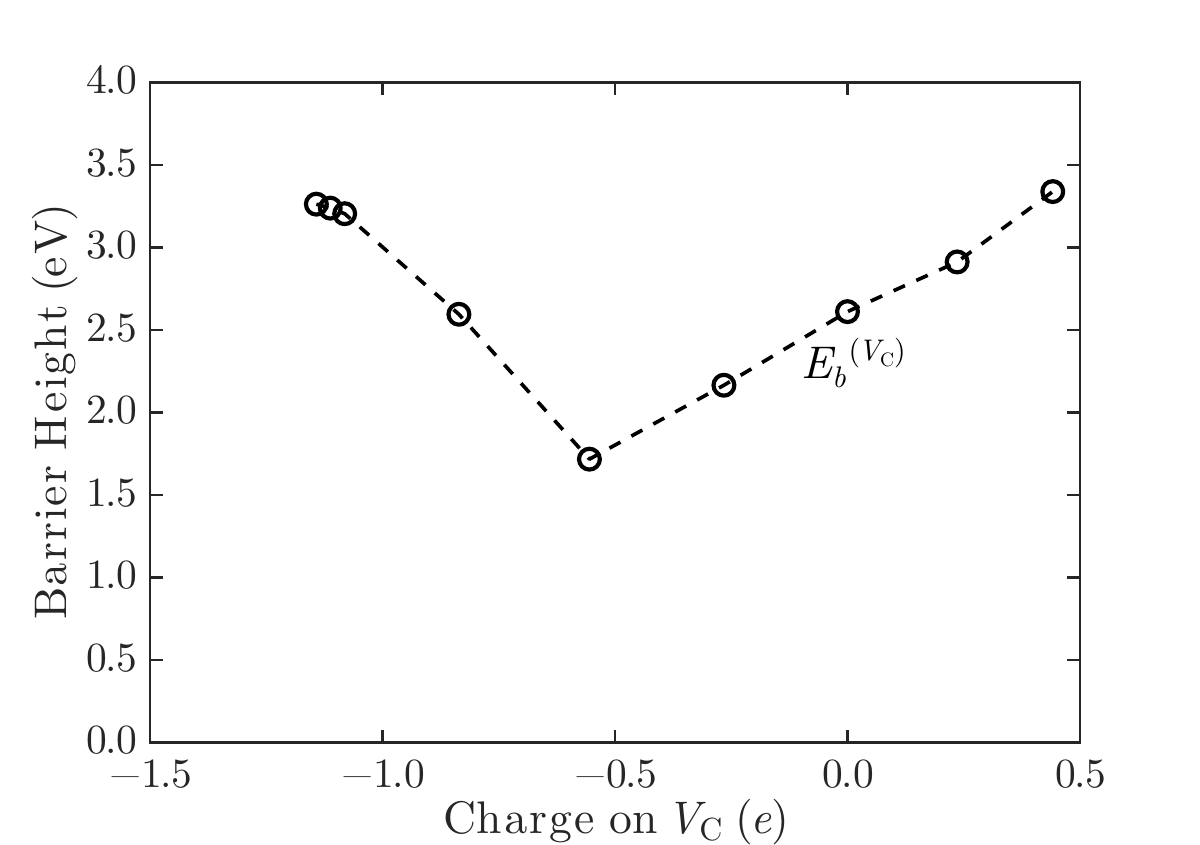}
\caption{Charge dependence of the barrier $E_b^{(V_{\rm C})}$ for diffusion for the isolated $V_{\rm C}$ (black open circles) in diamond. The lines between data points are a guide to the eye.} 
\label{fig:charge_diffVC}
\end{figure}

We now turn to the formation of complexes of the $V_{\rm C}$ with XV color centers in diamond. The first step in our model calculation is to compute the barrier height of diffusion for an isolated $V_{\rm C}$ in diamond as a function of the charge state of the vacancy $q$. We used the NEB method to compute the lowest energy diffusion pathway for an isolated $V_{\rm C}$ in diamond and then calculated the barrier height ($E_b^{V_{\rm C}}$) for that particular lowest energy pathway. We then hypothesized that this barrier height would also depend upon the charge state of the isolated $V_{\rm C}$. To test our hypothesis we performed a model calculation in which we computed the barrier height for diffusion of the $V_{\rm C}$ as a function of the charge state of the vacancy, $q$. It is important to realize that in our actual DFT calculation we cannot directly alter the charge state of the $V_{\rm C}$ in diamond. We can, however, modulate the total charge of the supercell which contains the isolated $V_{\rm C}$ and this total charge is redistributed throughout the supercell as a result of the DFT calculation. We used the DDEC6 method (see Section \ref{methods}) to compute the charge state of the $V_{\rm C}$ for a particular choice of the total charge of the supercell which contains the isolated $V_{\rm C}$ (we chose nine different values for the total charge of the system from $-6e$ to $+2e$). We note that our barriers for the neutral isolated $V_{\rm C}$ and the isolated $V_{\rm C}^-$ of about 2.6~eV and 3~eV, respectively, are in good agreement with other theoretical values of 2.6~eV and 3.5~eV.~\cite{Deak} Our results for the barrier height for an isolated $V_{\rm C}$ in diamond as a function of its charge state $q$ are illustrated in Fig. \ref{fig:charge_diffVC}. 

The V-shaped behavior of the barrier as a function of charge can be interpreted to first-order using very simple chemical and physical ideas. Starting at the right-hand side of Fig. \ref{fig:charge_diffVC} and moving towards the left corresponds to increasing the Fermi level of the host material. From a practical point of view this can be accomplished by doping diamond with electrons thus making more electrons available in the conduction band. These excess electrons are now available to the $V_{\rm C}$ and stabilize the dangling bonds surrounding it. This results in a lower barrier height since the presence of the passivating charge reduces the energy required to break and reform chemical bonds during the diffusion of $V_{\rm C}$ throughout the diamond host material. As the Fermi level is increased further, a point is reached where the $V_{\rm C}$ becomes saturated with charge and it can no longer acquire more electrons to passivate the neighboring dangling bonds. The excess charge interacts with the charge around the $V_{\rm C}$ and the ensuing Coulomb repulsion causes an increase in the diffusion barrier height. In summary, our model calculation shows that the charge of the $V_{\rm C}$ must be taken into account to determine how easy or hard it is for the defect to diffuse. This barrier height is implicitly a function of the Fermi level of the host semiconductor material since the expected charge state depends on the Fermi level position which can be altered either by doping or through the application of an external bias.

In the high-temperature limit, we obtain an attempt frequency $\nu_0 = 6.4\times10^{13}$~s$^{-1}$ for the neutral vacancy and a higher value of $\nu_0 = 16.2\times10^{13}$~s$^{-1}$ near a DDEC6 charge of $-1e$ for the vacancy, as expected from consideration of phonons traveling through an ionized medium.~\cite{stix1992waves} At $1200~^\circ$C, using $d = 1.6$~\AA, we find a diffusivity of $2.4\times10^{-15}$~m$^2\cdot$s$^{-1}$ for the neutral state using a barrier of $2.6$~eV, corresponding to a diffusion length of 49~nm$\cdot$s$^{-1}$, and a diffusivity of $2.6\times10^{-16}$~m$^2\cdot$s$^{-1}$ for the $-1$ charge state using a barrier of $3.0$~eV, corresponding to a diffusion length of 16~nm$\cdot$s$^{-1}$. Thus, diffusion is important in these systems for high temperature annealing.

\subsection{Diffusivity of $V_{\rm C}$ in the presence of X$V$ color centers in diamond}
\label{sec:vc_xv}
\begin{figure}[ht!] 
\centering
\includegraphics[width=0.99\textwidth]{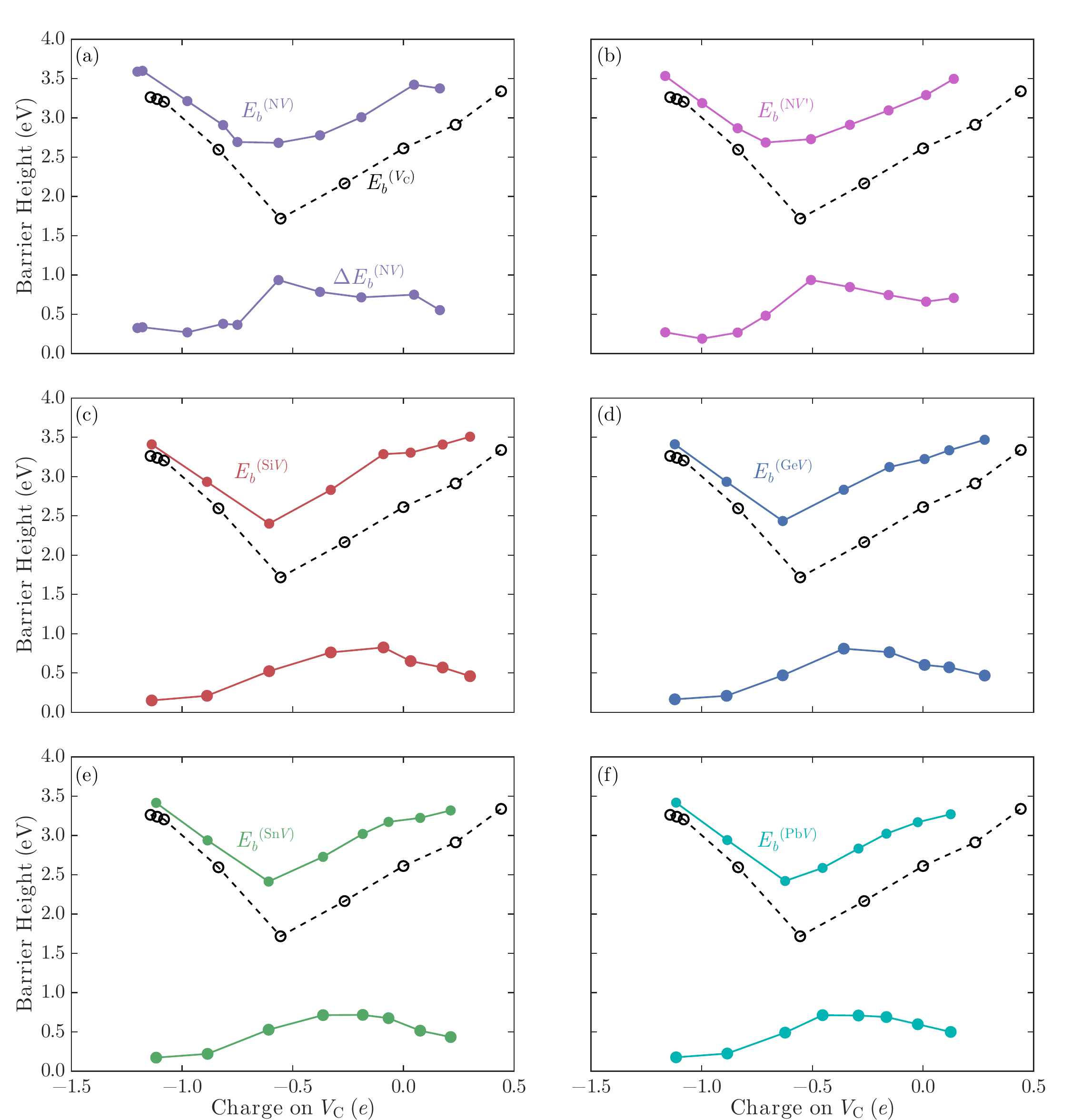}
\caption{Charge dependence of the barrier $E_b^{(V_{\rm C})}$ for diffusion of the isolated $V_{\rm C}$ (black open circles) and (colored dots) in the presence of (a) N$V$, (b) N$V'$, (c) Si$V$, (d) Ge$V$, (e) Sn$V$, and (f) Pb$V$. The barrier heights for $V_{\rm C}$ near these six latter species are $E_b^{({\rm X}V)}$ where X is a particular element. The horizontal axis is the charge on the $V_{\rm C}$. For N$V'$, the positions of the vacancy and the N atom have been switched to investigate the effect on the barrier of the orientation of the N$V$. The lines between data points are a guide to the eye. The curves at the bottom of each plot are the differences between the two barriers $\Delta E_b^{({\rm X}V)} = E_b^{({\rm X}V)} - E_b^{(V_{\rm C})}$. For the N$V$ (N$V'$), the total charge values were extended down to $-8e$ and $-7e$, respectively.} 
\label{fig:charge_diff}
\end{figure}

We next investigate the ease of diffusion of $V_{\rm C}$ if it is placed near a X$V$ color center. To address this question we used a simple model in which both the $V_{\rm C}$ and the X$V$ color center are placed in a $3\times3\times3$ supercell in such a manner as to maximize the distance between them by taking into account the periodicity of the supercell. Once this distance is determined we calculated the barrier height of diffusion for the $V_{\rm C}$ in the presence of the stationary X$V$ color center ($E_b^{({\rm X}V)}$). As previously discussed in Section III-C for the isolated $V_{\rm C}$, we expect this barrier height to depend on the charge state.  The total charge in the supercell containing both the X$V$ and the $V_{\rm C}$ was varied from $-6e$ to $+1e$ and these results are shown in Fig. \ref{fig:charge_diff}.

The barrier height for $V_{\rm C}$ in the presence of X$V$ as a function of the charge state of the $V_{\rm C}$ also has a characteristic V-shaped behavior which is similar to that of the isolated $V_{\rm C}$ for the reasons previously discussed, but it is always higher than that of the isolated $V_{\rm C}$. This can be explained by the fact that the X$V$ acts as a dopant, more precisely an as acceptor, and it removes charge from the local environment of the $V_{\rm C}$, thus making it harder to passivate the dangling bonds surrounding the $V_{\rm C}$ as it diffuses. The net effect is that the diffusion barrier height for the $V_{\rm C}$ is increased in the presence of the X$V$.

The results of Fig. \ref{fig:charge_diff} appear to suggest that carbon vacancies should tend to get stuck as they attempt to approach the X$V$, as a higher barrier for diffusion at a given location implies that it is harder to move away from that given location, and more so as the charge becomes more positive. In Section \ref{sec:therm_comp}, we provide calculations of the thermodynamics of the relevant systems which outline how isolation of color centers is favored under certain conditions.

We have considered the effect of increasing the size of the supercell from a $3\times3\times3$ multiple to a $4\times4\times4$ multiple of the conventional unit cell which corresponds to an increase in the distance between defect species from 7.62~\AA~to 9.18~\AA~(a change of about 20\%). These larger supercells did not produce any qualitative changes in the observed trends. The results nonetheless warrant some further discussion as for the case of the $4\times4\times4$ supercells, where the different defect species were placed slightly farther away from each other (and from their periodic images, as well), we found significantly higher barriers to diffusion for the cases with nonzero total charge than in the $3\times3\times3$ case. We explain this effect by noting that as defects approach one another their orbitals hybridize thus resulting in lower energies. Having explored the limiting cases of infinite distance between the species (the isolated case) and minimal distance between the species (discussed below) and two intermediate cases, we can conclude that the barrier for diffusion of the $V_{\rm C}$ depends on distance in the following manner: It increases from roughly zero as the distance between the species is increased until it reaches a maximum when the species are at the minimum distance where they no longer experience hybridization of orbitals, and then decreases until the barrier reaches the value for the isolated $V_{\rm C}$.

\subsection{Thermodynamics of forming complexes of X$V$ and $V_{\rm C}$}
\label{sec:therm_comp}
\begin{figure}[ht!] 
\centering
\includegraphics[width=0.8\textwidth]{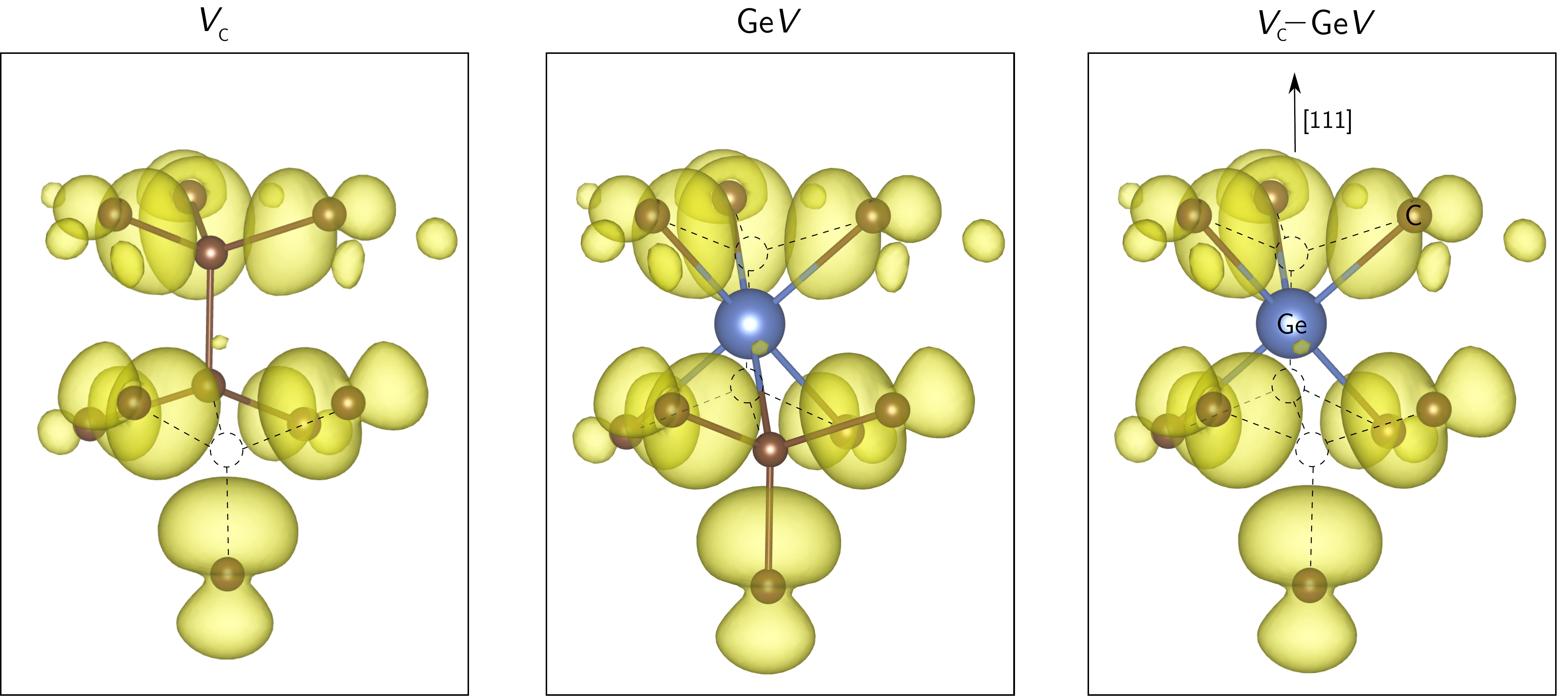}
\caption{Structure of the $V_{\rm C}$ defect (left), the isolated Ge$V$ center (middle) and the $V_{\rm C}-$Ge$V$ complex (right). Carbon atoms are indicated in brown and the Ge atom in blue. The associated charge density for the $V_{\rm C}-$Ge$V$ complex is shown to indicate where bonds are formed.} 
\label{fig:GeVCpres}
\end{figure}

We now determine the result of the reaction once a $V_{\rm C}$ in diamond reaches a color center (or is a small, but finite, distance away from one) since, as our calculations show, the $V_{\rm C}$ defects are quite mobile and can diffuse easily throughout the crystal. Based on the respective formation energies from our calculations, a $V_{\rm C}$ in close proximity to a color center is quite stable as compared to the case when the two constituents are isolated (a sample complex is shown for one of the heavier color centers, the Ge$V$, in Fig. \ref{fig:GeVCpres}), . 

To understand whether, for example, a $V_{\rm C}-$Ge$V$ complex will spontaneously form if the Ge$V$ and $V_{\rm C}$ species are brought very close together in the crystal, we compare the formation energy of the $V_{\rm C}$-Ge$V$ complex with the formation energy of a system consisting of the two isolated $V_{\rm C}$ and Ge$V$ species for the reaction below
\begin{equation}
\label{eq:reaction}
V_{\rm C} + {\rm Ge}V \rightarrow V_{\rm C}-{\rm Ge}V.
\end{equation}
Thermodynamically, if the formation energy of the system consisting of the two isolated species is higher than that of the complex, it follows that the complex should form spontaneously. Thus, the quantity of interest is $\Delta E_f$ for the $V_{\rm C} $-Ge$V$ complex relative to the isolated Ge$V$ and $V_{\rm C}$ species,
\begin{equation}
\label{eq:form_diff}
\Delta E_f = E^{V_{\rm C}-{\rm Ge}V}_f - E^{V_{\rm C}}_f - E^{{\rm Ge}V}_f,
\end{equation}
where $E^{V_{\rm C}-{\rm Ge}V}_f$, $E^{V_{\rm C}}_f$, and $E^{{\rm Ge}V}_f$ are the formation energies of the respective species indicated by the superscripts.
The calculation of this difference is complicated by the fact that the formation energies on the right side of Eq. \ref{eq:form_diff} also depend on the charge states for the individual species. Specifically,
\begin{equation}
E^{V_{\rm C}}_f = E^{V_{\rm C}}_f(q_{\scriptscriptstyle V_{\rm C}})
\end{equation}
\begin{equation}
E^{{\rm Ge}V}_f = E^{{\rm Ge}V}_f(q_{\scriptscriptstyle {\rm Ge}V})
\end{equation}
\begin{equation}
E^{V_{\rm C}-{\rm Ge}V}_f = E^{V_{\rm C}-{\rm Ge}V}_f(q_{\scriptscriptstyle V_{\rm C}-{\rm Ge}V})
\end{equation}
so that Eq. \ref{eq:form_diff} becomes
\begin{equation}
\label{eq:form_diff_chg}
\Delta E_f = E^{V_{\rm C}-{\rm Ge}V}_f(q_{\scriptscriptstyle V_{\rm C}-{\rm Ge}V}) - E^{V_{\rm C}}_f(q_{\scriptscriptstyle V_{\rm C}}) - E^{{\rm Ge}V}_f(q_{\scriptscriptstyle {\rm Ge}V}).
\end{equation}

Our goal is to calculate $\Delta E_f$ using Eq. \ref{eq:form_diff_chg}. The term $E^{V_{\rm C}}_f(q_{\scriptscriptstyle V_{\rm C}})$ is evaluated as follows. For a fixed value of $E_{\rm F}$ one simply calculates the formation energy for all charge states $q$ using Eq.~\ref{eq:form_eq} and takes the lowest value. The same procedure is applied for the Ge$V$ and the $V_{\rm C}-$Ge$V$. Let $q^*_{\scriptscriptstyle V_{\rm C}}$ be the charge state of the $V_{\rm C}$, $q^*_{\scriptscriptstyle {\rm Ge}V}$ be the charge state of the Ge$V$, and $q^*_{\scriptscriptstyle V_{\rm C}-{\rm Ge}V}$ be the charge state of the $V_{\rm C}-{\rm Ge}V$ after this first step. There is no reason why this first step should lead to charge conservation in Eq.~\ref{eq:reaction} as the two isolated species and the complex will necessarily form distinct bonding structures, which suggests different chemical behavior. Indeed, in general $q^*_{\scriptscriptstyle V_{\rm C}-{\rm Ge}V} \neq q^*_{\scriptscriptstyle V_{\rm C}}+q^*_{\scriptscriptstyle {\rm Ge}V}$. To rectify this imbalance, we add free electrons or holes to Eq. \ref{eq:reaction}. By analogy to the case of a binary compound AB where one must consider A-rich and B-rich preparation conditions in calculating the formation energy of defects, we determine whether to add free electrons or holes to the left or the right side of Eq. \ref{eq:reaction} by considering regimes where no energy is injected into the system (energy poor) or where the supply of energy is inexhaustible (energy rich). If no energy is injected into the system, the additional electrons or holes must go to the right side of Eq. \ref{eq:reaction}, while if the supply of energy is inexhaustible the additional electrons or holes go to the left side of Eq. \ref{eq:reaction}. We note that the procedure of adding free electrons or holes may reorder the formation energies, changing the species charge state corresponding to the minimum formation energy for the system. Thus, if we consider for example the regime where no energy is injected into the system, we need to consider all charge states of the $V_{\rm C}-{\rm Ge}V$ from $q_i = q^*_{\scriptscriptstyle V_{\rm C}-{\rm Ge}V}$ to $q_i = q^*_{\scriptscriptstyle V_{\rm C}}+q^*_{\scriptscriptstyle {\rm Ge}V}$ with the appropriate addition of free electrons or holes such that charge is always conserved in the reaction. We then calculate the formation energy of the corresponding systems and take the minimum of these energies. The procedure therefore ensures charge conservation and that only the thermodynamically favored systems or lowest energy systems are considered.      

We note that an energy poor forward reaction corresponds to the same set of formation energies as an energy rich reverse reaction and, similarly, an energy rich forward reaction is equivalent to an energy poor reverse reaction. Thus, if the system is energy poor the forward reaction is modeled with the energy poor forward reaction and the reverse reaction is modeled with the energy rich forward reaction. We argue that no ambiguity ensues regarding the results of the reaction as follows. Given an energy poor forward reaction where the products have energy higher than the reactants, the reaction should not proceed and there is no need to worry about whether the reverse reaction is energy rich or energy poor. If not, then the reaction will proceed. Indeed, if in the energy poor forward reaction the forward reaction product is of lower energy than the forward reaction reactants, it will necessarily be so in the energy rich forward reaction as well. Also, if the energy difference is quite great it may even be more correct to model the reverse reaction as an energy rich reaction so again there is absolutely no ambiguity. We then come to the question of energy rich reactions. If the reaction is energy rich and there is a sign flip in the difference in formation energies for the reaction when considering the energy poor counterpart (which is equivalent to the energy rich reverse reaction), then we would expect both forward and reverse reactions to occur (which is indeed exactly what we should observe if the system has enough energy). 

Let us generalize our discussion from the case of Ge$V$ to X$V$ and let $q_3 = q^*_{\scriptscriptstyle V_{\rm C}-{\rm X}V}$, $q_2 = q^*_{\scriptscriptstyle {\rm X}V}$ and $q_1 = q^*_{\scriptscriptstyle V_{\rm C}}$. Formally, for a given value of the Fermi level in energy poor conditions, let 
\begin{equation}
E^{(V_{\rm C}-{\rm X}V)}_{f_0} = \min\limits_{q_3}\left[E^{(V_{\rm C}-{\rm X}V)}_f(q_3)\right]+[q_3-(q_1+q_2)]\cdot(E_{\rm CBM}-E_{\rm F})
\end{equation}
where we are minimizing over the variable $q_3$ in the first term (yielding the charge state labeled in the same way), the last term is included to ensure charge conservation for the case $q_3 > q_1+q_2$ by adding the energy of $q_3-(q_1+q_2)$ free electrons to the right side of Eq. \ref{eq:form_diff_chg} and $E_{\rm CBM}$ is the absolute position of the conduction band minimum, or  
\begin{equation}
E^{(V_{\rm C}-{\rm X}V)}_{f_0} = \min\limits_{q_3}\left[E^{(V_{\rm C}-{\rm X}V)}_f(q_3)\right]+[(q_1+q_2)-q_3]\cdot(E_{\rm F}-E_{\rm VBM})
\end{equation}
where the last term is included to ensure charge conservation for the case $q_3 < q_1+q_2$ by adding the energy of $(q_1+q_2)-q_3$ free holes to the right side of Eq. \ref{eq:form_diff_chg}. In energy rich conditions, let 
\begin{equation}
E^{(V_{\rm C}+{\rm X}V)}_{f_0}  = \min\limits_{q_1,q_2}\left[E^{(V_{\rm C})}_f(q_1)+E^{({\rm X}V)}_f(q_2)\right]+[(q_1+q_2)-q_3]\cdot(E_{\rm CBM}-E_{\rm F})
\end{equation}
where we are minimizing over $q_1$ and $q_2$ in the first term (yielding similarly labeled charge states) and the last term is included to ensure charge conservation for the case $q_3 < q_1+q_2$ by subtracting the energy of $(q_1+q_2) - q_3$ free electrons from the right side of Eq. \ref{eq:form_diff_chg}, or
\begin{equation}
E^{(V_{\rm C}+{\rm X}V)}_{f_0}  = \min\limits_{q_1,q_2}\left[E^{(V_{\rm C})}_f(q_1)+E^{({\rm X}V)}_f(q_2)\right]+[q_3-(q_1+q_2)]\cdot(E_{\rm F}-E_{\rm VBM})
\end{equation}
where the last term is included to ensure charge conservation for the case $q_3 > q_1+q_2$ by subtracting the energy of $q_3-(q_1+q_2)$ free holes from the right side of Eq. \ref{eq:form_diff_chg}.
Given possible reordering of the formation energies with the addition of free electrons or holes, let $i = 0,...,\max\left[q_3-(q_1+q_2),0\right]$ and consider,
\begin{equation}
E^{(V_{\rm C}-{\rm X}V)}_{f_i} = E^{(V_{\rm C}-{\rm X}V)}_f(q_3-i)+[q_3-i-(q_1+q_2)]\cdot(E_{\rm CBM}-E_{\rm F})
\end{equation}
or let $i = 0,...,\max\left[(q_1+q_2)-q_3,0\right]$ and consider,
\begin{equation}
E^{(V_{\rm C}-{\rm X}V)}_{f_i} = E^{(V_{\rm C}-{\rm X}V)}_f(q_3+i)+[(q_1+q_2)-i-q_3]\cdot(E_{\rm F}-E_{\rm VBM})
\end{equation}
for energy poor conditions. For energy rich conditions, 
let $i = 0,...,\max\left[(q_1+q_2)-q_3,0\right]$ and consider, 
\begin{equation}
E^{(V_{\rm C}+{\rm X}V)}_{f_i}  = \min\limits_{q_1'+q_2' = q_1+q_2-i}\left[E^{(V_{\rm C})}_f(q'_1)+E^{({\rm X}V)}_f(q'_2)\right]+[(q_1+q_2)-i-q_3]\cdot(E_{\rm CBM}-E_{\rm F})
\end{equation}
where we are now minimizing over the variables $q_1'$ and $q_2'$ in the first term or let $i = 0,...,\max\left[q_3-(q_1+q_2),0\right]$ and consider, 
\begin{equation}
E^{(V_{\rm C}+{\rm X}V)}_{f_i}  = \min\limits_{q_1'+q_2' = q_1+q_2+i}\left[E^{(V_{\rm C})}_f(q'_1)+E^{({\rm X}V)}_f(q'_2)\right]+[q_3-i-(q_1+q_2)]\cdot(E_{\rm F}-E_{\rm VBM}).
\end{equation}
To ensure we have the lowest energy, for energy rich conditions we take 
\begin{align}
\varepsilon^{(V_{\rm C}+{\rm X}V)}_f &= \min\limits_{i}\left[E^{(V_{\rm C}+{\rm X}V)}_{f_i}\right],\label{eq:newepsmin1}\\
\varepsilon^{(V_{\rm C}-{\rm X}V)}_f &= E^{(V_{\rm C}-{\rm X}V)}_f(q_3).\label{eq:newepsplus1}
\end{align}
For energy poor conditions we take
\begin{align}
\varepsilon^{(V_{\rm C}-{\rm X}V)}_f &= \min\limits_{i}\left[E^{(V_{\rm C}-{\rm X}V)}_{f_i}\right], \label{eq:newepsmin2}\\
\varepsilon^{(V_{\rm C}+{\rm X}V)}_f &= E^{(V_{\rm C})}_f(q_1)+E^{({\rm X}V)}_f(q_2),\label{eq:newepsplus2}
\end{align}
where a dependence on the Fermi level is implied. Using Eqs. \ref{eq:newepsmin1}$-$\ref{eq:newepsplus2} ensures both charge conservation and the consideration of only the thermodynamically favored products and reactants. We have analogously also calculated the quantity $\varepsilon^{(V_{\rm C}--{\rm X}V)}_f$ for the separation of $V_{\rm C}$ and ${\rm X}V$ by a finite distance.

\begin{figure}[ht!] 
\centering
\includegraphics[width=0.9\textwidth]{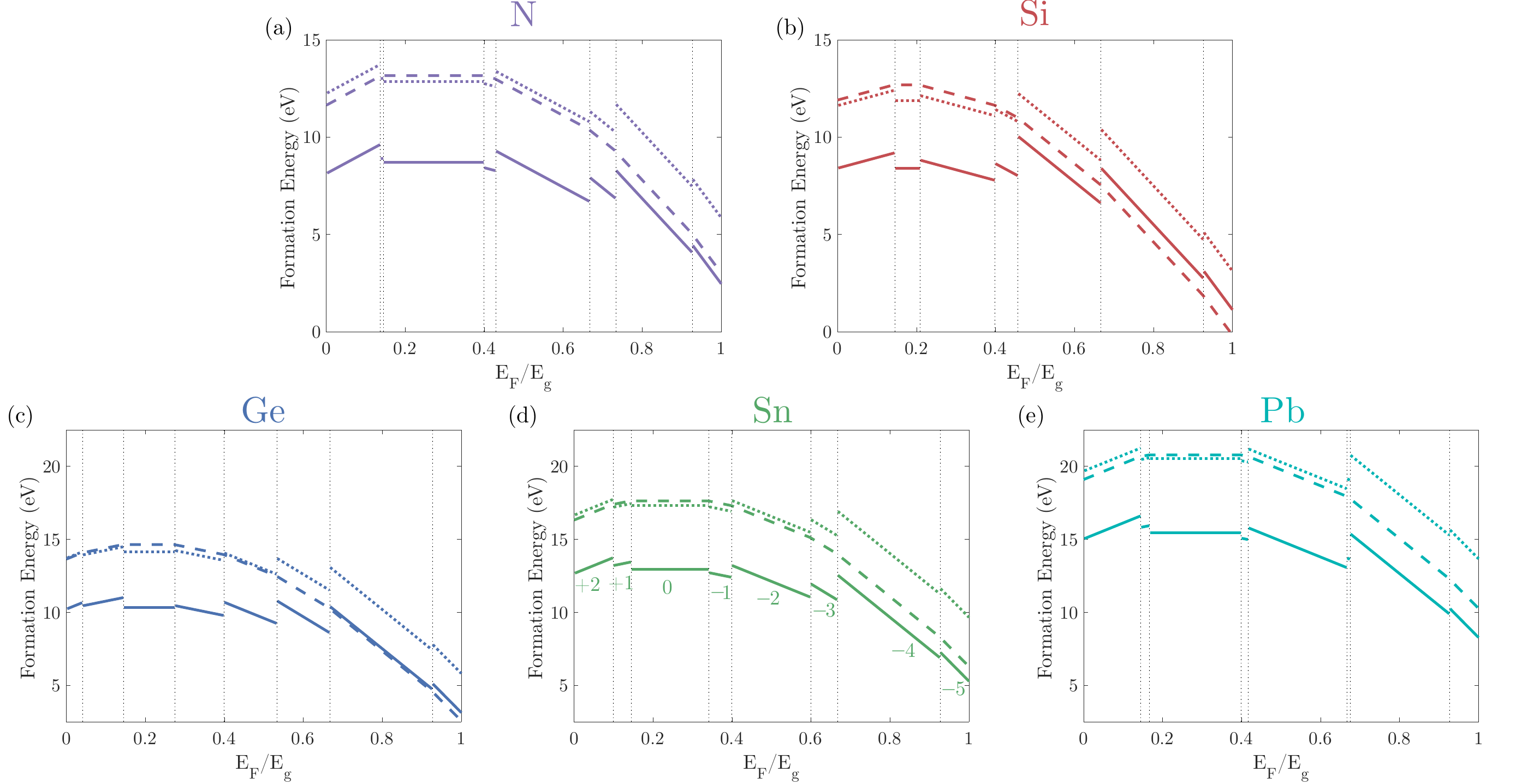}
\caption{ Formation energy $\varepsilon^{(V_{\rm C}-{\rm X}V)}_f$ for the $V_{\rm C}-$X$V$ complex (solid lines), $\varepsilon^{(V_{\rm C}+{\rm X}V)}_f$ for $V_{\rm C}$ and X$V$ separated by an infinite distance (dashed lines) and $\varepsilon^{(V_{\rm C}{\rm -- X}V)}_f$ for $V_{\rm C}$ and X$V$ separated by a finite distance (dotted lines), for (a) X = N, (b) X = Si, (c) X = Ge, (d) X = Sn, (e) X = Pb. Vertical dotted lines indicate transitions between charge states. Formation energies are shown for the regime where no energy is injected into the system. The inexhaustible supply of energy regime is not displayed.} 
\label{fig:cvpgev}
\end{figure}

In Fig. \ref{fig:cvpgev} we show the formation energy results for various $V_{\rm C}-$X$V$ complexes. These results show that the $V_{\rm C}$ easily forms complexes with the X$V$ for most Fermi level values, ultimately depleting the yield of isolated X$V$ color centers in diamond. We note that this spontaneity of complex formation assumes that the two species move from infinity to minimal separation without passing through any intermediate separation distances. We have only displayed results for the limiting regime where no energy is injected into the system, as that regime demonstrates the best chance of not forming complexes. The limiting regime where the supply of energy is inexhaustible has uniformly higher formation energies for the system consisting of the separated X$V$ and $V_{\rm C}$ than for the system of the X$V-V_{\rm C}$ for all species X and all values of the Fermi level. We explain the propensity to form complexes by the fact that hybridization is generally energetically favorable process, though, for the atoms with fewer electronic orbitals such as Si and Ge, the bonding orbitals do not appear to be extended enough to support the formation of complexes once they have already been saturated with sufficient negative charge (a non-intuitive result as one might expect complexes to always form). This inability to support the formation of complexes for high Fermi level values is reflected in the fact that the solid lines jump above the dashed lines in Fig. \ref{fig:cvpgev}(b) and (c). The significantly larger electronegativity of nitrogen explains its ability to continue forming complexes at high Fermi level values despite having few electronic orbitals. 

Given that the two species must pass through intermediate distances in approaching one another to minimal separation from infinity, in Fig. \ref{fig:cvpgev} we have also explored the case of finite separation by a distance of 7.62~\AA~(the maximum separation between defects in a $3\times3\times3$ supercell). We emphasize that it is the difference between the lines of interest and the dashed lines in a given plot that is physically relevant to the process of the X$V$ and $V_{\rm C}$ approaching each other to some distance given the existence of the X$V$ and $V_{\rm C}$ species at infinity. We find a very intuitive result which may explain why we can actually experimentally observe isolated charged color centers. The result is that for Fermi level values corresponding to two similarly charged species, the energy associated with the system of the two species increases as the two species approach the finite separation from infinity. We can understand this result by recognizing the influence of repulsion of similarly charged species. Indeed, for Fermi level values where one or both of the species are neutral, the energy is no longer higher as the two species approach the finite separation from infinity. As above, we expect the energies to increase up until the minimum distance where hybridization of orbitals is no longer experienced and then decrease thereafter, other than when one or more of the species is neutral in which case polarization effects must be taken into account. Based on Fig. \ref{fig:charge_diff}, the fact that the $V_{\rm C}$ are more likely to get stuck as they approach the color center for low Fermi level values and thus be impeded in further motion may explain the ability to form neutral color centers despite thermodynamic results suggesting the contrary. Given that the justification for the $V_{\rm C}$ getting stuck relates to the removal of the charge needed to passivate dangling bonds as the $V_{\rm C}$ diffuses, the isolation of negatively charged color centers near neutral $V_{\rm C}$ should not be as favored as the isolation of neutral color centers near neutral $V_{\rm C}$ since the negatively charged color centers could donate electrons to passivate the $V_{\rm C}$ during diffusion.

\subsection{The effect of $V_{\rm C}$ on acoustic phonons}
\label{sec:phons}
We finally turn to the question of how best to deal with a mechanism that can be detrimental to the performance of X$V^-$ color centers in diamond, namely the effect of acoustic phonons. As the mass of the X element in a X$V^-$ color center is increased in going from Si to Ge to Sn, and finally to Pb, the enhanced spin-orbit coupling of the heavier elements causes a larger energy level splitting in the ground state.~\cite{Iwasaki2} This observation is relevant as it is well known that acoustic phonons can interact with the lighter X$V^-$ color centers on the energy scale of the ground state or orbital splitting frequency thus disrupting the coherence times of the spin states of these X$V^-$ color centers at higher temperatures.~\cite{Jahnke,Pingault} Any proposal to increase the spin coherence time of color centers must clearly address the matter of the prevalence of acoustic phonons in the host material. It has been suggested that by decreasing the phonon density of states (PDOS) $g(\omega)$ about the orbital splitting frequency of X$V^-$ color centers, one can enhance their spin coherence times.~\cite{Pingault} For the X$V^-$ color centers under investigation, these frequencies lie below 2.5 THz~\cite{Pingault,Gali4,Iwasaki2,Siyushev,Trusheim}. 

Let us consider a simple model in which we propose a way to decrease the number of acoustic phonons in the bulk material by modulating the $V_{\rm C}$ density in the host material. We will first show that for a simple cubic lattice the PDOS is proportional to the inverse of the primitive unit cell volume V$_{pc}$ as the frequency goes to zero. Explicitly, using the Debye model for a 3D crystal, the PDOS per unit volume $\tilde{g}(\omega)$ is~\cite{kaxiras2003atomic,ashcroft1976solid}
\begin{equation}
    \tilde{g}(\omega) = \frac{3}{2\pi^2}\frac{\omega^2}{v^3},
    \label{eq:phon_dens}
\end{equation}
where $v = \omega_{\rm D}\left(\frac{\Omega}{2\pi^2N_{ph}}\right)^{(1/3)}$ is the speed of sound, $\omega_{\rm D}$ is the Debye frequency, $N_{ph}$ is the number of available acoustic phonon modes ($N_{ph} = 3N$, where $N$ is the number of primitive unit cells in the crystal), and $\Omega$ is the volume of the crystal. Upon multiplying Eq. \ref{eq:phon_dens} by the volume of the crystal $\Omega$, we obtain an expression for the PDOS,
\begin{equation}
    g(\omega) = 9N\frac{\omega^2}{\omega_{\rm D}^3}.
\end{equation}
Since the number of primitive unit cells, $N$, is inversely proportional to V$_{pc}$ for a fixed crystal size, we have proved our desired result. We note here that such an approximation is valid for small frequencies, where the phonon dispersion relation can be approximated as linear. 

Then, if the size of the primitive unit cell can be increased relative to the size of the crystal, the density of phonons for low frequencies would decrease. A simple way to increase V$_{pc}$ would be to have a diamond lattice with a very small but nonzero number of defects. As the color center already constitutes one defect, the ideal scenario would be to have no $V_{\rm C}$. Essentially, if the number of defects were zero, the unit cell would reduce to the proper unit cell of diamond, but if there is a single defect in the crystal, the unit cell should then increase to the size of the crystal. Roughly speaking, for uniformly distributed defects in the crystal the size of the unit cell should be about the size of the crystal divided by the number of defects and the phonon density of states near zero frequency should then increase or decrease correspondingly. In a different problem entirely equivalent to the one discussed here, it has been shown that the waves scattered off a system with subwavelength patterning can be solved with good accuracy by modeling each region of the system as one being reproduced periodically throughout the system, solving for the waves from that region, and then stitching the solutions together.~\cite{Pestourie} We therefore argue that the phonon behavior should be determined to acceptable accuracy based on local densities of defects in the physically realized systems.  

\begin{figure}[ht!] 
\centering
\includegraphics[width=0.5\textwidth]{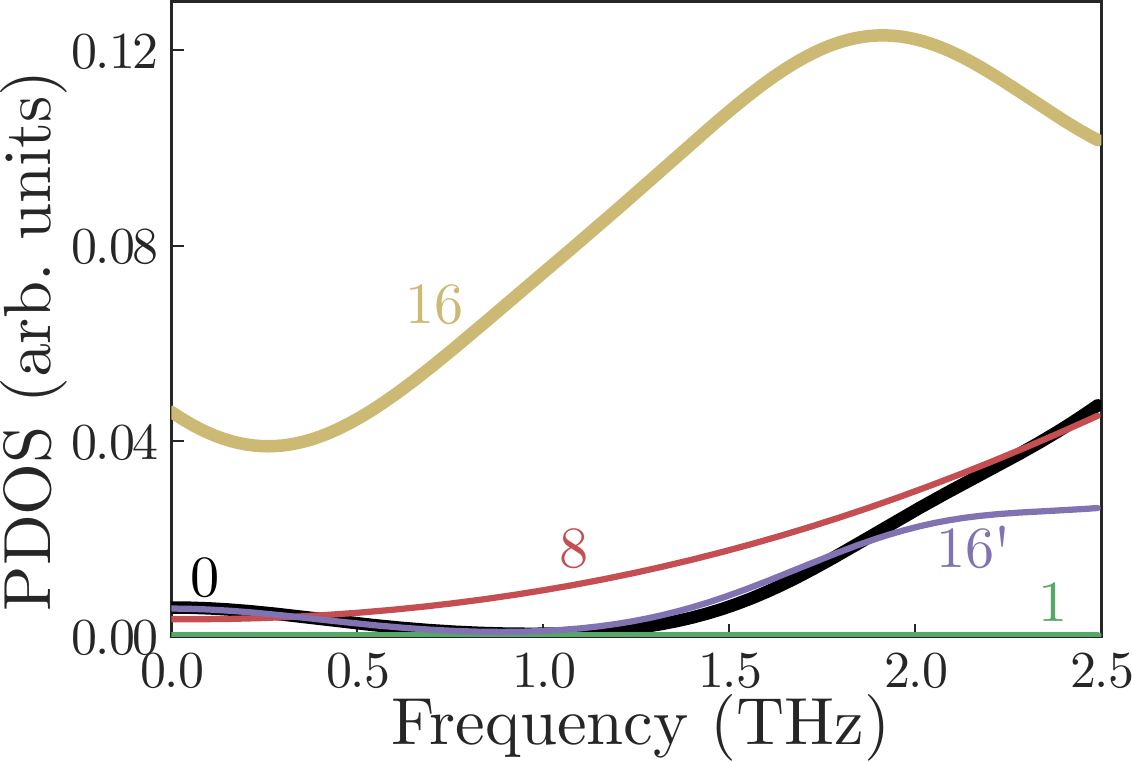}
\caption{Phonon density of states for various $V_{\rm C}$ concentrations and positions in the diamond lattice, as described in the text. The labels of the various curves correspond to the number of $V_{\rm C}$ defects in a 64-atom supercell. For supercell configurations that have a nonzero density of $V_{\rm C}$ and that have the $V_{\rm C}$ roughly uniformly distributed in the supercell, we observe an increase in the PDOS at low frequency as the density is increased. The primed number indicates a configuration where the distribution of $V_{\rm C}$ deviated significantly from uniform.} 
\label{fig:phon}
\end{figure}

In Fig. \ref{fig:phon}, using the Phonopy code,~\cite{Togo} we investigate the effect of changing the density of the $V_{\rm C}$ in the diamond lattice on the PDOS by considering five different 64-atom supercells of varying defect densities: a supercell with no vacancies, two distinct supercells each containing 16 vacancies located in two different configurations, a supercell with 8 vacancies, and a supercell with 1 vacancy. For the supercell with 1 vacancy (green), the vacancy was placed at the carbon site located at (0, 0, 0) in the supercell. For the supercell with 8 vacancies (red), the vacancies were generated by placing a vacancy at the carbon site located at the (0, 0, 0) position of the conventional unit cell of diamond which was then periodically repeated throughout the supercell to generate all eight vacancies. As a representative test of the idea that it is indeed the size of the smallest effectively periodic unit that affects the PDOS, we also created two supercells with 16 vacancies. In one, the vacancies were placed at the carbon sites located at (0, 0, 0) and (0.25, 0.25, 0.25) (purple) positions of the conventional unit cell of diamond and in the other the vacancies were placed at the carbon sites located at  (0, 0, 0) and (0.5, 0.5, 0) (gold) positions of the conventional unit cell of diamond, both in units of the lattice constant of the conventional unit cell. These two conventional unit cells were then periodically repeated throughout the supercell to generate two distinct configurations of 16 vacancies. We see that the phonon density of states is effectively zero at low frequencies for the smallest nonzero defect density $\frac{1~V_{\rm C}}{\rm supercell}$ and increases as the density of defects is increased. The pronounced increase for a curve corresponding to $\frac{16~V_{\rm C}}{\rm supercell}$ (in gold) over the other curve corresponding to $\frac{16~V_{\rm C}}{\rm supercell}$ (in purple) is due to the following effect: For the latter structure, the second vacancy in a conventional unit cell was placed as close as possible to the first vacancy resulting in very little change in the effective periodicity of the cell, while for the former structure it was placed as far as possible from the first vacancy, thus roughly halving the conventional unit cell along certain directions with a corresponding increase in the phonon density of states at low frequencies. 

\begin{figure}[ht!] 
\centering
\includegraphics[width=0.5\textwidth]{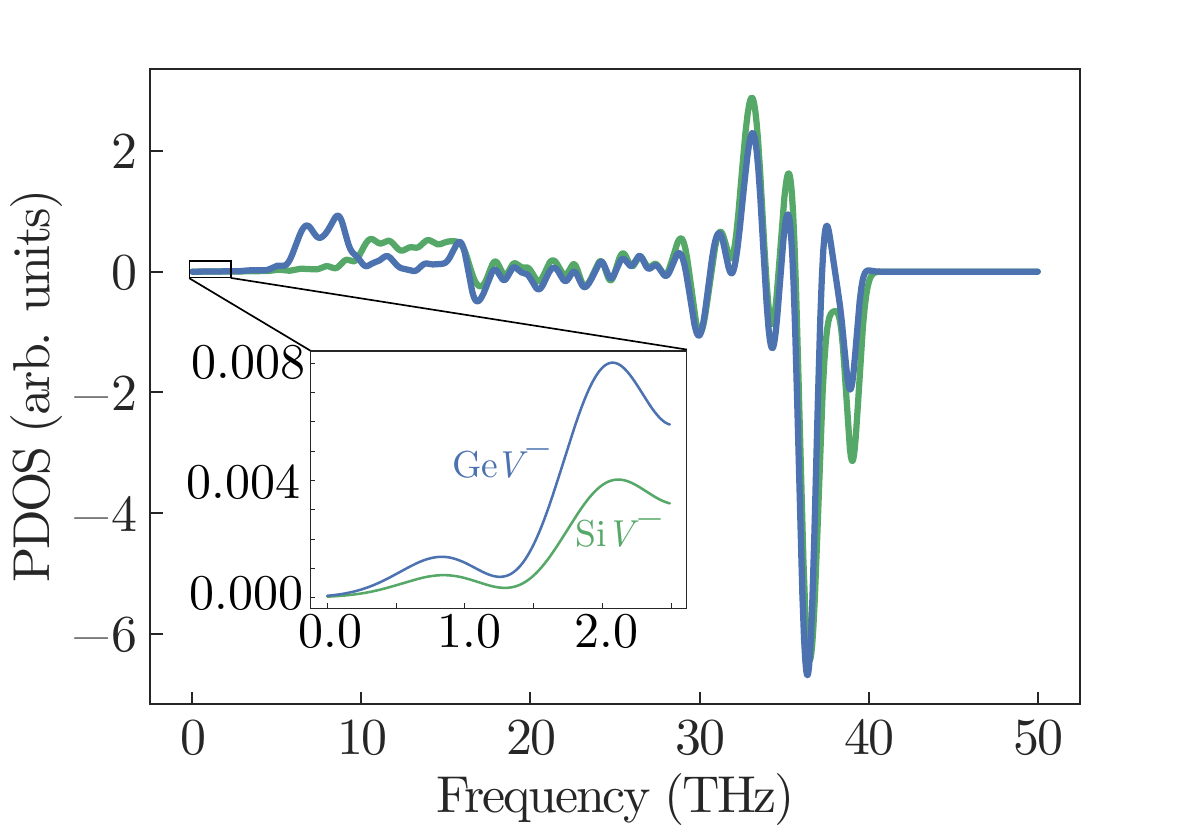}
\caption{ Difference between the phonon density of states for a $3\times3\times3$ supercell system containing a single Si$V^-$ and for stoichiometric diamond (green) and for a $3\times3\times3$ supercell system containing a single Ge$V^-$ and for stoichiometric diamond (blue). One notes that for the low frequencies (until roughly 10 THz), the Ge$V^-$ curve is consistently higher than the Si$V^-$ curve.} 
\label{fig:phon2}
\end{figure}

In Fig. \ref{fig:phon2}, again using the Phonopy code,~\cite{Togo} we have examined the phonon density of states using a supercell consisting of either a single Si$V^-$ or a single Ge$V^-$. We propose to use the results of Fig. \ref{fig:phon2} to explain the fact that the Ge$V^-$ has a spin coherence time equal to or less than that of the Si$V^-$ despite having a larger ground state splitting.~\cite{Zhou_2018} Given that the systems we are dealing with consist of electronic spins, for a total spin $S$, the corresponding system will involve a number of entangled electrons $n_{e}$ at least equal to $2S$. Thus, since the Larmor frequency is inversely proportional to the mass of the system, under an external field the spin of the system will precess with a Larmor frequency suppressed by a factor of $n_{e}m_e$, where $m_e$ is the mass of the electron. As the total spin increases, $n_{e}$ increases implying that the spin will rotate out of phase more slowly. Indeed, Childress \textit{et al.}~\cite{Childress} found suppressed decoherence when N$V^-$ spins were entangled with the much more massive $^{13}$C nuclear spins. One might then imagine that for systems with the same total spin and the same environment, the bulk of the difference in the spin coherence time might be explained by deferring to ground state splitting. However, as alluded to above, this explanation does not work for the Si$V^-$ and Ge$V^-$. We make a further simple observation that for a field that is periodically varying with frequency $\omega_{\rm B}$, the change in the angle of the spin with time will be suppressed by a factor of $\omega_{\rm B}$, obtained by integrating the expression for the Larmor frequency. Thus, the fluctuating fields associated with phonons with high enough frequencies will be unable to rotate the color center spin completely out of phase, suggesting we should predominantly consider the low frequency regime in comparing the Ge$V^-$ and Si$V^-$ phonon densities. Indeed, in that regime, Ge$V^-$ has a higher phonon density than Si$V^-$ which we propose would explain its equal or shorter spin coherence time despite the larger ground state splitting.  


\section{\label{diffusivity}CONCLUSIONS}
Single carbon vacancies can have a deleterious effect on the spin and optical properties of isolated Group IV color centers (X$V$) in diamond, where X = Si, Ge, Sn, and Pb and $V$ is a carbon vacancy. In an effort to understand the thermodynamics and kinetics of how a single carbon vacancy-X$V$ complex ($V_{\rm C}-$X$V$) is created, we used DFT to compute the formation energies and charge transition levels for both $V_{\rm C}$ and each the four individual Group IV color centers, the diffusion barrier height $E_b^{(V_{\rm C})}$ for isolated $V_{\rm C}$, and the diffusion barrier $E_b^{({\rm X}V)}$ for $V_{\rm C}$ in the proximity in each of these four X$V$ color centers. We have also demonstrated the dependence of both $E_b^{(V_{\rm C})}$and $E_b^{({\rm X}V)}$ on the Fermi level of the host diamond material and have computed the formation energy for a $V_{\rm C}-$X$V$ complex while conserving charge. All of this information can be used as an experimental guide on how to dope or apply an electric bias to samples in order to reversibly tune the Fermi level to the appropriate regime where diffusion of $V_{\rm C}$ towards a given color center would be impeded or rendered energetically unfavorable for the formation of a $V_{\rm C}-$X$V$ complex. By better isolating color centers from the $V_{\rm C}$ defects and thus reducing the prevalence of acoustic phonons, we expect that the spin coherence times for X$V$ color centers would be enhanced.  Finally, we note that all of these results involving X$V$ color centers are applicable to N$V$ centers in diamond. 

{\bf ACKNOWLEDGMENTS:}
RKD gratefully acknowledges financial support from the IACS Student Scholarship. We also acknowledge support by the STC Center for Integrated Quantum Materials, NSF Grant No. DMR-1231319, and the NSF-ECCS-1748106 EAGER grant. This work used computational resources of the Extreme Science and Engineering Discovery Environment (XSEDE), which is supported by National Science Foundation Grant Number ACI-1548562,~\cite{Towns} on Stampede2 at TACC through allocation TG-DMR120073, and of the National Energy Research Scientific Computing Center (NERSC), a U.S. Department of Energy Office of Science User Facility operated under Contract No. DE-AC02-05CH11231. We would like to thank Marko Lon\ifmmode \check{c}\else \v{c}\fi{}ar for bringing our attention to Ref.~\cite{Trusheim} when it was in the form of a preprint. We also thank Diana Prado Lopes Aude Craik, Smarak Maity, Linbo Shao, and Michael Walsh for very useful experimental discussions on the creation of $V_{\rm C}$ in diamond by ion implantation and on the properties of N$V$ and Group IV color centers as well. Finally, we thank Pratibha Dev for useful theoretical discussions regarding the convergence of the results. 

\bibliography{refs_GeV}

\begin{thebibliography}{94}%
\makeatletter
\providecommand \@ifxundefined [1]{%
 \@ifx{#1\undefined}
}%
\providecommand \@ifnum [1]{%
 \ifnum #1\expandafter \@firstoftwo
 \else \expandafter \@secondoftwo
 \fi
}%
\providecommand \@ifx [1]{%
 \ifx #1\expandafter \@firstoftwo
 \else \expandafter \@secondoftwo
 \fi
}%
\providecommand \natexlab [1]{#1}%
\providecommand \enquote  [1]{``#1''}%
\providecommand \bibnamefont  [1]{#1}%
\providecommand \bibfnamefont [1]{#1}%
\providecommand \citenamefont [1]{#1}%
\providecommand \href@noop [0]{\@secondoftwo}%
\providecommand \href [0]{\begingroup \@sanitize@url \@href}%
\providecommand \@href[1]{\@@startlink{#1}\@@href}%
\providecommand \@@href[1]{\endgroup#1\@@endlink}%
\providecommand \@sanitize@url [0]{\catcode `\\12\catcode `\$12\catcode
  `\&12\catcode `\#12\catcode `\^12\catcode `\_12\catcode `\%12\relax}%
\providecommand \@@startlink[1]{}%
\providecommand \@@endlink[0]{}%
\providecommand \url  [0]{\begingroup\@sanitize@url \@url }%
\providecommand \@url [1]{\endgroup\@href {#1}{\urlprefix }}%
\providecommand \urlprefix  [0]{URL }%
\providecommand \Eprint [0]{\href }%
\providecommand \doibase [0]{http://dx.doi.org/}%
\providecommand \selectlanguage [0]{\@gobble}%
\providecommand \bibinfo  [0]{\@secondoftwo}%
\providecommand \bibfield  [0]{\@secondoftwo}%
\providecommand \translation [1]{[#1]}%
\providecommand \BibitemOpen [0]{}%
\providecommand \bibitemStop [0]{}%
\providecommand \bibitemNoStop [0]{.\EOS\space}%
\providecommand \EOS [0]{\spacefactor3000\relax}%
\providecommand \BibitemShut  [1]{\csname bibitem#1\endcsname}%
\let\auto@bib@innerbib\@empty
\bibitem [{\citenamefont {Doherty}\ \emph {et~al.}(2013)\citenamefont
  {Doherty}, \citenamefont {Manson}, \citenamefont {Delaney}, \citenamefont
  {Jelezko}, \citenamefont {Wrachtrup},\ and\ \citenamefont
  {Hollenberg}}]{Doherty2013nitrogen}%
  \BibitemOpen
  \bibfield  {author} {\bibinfo {author} {\bibfnamefont {M.~W.}\ \bibnamefont
  {Doherty}}, \bibinfo {author} {\bibfnamefont {N.~B.}\ \bibnamefont {Manson}},
  \bibinfo {author} {\bibfnamefont {P.}~\bibnamefont {Delaney}}, \bibinfo
  {author} {\bibfnamefont {F.}~\bibnamefont {Jelezko}}, \bibinfo {author}
  {\bibfnamefont {J.}~\bibnamefont {Wrachtrup}}, \ and\ \bibinfo {author}
  {\bibfnamefont {L.~C.}\ \bibnamefont {Hollenberg}},\ }\bibfield  {title}
  {\enquote {\bibinfo {title} {The nitrogen-vacancy colour centre in
  diamond},}\ }\href {\doibase https://doi.org/10.1016/j.physrep.2013.02.001}
  {\bibfield  {journal} {\bibinfo  {journal} {Phys. Rep.}\ }\textbf {\bibinfo
  {volume} {528}},\ \bibinfo {pages} {1} (\bibinfo {year} {2013})}\BibitemShut
  {NoStop}%
\bibitem [{\citenamefont {De\'ak}\ \emph {et~al.}(2014)\citenamefont {De\'ak},
  \citenamefont {Aradi}, \citenamefont {Kaviani}, \citenamefont {Frauenheim},\
  and\ \citenamefont {Gali}}]{Deak}%
  \BibitemOpen
  \bibfield  {author} {\bibinfo {author} {\bibfnamefont {P.}~\bibnamefont
  {De\'ak}}, \bibinfo {author} {\bibfnamefont {B.}~\bibnamefont {Aradi}},
  \bibinfo {author} {\bibfnamefont {M.}~\bibnamefont {Kaviani}}, \bibinfo
  {author} {\bibfnamefont {T.}~\bibnamefont {Frauenheim}}, \ and\ \bibinfo
  {author} {\bibfnamefont {A.}~\bibnamefont {Gali}},\ }\bibfield  {title}
  {\enquote {\bibinfo {title} {{Formation of NV centers in diamond: A
  theoretical study based on calculated transitions and migration of nitrogen
  and vacancy related defects}},}\ }\href {\doibase 10.1103/PhysRevB.89.075203}
  {\bibfield  {journal} {\bibinfo  {journal} {Phys. Rev. B}\ }\textbf {\bibinfo
  {volume} {89}},\ \bibinfo {pages} {075203} (\bibinfo {year}
  {2014})}\BibitemShut {NoStop}%
\bibitem [{\citenamefont {Gali}\ \emph {et~al.}(2009)\citenamefont {Gali},
  \citenamefont {Janz\'en}, \citenamefont {De\'ak}, \citenamefont {Kresse},\
  and\ \citenamefont {Kaxiras}}]{Gali2}%
  \BibitemOpen
  \bibfield  {author} {\bibinfo {author} {\bibfnamefont {A.}~\bibnamefont
  {Gali}}, \bibinfo {author} {\bibfnamefont {E.}~\bibnamefont {Janz\'en}},
  \bibinfo {author} {\bibfnamefont {P.}~\bibnamefont {De\'ak}}, \bibinfo
  {author} {\bibfnamefont {G.}~\bibnamefont {Kresse}}, \ and\ \bibinfo {author}
  {\bibfnamefont {E.}~\bibnamefont {Kaxiras}},\ }\bibfield  {title} {\enquote
  {\bibinfo {title} {{Theory of Spin-Conserving Excitation of the
  $N\ensuremath{-}{V}^{\ensuremath{-}}$ Center in Diamond}},}\ }\href {\doibase
  10.1103/PhysRevLett.103.186404} {\bibfield  {journal} {\bibinfo  {journal}
  {Phys. Rev. Lett.}\ }\textbf {\bibinfo {volume} {103}},\ \bibinfo {pages}
  {186404} (\bibinfo {year} {2009})}\BibitemShut {NoStop}%
\bibitem [{\citenamefont {Awschalom}\ \emph {et~al.}(2018)\citenamefont
  {Awschalom}, \citenamefont {Hanson}, \citenamefont {Wrachtrup},\ and\
  \citenamefont {Zhou}}]{awschalom2018quant}%
  \BibitemOpen
  \bibfield  {author} {\bibinfo {author} {\bibfnamefont {D.~D.}\ \bibnamefont
  {Awschalom}}, \bibinfo {author} {\bibfnamefont {R.}~\bibnamefont {Hanson}},
  \bibinfo {author} {\bibfnamefont {J.}~\bibnamefont {Wrachtrup}}, \ and\
  \bibinfo {author} {\bibfnamefont {B.~B.}\ \bibnamefont {Zhou}},\ }\bibfield
  {title} {\enquote {\bibinfo {title} {{Quantum technologies with optically
  interfaced solid-state spins}},}\ }\href {\doibase 10.1038/s41566-018-0232-2}
  {\bibfield  {journal} {\bibinfo  {journal} {Nat. Photon.}\ }\textbf {\bibinfo
  {volume} {12}},\ \bibinfo {pages} {516} (\bibinfo {year} {2018})}\BibitemShut
  {NoStop}%
\bibitem [{\citenamefont {Kuate~Defo}, \citenamefont {Wang},\ and\
  \citenamefont {Manjunathaiah}(2019)}]{DEFO2019}%
  \BibitemOpen
  \bibfield  {author} {\bibinfo {author} {\bibfnamefont {R.}~\bibnamefont
  {Kuate~Defo}}, \bibinfo {author} {\bibfnamefont {R.}~\bibnamefont {Wang}}, \
  and\ \bibinfo {author} {\bibfnamefont {M.}~\bibnamefont {Manjunathaiah}},\
  }\bibfield  {title} {\enquote {\bibinfo {title} {{Parallel BFS Implementing
  Optimized Decomposition of Space and kMC simulations for Diffusion of
  Vacancies for Quantum Storage}},}\ }\href
  {http://www.sciencedirect.com/science/article/pii/S1877750318304320}
  {\bibfield  {journal} {\bibinfo  {journal} {Journal of Computational
  Science}\ } (\bibinfo {year} {2019})}\BibitemShut {NoStop}%
\bibitem [{\citenamefont {Kuate~Defo}\ \emph {et~al.}(2018)\citenamefont
  {Kuate~Defo}, \citenamefont {Zhang}, \citenamefont {Bracher}, \citenamefont
  {Kim}, \citenamefont {Hu},\ and\ \citenamefont {Kaxiras}}]{Kuate}%
  \BibitemOpen
  \bibfield  {author} {\bibinfo {author} {\bibfnamefont {R.}~\bibnamefont
  {Kuate~Defo}}, \bibinfo {author} {\bibfnamefont {X.}~\bibnamefont {Zhang}},
  \bibinfo {author} {\bibfnamefont {D.}~\bibnamefont {Bracher}}, \bibinfo
  {author} {\bibfnamefont {G.}~\bibnamefont {Kim}}, \bibinfo {author}
  {\bibfnamefont {E.}~\bibnamefont {Hu}}, \ and\ \bibinfo {author}
  {\bibfnamefont {E.}~\bibnamefont {Kaxiras}},\ }\bibfield  {title} {\enquote
  {\bibinfo {title} {{Energetics and kinetics of vacancy defects in
  $4H$-SiC}},}\ }\href {\doibase 10.1103/PhysRevB.98.104103} {\bibfield
  {journal} {\bibinfo  {journal} {Phys. Rev. B}\ }\textbf {\bibinfo {volume}
  {98}},\ \bibinfo {pages} {104103} (\bibinfo {year} {2018})}\BibitemShut
  {NoStop}%
\bibitem [{\citenamefont {Bar-Gill}\ \emph {et~al.}(2013)\citenamefont
  {Bar-Gill}, \citenamefont {Pham}, \citenamefont {Jarmola}, \citenamefont
  {Budker},\ and\ \citenamefont {Walsworth}}]{Walsworth}%
  \BibitemOpen
  \bibfield  {author} {\bibinfo {author} {\bibfnamefont {N.}~\bibnamefont
  {Bar-Gill}}, \bibinfo {author} {\bibfnamefont {L.~M.}\ \bibnamefont {Pham}},
  \bibinfo {author} {\bibfnamefont {A.}~\bibnamefont {Jarmola}}, \bibinfo
  {author} {\bibfnamefont {D.}~\bibnamefont {Budker}}, \ and\ \bibinfo {author}
  {\bibfnamefont {R.~L.}\ \bibnamefont {Walsworth}},\ }\bibfield  {title}
  {\enquote {\bibinfo {title} {Solid-state electronic spin coherence time
  approaching one second},}\ }\href {https://doi.org/10.1038/ncomms2771}
  {\bibfield  {journal} {\bibinfo  {journal} {Nat. Commun.}\ }\textbf {\bibinfo
  {volume} {4}},\ \bibinfo {pages} {1743} (\bibinfo {year} {2013})}\BibitemShut
  {NoStop}%
\bibitem [{\citenamefont {Balasubramanian}\ \emph {et~al.}(2009)\citenamefont
  {Balasubramanian}, \citenamefont {Neumann}, \citenamefont {Twitchen},
  \citenamefont {Markham}, \citenamefont {Kolesov}, \citenamefont {Mizuochi},
  \citenamefont {Isoya}, \citenamefont {Achard}, \citenamefont {Beck},
  \citenamefont {Tissler}, \citenamefont {Jacques}, \citenamefont {Hemmer},
  \citenamefont {Jelezko},\ and\ \citenamefont {Wrachtrup}}]{Balasubramanian}%
  \BibitemOpen
  \bibfield  {author} {\bibinfo {author} {\bibfnamefont {G.}~\bibnamefont
  {Balasubramanian}}, \bibinfo {author} {\bibfnamefont {P.}~\bibnamefont
  {Neumann}}, \bibinfo {author} {\bibfnamefont {D.}~\bibnamefont {Twitchen}},
  \bibinfo {author} {\bibfnamefont {M.}~\bibnamefont {Markham}}, \bibinfo
  {author} {\bibfnamefont {R.}~\bibnamefont {Kolesov}}, \bibinfo {author}
  {\bibfnamefont {N.}~\bibnamefont {Mizuochi}}, \bibinfo {author}
  {\bibfnamefont {J.}~\bibnamefont {Isoya}}, \bibinfo {author} {\bibfnamefont
  {J.}~\bibnamefont {Achard}}, \bibinfo {author} {\bibfnamefont
  {J.}~\bibnamefont {Beck}}, \bibinfo {author} {\bibfnamefont {J.}~\bibnamefont
  {Tissler}}, \bibinfo {author} {\bibfnamefont {V.}~\bibnamefont {Jacques}},
  \bibinfo {author} {\bibfnamefont {P.~R.}\ \bibnamefont {Hemmer}}, \bibinfo
  {author} {\bibfnamefont {F.}~\bibnamefont {Jelezko}}, \ and\ \bibinfo
  {author} {\bibfnamefont {J.}~\bibnamefont {Wrachtrup}},\ }\bibfield  {title}
  {\enquote {\bibinfo {title} {Ultralong spin coherence time in isotopically
  engineered diamond},}\ }\href {https://doi.org/10.1038/nmat2420} {\bibfield
  {journal} {\bibinfo  {journal} {Nat. Mater.}\ }\textbf {\bibinfo {volume}
  {8}},\ \bibinfo {pages} {383} (\bibinfo {year} {2009})}\BibitemShut {NoStop}%
\bibitem [{\citenamefont {Abobeih}\ \emph {et~al.}(2018)\citenamefont
  {Abobeih}, \citenamefont {Cramer}, \citenamefont {Bakker}, \citenamefont
  {Kalb}, \citenamefont {Markham}, \citenamefont {Twitchen},\ and\
  \citenamefont {Taminiau}}]{Abobeih}%
  \BibitemOpen
  \bibfield  {author} {\bibinfo {author} {\bibfnamefont {M.~H.}\ \bibnamefont
  {Abobeih}}, \bibinfo {author} {\bibfnamefont {J.}~\bibnamefont {Cramer}},
  \bibinfo {author} {\bibfnamefont {M.~A.}\ \bibnamefont {Bakker}}, \bibinfo
  {author} {\bibfnamefont {N.}~\bibnamefont {Kalb}}, \bibinfo {author}
  {\bibfnamefont {M.}~\bibnamefont {Markham}}, \bibinfo {author} {\bibfnamefont
  {D.~J.}\ \bibnamefont {Twitchen}}, \ and\ \bibinfo {author} {\bibfnamefont
  {T.~H.}\ \bibnamefont {Taminiau}},\ }\bibfield  {title} {\enquote {\bibinfo
  {title} {One-second coherence for a single electron spin coupled to a
  multi-qubit nuclear-spin environment},}\ }\href {\doibase
  10.1038/s41467-018-04916-z} {\bibfield  {journal} {\bibinfo  {journal} {Nat.
  Commun.}\ }\textbf {\bibinfo {volume} {9}},\ \bibinfo {pages} {2552}
  (\bibinfo {year} {2018})}\BibitemShut {NoStop}%
\bibitem [{\citenamefont {Jelezko}\ \emph {et~al.}(2001)\citenamefont
  {Jelezko}, \citenamefont {Tietz}, \citenamefont {Gruber}, \citenamefont
  {Popa}, \citenamefont {Nizovtsev}, \citenamefont {Kilin},\ and\ \citenamefont
  {Wrachtrup}}]{Jelezko}%
  \BibitemOpen
  \bibfield  {author} {\bibinfo {author} {\bibfnamefont {F.}~\bibnamefont
  {Jelezko}}, \bibinfo {author} {\bibfnamefont {C.}~\bibnamefont {Tietz}},
  \bibinfo {author} {\bibfnamefont {A.}~\bibnamefont {Gruber}}, \bibinfo
  {author} {\bibfnamefont {I.}~\bibnamefont {Popa}}, \bibinfo {author}
  {\bibfnamefont {A.}~\bibnamefont {Nizovtsev}}, \bibinfo {author}
  {\bibfnamefont {S.}~\bibnamefont {Kilin}}, \ and\ \bibinfo {author}
  {\bibfnamefont {J.}~\bibnamefont {Wrachtrup}},\ }\bibfield  {title} {\enquote
  {\bibinfo {title} {{Spectroscopy of Single N-V Centers in Diamond}},}\ }\href
  {\doibase 10.1002/1438-5171(200112)2:4<255::AID-SIMO255>3.0.CO;2-D}
  {\bibfield  {journal} {\bibinfo  {journal} {Single Mol.}\ }\textbf {\bibinfo
  {volume} {2}},\ \bibinfo {pages} {255} (\bibinfo {year} {2001})}\BibitemShut
  {NoStop}%
\bibitem [{\citenamefont {Zaitsev}, \citenamefont {Vavilov},\ and\
  \citenamefont {Gippius}(1981)}]{zaitsev1981cathodoluminescence}%
  \BibitemOpen
  \bibfield  {author} {\bibinfo {author} {\bibfnamefont {A.}~\bibnamefont
  {Zaitsev}}, \bibinfo {author} {\bibfnamefont {V.}~\bibnamefont {Vavilov}}, \
  and\ \bibinfo {author} {\bibfnamefont {A.}~\bibnamefont {Gippius}},\
  }\bibfield  {title} {\enquote {\bibinfo {title} {Cathodoluminescence of
  diamond associated with silicon impurity},}\ }\href@noop {} {\bibfield
  {journal} {\bibinfo  {journal} {Sov. Phys. Leb. Inst. Rep}\ }\textbf
  {\bibinfo {volume} {10}},\ \bibinfo {pages} {15} (\bibinfo {year}
  {1981})}\BibitemShut {NoStop}%
\bibitem [{\citenamefont {Goss}\ \emph {et~al.}(1996)\citenamefont {Goss},
  \citenamefont {Jones}, \citenamefont {Breuer}, \citenamefont {Briddon},\ and\
  \citenamefont {\"Oberg}}]{goss1996twelve}%
  \BibitemOpen
  \bibfield  {author} {\bibinfo {author} {\bibfnamefont {J.~P.}\ \bibnamefont
  {Goss}}, \bibinfo {author} {\bibfnamefont {R.}~\bibnamefont {Jones}},
  \bibinfo {author} {\bibfnamefont {S.~J.}\ \bibnamefont {Breuer}}, \bibinfo
  {author} {\bibfnamefont {P.~R.}\ \bibnamefont {Briddon}}, \ and\ \bibinfo
  {author} {\bibfnamefont {S.}~\bibnamefont {\"Oberg}},\ }\bibfield  {title}
  {\enquote {\bibinfo {title} {{The Twelve-Line 1.682 eV Luminescence Center in
  Diamond and the Vacancy-Silicon Complex}},}\ }\href {\doibase
  10.1103/PhysRevLett.77.3041} {\bibfield  {journal} {\bibinfo  {journal}
  {Phys. Rev. Lett.}\ }\textbf {\bibinfo {volume} {77}},\ \bibinfo {pages}
  {3041} (\bibinfo {year} {1996})}\BibitemShut {NoStop}%
\bibitem [{\citenamefont {Clark}\ \emph {et~al.}(1995)\citenamefont {Clark},
  \citenamefont {Kanda}, \citenamefont {Kiflawi},\ and\ \citenamefont
  {Sittas}}]{clark1995silicon}%
  \BibitemOpen
  \bibfield  {author} {\bibinfo {author} {\bibfnamefont {C.~D.}\ \bibnamefont
  {Clark}}, \bibinfo {author} {\bibfnamefont {H.}~\bibnamefont {Kanda}},
  \bibinfo {author} {\bibfnamefont {I.}~\bibnamefont {Kiflawi}}, \ and\
  \bibinfo {author} {\bibfnamefont {G.}~\bibnamefont {Sittas}},\ }\bibfield
  {title} {\enquote {\bibinfo {title} {Silicon defects in diamond},}\ }\href
  {\doibase 10.1103/PhysRevB.51.16681} {\bibfield  {journal} {\bibinfo
  {journal} {Phys. Rev. B}\ }\textbf {\bibinfo {volume} {51}},\ \bibinfo
  {pages} {16681} (\bibinfo {year} {1995})}\BibitemShut {NoStop}%
\bibitem [{\citenamefont {Gali}\ and\ \citenamefont {Maze}(2013)}]{gali2013ab}%
  \BibitemOpen
  \bibfield  {author} {\bibinfo {author} {\bibfnamefont {A.}~\bibnamefont
  {Gali}}\ and\ \bibinfo {author} {\bibfnamefont {J.~R.}\ \bibnamefont
  {Maze}},\ }\bibfield  {title} {\enquote {\bibinfo {title} {{Ab initio study
  of the split silicon-vacancy defect in diamond: Electronic structure and
  related properties}},}\ }\href {\doibase 10.1103/PhysRevB.88.235205}
  {\bibfield  {journal} {\bibinfo  {journal} {Phys. Rev. B}\ }\textbf {\bibinfo
  {volume} {88}},\ \bibinfo {pages} {235205} (\bibinfo {year}
  {2013})}\BibitemShut {NoStop}%
\bibitem [{\citenamefont {Neu}\ \emph {et~al.}(2011)\citenamefont {Neu},
  \citenamefont {Steinmetz}, \citenamefont {Riedrich-M\"{o}ller}, \citenamefont
  {Gsell}, \citenamefont {Fischer}, \citenamefont {Schreck},\ and\
  \citenamefont {Becher}}]{neu2011single}%
  \BibitemOpen
  \bibfield  {author} {\bibinfo {author} {\bibfnamefont {E.}~\bibnamefont
  {Neu}}, \bibinfo {author} {\bibfnamefont {D.}~\bibnamefont {Steinmetz}},
  \bibinfo {author} {\bibfnamefont {J.}~\bibnamefont {Riedrich-M\"{o}ller}},
  \bibinfo {author} {\bibfnamefont {S.}~\bibnamefont {Gsell}}, \bibinfo
  {author} {\bibfnamefont {M.}~\bibnamefont {Fischer}}, \bibinfo {author}
  {\bibfnamefont {M.}~\bibnamefont {Schreck}}, \ and\ \bibinfo {author}
  {\bibfnamefont {C.}~\bibnamefont {Becher}},\ }\bibfield  {title} {\enquote
  {\bibinfo {title} {Single photon emission from silicon-vacancy colour centres
  in chemical vapour deposition nano-diamonds on iridium},}\ }\href {\doibase
  10.1088/1367-2630/13/2/025012} {\bibfield  {journal} {\bibinfo  {journal}
  {New J. Phys.}\ }\textbf {\bibinfo {volume} {13}},\ \bibinfo {pages} {025012}
  (\bibinfo {year} {2011})}\BibitemShut {NoStop}%
\bibitem [{\citenamefont {Hepp}\ \emph {et~al.}(2014)\citenamefont {Hepp},
  \citenamefont {M\"uller}, \citenamefont {Waselowski}, \citenamefont {Becker},
  \citenamefont {Pingault}, \citenamefont {Sternschulte}, \citenamefont
  {Steinm\"uller-Nethl}, \citenamefont {Gali}, \citenamefont {Maze},
  \citenamefont {Atat\"ure},\ and\ \citenamefont
  {Becher}}]{hepp2014electronic}%
  \BibitemOpen
  \bibfield  {author} {\bibinfo {author} {\bibfnamefont {C.}~\bibnamefont
  {Hepp}}, \bibinfo {author} {\bibfnamefont {T.}~\bibnamefont {M\"uller}},
  \bibinfo {author} {\bibfnamefont {V.}~\bibnamefont {Waselowski}}, \bibinfo
  {author} {\bibfnamefont {J.~N.}\ \bibnamefont {Becker}}, \bibinfo {author}
  {\bibfnamefont {B.}~\bibnamefont {Pingault}}, \bibinfo {author}
  {\bibfnamefont {H.}~\bibnamefont {Sternschulte}}, \bibinfo {author}
  {\bibfnamefont {D.}~\bibnamefont {Steinm\"uller-Nethl}}, \bibinfo {author}
  {\bibfnamefont {A.}~\bibnamefont {Gali}}, \bibinfo {author} {\bibfnamefont
  {J.~R.}\ \bibnamefont {Maze}}, \bibinfo {author} {\bibfnamefont
  {M.}~\bibnamefont {Atat\"ure}}, \ and\ \bibinfo {author} {\bibfnamefont
  {C.}~\bibnamefont {Becher}},\ }\bibfield  {title} {\enquote {\bibinfo {title}
  {{Electronic Structure of the Silicon Vacancy Color Center in Diamond}},}\
  }\href {\doibase 10.1103/PhysRevLett.112.036405} {\bibfield  {journal}
  {\bibinfo  {journal} {Phys. Rev. Lett.}\ }\textbf {\bibinfo {volume} {112}},\
  \bibinfo {pages} {036405} (\bibinfo {year} {2014})}\BibitemShut {NoStop}%
\bibitem [{\citenamefont {Rogers}\ \emph {et~al.}(2014)\citenamefont {Rogers},
  \citenamefont {Jahnke}, \citenamefont {Metsch}, \citenamefont {Sipahigil},
  \citenamefont {Binder}, \citenamefont {Teraji}, \citenamefont {Sumiya},
  \citenamefont {Isoya}, \citenamefont {Lukin}, \citenamefont {Hemmer},\ and\
  \citenamefont {Jelezko}}]{rogers2014all}%
  \BibitemOpen
  \bibfield  {author} {\bibinfo {author} {\bibfnamefont {L.~J.}\ \bibnamefont
  {Rogers}}, \bibinfo {author} {\bibfnamefont {K.~D.}\ \bibnamefont {Jahnke}},
  \bibinfo {author} {\bibfnamefont {M.~H.}\ \bibnamefont {Metsch}}, \bibinfo
  {author} {\bibfnamefont {A.}~\bibnamefont {Sipahigil}}, \bibinfo {author}
  {\bibfnamefont {J.~M.}\ \bibnamefont {Binder}}, \bibinfo {author}
  {\bibfnamefont {T.}~\bibnamefont {Teraji}}, \bibinfo {author} {\bibfnamefont
  {H.}~\bibnamefont {Sumiya}}, \bibinfo {author} {\bibfnamefont
  {J.}~\bibnamefont {Isoya}}, \bibinfo {author} {\bibfnamefont {M.~D.}\
  \bibnamefont {Lukin}}, \bibinfo {author} {\bibfnamefont {P.}~\bibnamefont
  {Hemmer}}, \ and\ \bibinfo {author} {\bibfnamefont {F.}~\bibnamefont
  {Jelezko}},\ }\bibfield  {title} {\enquote {\bibinfo {title} {{All-Optical
  Initialization, Readout, and Coherent Preparation of Single Silicon-Vacancy
  Spins in Diamond}},}\ }\href {\doibase 10.1103/PhysRevLett.113.263602}
  {\bibfield  {journal} {\bibinfo  {journal} {Phys. Rev. Lett.}\ }\textbf
  {\bibinfo {volume} {113}},\ \bibinfo {pages} {263602} (\bibinfo {year}
  {2014})}\BibitemShut {NoStop}%
\bibitem [{\citenamefont {Sipahigil}\ \emph {et~al.}(2014)\citenamefont
  {Sipahigil}, \citenamefont {Jahnke}, \citenamefont {Rogers}, \citenamefont
  {Teraji}, \citenamefont {Isoya}, \citenamefont {Zibrov}, \citenamefont
  {Jelezko},\ and\ \citenamefont {Lukin}}]{sipahigil2014indistinguishable}%
  \BibitemOpen
  \bibfield  {author} {\bibinfo {author} {\bibfnamefont {A.}~\bibnamefont
  {Sipahigil}}, \bibinfo {author} {\bibfnamefont {K.~D.}\ \bibnamefont
  {Jahnke}}, \bibinfo {author} {\bibfnamefont {L.~J.}\ \bibnamefont {Rogers}},
  \bibinfo {author} {\bibfnamefont {T.}~\bibnamefont {Teraji}}, \bibinfo
  {author} {\bibfnamefont {J.}~\bibnamefont {Isoya}}, \bibinfo {author}
  {\bibfnamefont {A.~S.}\ \bibnamefont {Zibrov}}, \bibinfo {author}
  {\bibfnamefont {F.}~\bibnamefont {Jelezko}}, \ and\ \bibinfo {author}
  {\bibfnamefont {M.~D.}\ \bibnamefont {Lukin}},\ }\bibfield  {title} {\enquote
  {\bibinfo {title} {{Indistinguishable Photons from Separated Silicon-Vacancy
  Centers in Diamond}},}\ }\href {\doibase 10.1103/PhysRevLett.113.113602}
  {\bibfield  {journal} {\bibinfo  {journal} {Phys. Rev. Lett.}\ }\textbf
  {\bibinfo {volume} {113}},\ \bibinfo {pages} {113602} (\bibinfo {year}
  {2014})}\BibitemShut {NoStop}%
\bibitem [{\citenamefont {Pingault}\ \emph {et~al.}(2017)\citenamefont
  {Pingault}, \citenamefont {Jarausch}, \citenamefont {Hepp}, \citenamefont
  {Klintberg}, \citenamefont {Becker}, \citenamefont {Markham}, \citenamefont
  {Becher},\ and\ \citenamefont {Atat{\"u}re}}]{Pingault}%
  \BibitemOpen
  \bibfield  {author} {\bibinfo {author} {\bibfnamefont {B.}~\bibnamefont
  {Pingault}}, \bibinfo {author} {\bibfnamefont {D.-D.}\ \bibnamefont
  {Jarausch}}, \bibinfo {author} {\bibfnamefont {C.}~\bibnamefont {Hepp}},
  \bibinfo {author} {\bibfnamefont {L.}~\bibnamefont {Klintberg}}, \bibinfo
  {author} {\bibfnamefont {J.~N.}\ \bibnamefont {Becker}}, \bibinfo {author}
  {\bibfnamefont {M.}~\bibnamefont {Markham}}, \bibinfo {author} {\bibfnamefont
  {C.}~\bibnamefont {Becher}}, \ and\ \bibinfo {author} {\bibfnamefont
  {M.}~\bibnamefont {Atat{\"u}re}},\ }\bibfield  {title} {\enquote {\bibinfo
  {title} {Coherent control of the silicon-vacancy spin in diamond},}\ }\href
  {https://doi.org/10.1038/ncomms15579} {\bibfield  {journal} {\bibinfo
  {journal} {Nat. Commun.}\ }\textbf {\bibinfo {volume} {8}},\ \bibinfo {pages}
  {15579} (\bibinfo {year} {2017})}\BibitemShut {NoStop}%
\bibitem [{\citenamefont {Pingault}\ \emph {et~al.}(2014)\citenamefont
  {Pingault}, \citenamefont {Becker}, \citenamefont {Schulte}, \citenamefont
  {Arend}, \citenamefont {Hepp}, \citenamefont {Godde}, \citenamefont
  {Tartakovskii}, \citenamefont {Markham}, \citenamefont {Becher},\ and\
  \citenamefont {Atat\"ure}}]{pingault2014all}%
  \BibitemOpen
  \bibfield  {author} {\bibinfo {author} {\bibfnamefont {B.}~\bibnamefont
  {Pingault}}, \bibinfo {author} {\bibfnamefont {J.~N.}\ \bibnamefont
  {Becker}}, \bibinfo {author} {\bibfnamefont {C.~H.~H.}\ \bibnamefont
  {Schulte}}, \bibinfo {author} {\bibfnamefont {C.}~\bibnamefont {Arend}},
  \bibinfo {author} {\bibfnamefont {C.}~\bibnamefont {Hepp}}, \bibinfo {author}
  {\bibfnamefont {T.}~\bibnamefont {Godde}}, \bibinfo {author} {\bibfnamefont
  {A.~I.}\ \bibnamefont {Tartakovskii}}, \bibinfo {author} {\bibfnamefont
  {M.}~\bibnamefont {Markham}}, \bibinfo {author} {\bibfnamefont
  {C.}~\bibnamefont {Becher}}, \ and\ \bibinfo {author} {\bibfnamefont
  {M.}~\bibnamefont {Atat\"ure}},\ }\bibfield  {title} {\enquote {\bibinfo
  {title} {{All-Optical Formation of Coherent Dark States of Silicon-Vacancy
  Spins in Diamond}},}\ }\href {\doibase 10.1103/PhysRevLett.113.263601}
  {\bibfield  {journal} {\bibinfo  {journal} {Phys. Rev. Lett.}\ }\textbf
  {\bibinfo {volume} {113}},\ \bibinfo {pages} {263601} (\bibinfo {year}
  {2014})}\BibitemShut {NoStop}%
\bibitem [{\citenamefont {Dietrich}\ \emph {et~al.}(2014)\citenamefont
  {Dietrich}, \citenamefont {Jahnke}, \citenamefont {Binder}, \citenamefont
  {Teraji}, \citenamefont {Isoya}, \citenamefont {Rogers},\ and\ \citenamefont
  {Jelezko}}]{dietrich2014isotopically}%
  \BibitemOpen
  \bibfield  {author} {\bibinfo {author} {\bibfnamefont {A.}~\bibnamefont
  {Dietrich}}, \bibinfo {author} {\bibfnamefont {K.~D.}\ \bibnamefont
  {Jahnke}}, \bibinfo {author} {\bibfnamefont {J.~M.}\ \bibnamefont {Binder}},
  \bibinfo {author} {\bibfnamefont {T.}~\bibnamefont {Teraji}}, \bibinfo
  {author} {\bibfnamefont {J.}~\bibnamefont {Isoya}}, \bibinfo {author}
  {\bibfnamefont {L.~J.}\ \bibnamefont {Rogers}}, \ and\ \bibinfo {author}
  {\bibfnamefont {F.}~\bibnamefont {Jelezko}},\ }\bibfield  {title} {\enquote
  {\bibinfo {title} {Isotopically varying spectral features of silicon-vacancy
  in diamond},}\ }\href {\doibase 10.1088/1367-2630/16/11/113019} {\bibfield
  {journal} {\bibinfo  {journal} {New J. Phys.}\ }\textbf {\bibinfo {volume}
  {16}},\ \bibinfo {pages} {113019} (\bibinfo {year} {2014})}\BibitemShut
  {NoStop}%
\bibitem [{\citenamefont {Becker}\ \emph {et~al.}(2016)\citenamefont {Becker},
  \citenamefont {G{\"o}rlitz}, \citenamefont {Arend}, \citenamefont {Markham},\
  and\ \citenamefont {Becher}}]{becker2016ultrafast}%
  \BibitemOpen
  \bibfield  {author} {\bibinfo {author} {\bibfnamefont {J.~N.}\ \bibnamefont
  {Becker}}, \bibinfo {author} {\bibfnamefont {J.}~\bibnamefont {G{\"o}rlitz}},
  \bibinfo {author} {\bibfnamefont {C.}~\bibnamefont {Arend}}, \bibinfo
  {author} {\bibfnamefont {M.}~\bibnamefont {Markham}}, \ and\ \bibinfo
  {author} {\bibfnamefont {C.}~\bibnamefont {Becher}},\ }\bibfield  {title}
  {\enquote {\bibinfo {title} {Ultrafast all-optical coherent control of single
  silicon vacancy colour centres in diamond},}\ }\href
  {https://doi.org/10.1038/ncomms13512} {\bibfield  {journal} {\bibinfo
  {journal} {Nat. Commun.}\ }\textbf {\bibinfo {volume} {7}},\ \bibinfo {pages}
  {13512} (\bibinfo {year} {2016})}\BibitemShut {NoStop}%
\bibitem [{\citenamefont {Jahnke}\ \emph {et~al.}(2015)\citenamefont {Jahnke},
  \citenamefont {Sipahigil}, \citenamefont {Binder}, \citenamefont {Doherty},
  \citenamefont {Metsch}, \citenamefont {Rogers}, \citenamefont {Manson},
  \citenamefont {Lukin},\ and\ \citenamefont {Jelezko}}]{Jahnke}%
  \BibitemOpen
  \bibfield  {author} {\bibinfo {author} {\bibfnamefont {K.~D.}\ \bibnamefont
  {Jahnke}}, \bibinfo {author} {\bibfnamefont {A.}~\bibnamefont {Sipahigil}},
  \bibinfo {author} {\bibfnamefont {J.~M.}\ \bibnamefont {Binder}}, \bibinfo
  {author} {\bibfnamefont {M.~W.}\ \bibnamefont {Doherty}}, \bibinfo {author}
  {\bibfnamefont {M.}~\bibnamefont {Metsch}}, \bibinfo {author} {\bibfnamefont
  {L.~J.}\ \bibnamefont {Rogers}}, \bibinfo {author} {\bibfnamefont {N.~B.}\
  \bibnamefont {Manson}}, \bibinfo {author} {\bibfnamefont {M.~D.}\
  \bibnamefont {Lukin}}, \ and\ \bibinfo {author} {\bibfnamefont
  {F.}~\bibnamefont {Jelezko}},\ }\bibfield  {title} {\enquote {\bibinfo
  {title} {Electron{\textendash}phonon processes of the silicon-vacancy centre
  in diamond},}\ }\href {\doibase 10.1088/1367-2630/17/4/043011} {\bibfield
  {journal} {\bibinfo  {journal} {New J. Phys.}\ }\textbf {\bibinfo {volume}
  {17}},\ \bibinfo {pages} {043011} (\bibinfo {year} {2015})}\BibitemShut
  {NoStop}%
\bibitem [{\citenamefont {Sukachev}\ \emph {et~al.}(2017)\citenamefont
  {Sukachev}, \citenamefont {Sipahigil}, \citenamefont {Nguyen}, \citenamefont
  {Bhaskar}, \citenamefont {Evans}, \citenamefont {Jelezko},\ and\
  \citenamefont {Lukin}}]{sukachev2017silicon}%
  \BibitemOpen
  \bibfield  {author} {\bibinfo {author} {\bibfnamefont {D.~D.}\ \bibnamefont
  {Sukachev}}, \bibinfo {author} {\bibfnamefont {A.}~\bibnamefont {Sipahigil}},
  \bibinfo {author} {\bibfnamefont {C.~T.}\ \bibnamefont {Nguyen}}, \bibinfo
  {author} {\bibfnamefont {M.~K.}\ \bibnamefont {Bhaskar}}, \bibinfo {author}
  {\bibfnamefont {R.~E.}\ \bibnamefont {Evans}}, \bibinfo {author}
  {\bibfnamefont {F.}~\bibnamefont {Jelezko}}, \ and\ \bibinfo {author}
  {\bibfnamefont {M.~D.}\ \bibnamefont {Lukin}},\ }\bibfield  {title} {\enquote
  {\bibinfo {title} {{Silicon-Vacancy Spin Qubit in Diamond: A Quantum Memory
  Exceeding 10 ms with Single-Shot State Readout}},}\ }\href {\doibase
  10.1103/PhysRevLett.119.223602} {\bibfield  {journal} {\bibinfo  {journal}
  {Phys. Rev. Lett.}\ }\textbf {\bibinfo {volume} {119}},\ \bibinfo {pages}
  {223602} (\bibinfo {year} {2017})}\BibitemShut {NoStop}%
\bibitem [{\citenamefont {Becker}\ \emph {et~al.}(2018)\citenamefont {Becker},
  \citenamefont {Pingault}, \citenamefont {Gro\ss{}}, \citenamefont
  {G\"undo\ifmmode~\breve{g}\else \u{g}\fi{}an}, \citenamefont {Kukharchyk},
  \citenamefont {Markham}, \citenamefont {Edmonds}, \citenamefont {Atat\"ure},
  \citenamefont {Bushev},\ and\ \citenamefont {Becher}}]{becker2018all}%
  \BibitemOpen
  \bibfield  {author} {\bibinfo {author} {\bibfnamefont {J.~N.}\ \bibnamefont
  {Becker}}, \bibinfo {author} {\bibfnamefont {B.}~\bibnamefont {Pingault}},
  \bibinfo {author} {\bibfnamefont {D.}~\bibnamefont {Gro\ss{}}}, \bibinfo
  {author} {\bibfnamefont {M.}~\bibnamefont {G\"undo\ifmmode~\breve{g}\else
  \u{g}\fi{}an}}, \bibinfo {author} {\bibfnamefont {N.}~\bibnamefont
  {Kukharchyk}}, \bibinfo {author} {\bibfnamefont {M.}~\bibnamefont {Markham}},
  \bibinfo {author} {\bibfnamefont {A.}~\bibnamefont {Edmonds}}, \bibinfo
  {author} {\bibfnamefont {M.}~\bibnamefont {Atat\"ure}}, \bibinfo {author}
  {\bibfnamefont {P.}~\bibnamefont {Bushev}}, \ and\ \bibinfo {author}
  {\bibfnamefont {C.}~\bibnamefont {Becher}},\ }\bibfield  {title} {\enquote
  {\bibinfo {title} {{All-Optical Control of the Silicon-Vacancy Spin in
  Diamond at Millikelvin Temperatures}},}\ }\href {\doibase
  10.1103/PhysRevLett.120.053603} {\bibfield  {journal} {\bibinfo  {journal}
  {Phys. Rev. Lett.}\ }\textbf {\bibinfo {volume} {120}},\ \bibinfo {pages}
  {053603} (\bibinfo {year} {2018})}\BibitemShut {NoStop}%
\bibitem [{\citenamefont {Nguyen}\ \emph {et~al.}(2018)\citenamefont {Nguyen},
  \citenamefont {Evans}, \citenamefont {Sipahigil}, \citenamefont {Bhaskar},
  \citenamefont {Sukachev}, \citenamefont {Agafonov}, \citenamefont {Davydov},
  \citenamefont {Kulikova}, \citenamefont {Jelezko},\ and\ \citenamefont
  {Lukin}}]{nguyen2018all}%
  \BibitemOpen
  \bibfield  {author} {\bibinfo {author} {\bibfnamefont {C.~T.}\ \bibnamefont
  {Nguyen}}, \bibinfo {author} {\bibfnamefont {R.~E.}\ \bibnamefont {Evans}},
  \bibinfo {author} {\bibfnamefont {A.}~\bibnamefont {Sipahigil}}, \bibinfo
  {author} {\bibfnamefont {M.~K.}\ \bibnamefont {Bhaskar}}, \bibinfo {author}
  {\bibfnamefont {D.~D.}\ \bibnamefont {Sukachev}}, \bibinfo {author}
  {\bibfnamefont {V.~N.}\ \bibnamefont {Agafonov}}, \bibinfo {author}
  {\bibfnamefont {V.~A.}\ \bibnamefont {Davydov}}, \bibinfo {author}
  {\bibfnamefont {L.~F.}\ \bibnamefont {Kulikova}}, \bibinfo {author}
  {\bibfnamefont {F.}~\bibnamefont {Jelezko}}, \ and\ \bibinfo {author}
  {\bibfnamefont {M.~D.}\ \bibnamefont {Lukin}},\ }\bibfield  {title} {\enquote
  {\bibinfo {title} {All-optical nanoscale thermometry with silicon-vacancy
  centers in diamond},}\ }\href {\doibase 10.1063/1.5029904} {\bibfield
  {journal} {\bibinfo  {journal} {Appl. Phys. Lett.}\ }\textbf {\bibinfo
  {volume} {112}},\ \bibinfo {pages} {203102} (\bibinfo {year}
  {2018})}\BibitemShut {NoStop}%
\bibitem [{\citenamefont {D'Haenens-Johansson}\ \emph
  {et~al.}(2011)\citenamefont {D'Haenens-Johansson}, \citenamefont {Edmonds},
  \citenamefont {Green}, \citenamefont {Newton}, \citenamefont {Davies},
  \citenamefont {Martineau}, \citenamefont {Khan},\ and\ \citenamefont
  {Twitchen}}]{d2011optical}%
  \BibitemOpen
  \bibfield  {author} {\bibinfo {author} {\bibfnamefont {U.~F.~S.}\
  \bibnamefont {D'Haenens-Johansson}}, \bibinfo {author} {\bibfnamefont
  {A.~M.}\ \bibnamefont {Edmonds}}, \bibinfo {author} {\bibfnamefont {B.~L.}\
  \bibnamefont {Green}}, \bibinfo {author} {\bibfnamefont {M.~E.}\ \bibnamefont
  {Newton}}, \bibinfo {author} {\bibfnamefont {G.}~\bibnamefont {Davies}},
  \bibinfo {author} {\bibfnamefont {P.~M.}\ \bibnamefont {Martineau}}, \bibinfo
  {author} {\bibfnamefont {R.~U.~A.}\ \bibnamefont {Khan}}, \ and\ \bibinfo
  {author} {\bibfnamefont {D.~J.}\ \bibnamefont {Twitchen}},\ }\bibfield
  {title} {\enquote {\bibinfo {title} {Optical properties of the neutral
  silicon split-vacancy center in diamond},}\ }\href {\doibase
  10.1103/PhysRevB.84.245208} {\bibfield  {journal} {\bibinfo  {journal} {Phys.
  Rev. B}\ }\textbf {\bibinfo {volume} {84}},\ \bibinfo {pages} {245208}
  (\bibinfo {year} {2011})}\BibitemShut {NoStop}%
\bibitem [{\citenamefont {Rose}\ \emph {et~al.}(2018)\citenamefont {Rose},
  \citenamefont {Huang}, \citenamefont {Zhang}, \citenamefont {Stevenson},
  \citenamefont {Tyryshkin}, \citenamefont {Sangtawesin}, \citenamefont
  {Srinivasan}, \citenamefont {Loudin}, \citenamefont {Markham}, \citenamefont
  {Edmonds}, \citenamefont {Twitchen}, \citenamefont {Lyon},\ and\
  \citenamefont {de~Leon}}]{rose2018observation}%
  \BibitemOpen
  \bibfield  {author} {\bibinfo {author} {\bibfnamefont {B.~C.}\ \bibnamefont
  {Rose}}, \bibinfo {author} {\bibfnamefont {D.}~\bibnamefont {Huang}},
  \bibinfo {author} {\bibfnamefont {Z.-H.}\ \bibnamefont {Zhang}}, \bibinfo
  {author} {\bibfnamefont {P.}~\bibnamefont {Stevenson}}, \bibinfo {author}
  {\bibfnamefont {A.~M.}\ \bibnamefont {Tyryshkin}}, \bibinfo {author}
  {\bibfnamefont {S.}~\bibnamefont {Sangtawesin}}, \bibinfo {author}
  {\bibfnamefont {S.}~\bibnamefont {Srinivasan}}, \bibinfo {author}
  {\bibfnamefont {L.}~\bibnamefont {Loudin}}, \bibinfo {author} {\bibfnamefont
  {M.~L.}\ \bibnamefont {Markham}}, \bibinfo {author} {\bibfnamefont {A.~M.}\
  \bibnamefont {Edmonds}}, \bibinfo {author} {\bibfnamefont {D.~J.}\
  \bibnamefont {Twitchen}}, \bibinfo {author} {\bibfnamefont {S.~A.}\
  \bibnamefont {Lyon}}, \ and\ \bibinfo {author} {\bibfnamefont {N.~P.}\
  \bibnamefont {de~Leon}},\ }\bibfield  {title} {\enquote {\bibinfo {title}
  {Observation of an environmentally insensitive solid-state spin defect in
  diamond},}\ }\href {\doibase 10.1126/science.aao0290} {\bibfield  {journal}
  {\bibinfo  {journal} {Science}\ }\textbf {\bibinfo {volume} {361}},\ \bibinfo
  {pages} {60} (\bibinfo {year} {2018})}\BibitemShut {NoStop}%
\bibitem [{\citenamefont {Iakoubovskii}\ and\ \citenamefont
  {Stesmans}(2002)}]{iakoubovskii2002characterization}%
  \BibitemOpen
  \bibfield  {author} {\bibinfo {author} {\bibfnamefont {K.}~\bibnamefont
  {Iakoubovskii}}\ and\ \bibinfo {author} {\bibfnamefont {A.}~\bibnamefont
  {Stesmans}},\ }\bibfield  {title} {\enquote {\bibinfo {title}
  {{Characterization of hydrogen and silicon-related defects in CVD diamond by
  electron spin resonance}},}\ }\href {\doibase 10.1103/PhysRevB.66.195207}
  {\bibfield  {journal} {\bibinfo  {journal} {Phys. Rev. B}\ }\textbf {\bibinfo
  {volume} {66}},\ \bibinfo {pages} {195207} (\bibinfo {year}
  {2002})}\BibitemShut {NoStop}%
\bibitem [{\citenamefont {Edmonds}\ \emph {et~al.}(2008)\citenamefont
  {Edmonds}, \citenamefont {Newton}, \citenamefont {Martineau}, \citenamefont
  {Twitchen},\ and\ \citenamefont {Williams}}]{edmonds2008electron}%
  \BibitemOpen
  \bibfield  {author} {\bibinfo {author} {\bibfnamefont {A.~M.}\ \bibnamefont
  {Edmonds}}, \bibinfo {author} {\bibfnamefont {M.~E.}\ \bibnamefont {Newton}},
  \bibinfo {author} {\bibfnamefont {P.~M.}\ \bibnamefont {Martineau}}, \bibinfo
  {author} {\bibfnamefont {D.~J.}\ \bibnamefont {Twitchen}}, \ and\ \bibinfo
  {author} {\bibfnamefont {S.~D.}\ \bibnamefont {Williams}},\ }\bibfield
  {title} {\enquote {\bibinfo {title} {Electron paramagnetic resonance studies
  of silicon-related defects in diamond},}\ }\href {\doibase
  10.1103/PhysRevB.77.245205} {\bibfield  {journal} {\bibinfo  {journal} {Phys.
  Rev. B}\ }\textbf {\bibinfo {volume} {77}},\ \bibinfo {pages} {245205}
  (\bibinfo {year} {2008})}\BibitemShut {NoStop}%
\bibitem [{\citenamefont {Thiering}\ and\ \citenamefont
  {Gali}(2019)}]{thiering2019the}%
  \BibitemOpen
  \bibfield  {author} {\bibinfo {author} {\bibfnamefont {G.}~\bibnamefont
  {Thiering}}\ and\ \bibinfo {author} {\bibfnamefont {A.}~\bibnamefont
  {Gali}},\ }\bibfield  {title} {\enquote {\bibinfo {title} {{The ($e_g \otimes
  e_u$) $\otimes$ E$_g$ product Jahn--Teller effect in the neutral group-IV
  vacancy quantum bits in diamond}},}\ }\href {\doibase
  10.1038/s41524-019-0158-3} {\bibfield  {journal} {\bibinfo  {journal} {npj
  Comput. Mater.}\ }\textbf {\bibinfo {volume} {5}},\ \bibinfo {pages} {18}
  (\bibinfo {year} {2019})}\BibitemShut {NoStop}%
\bibitem [{\citenamefont {Metsch}\ \emph {et~al.}(2019)\citenamefont {Metsch},
  \citenamefont {Senkalla}, \citenamefont {Tratzmiller}, \citenamefont
  {Scheuer}, \citenamefont {Kern}, \citenamefont {Achard}, \citenamefont
  {Tallaire}, \citenamefont {Plenio}, \citenamefont {Siyushev},\ and\
  \citenamefont {Jelezko}}]{metsch2019init}%
  \BibitemOpen
  \bibfield  {author} {\bibinfo {author} {\bibfnamefont {M.~H.}\ \bibnamefont
  {Metsch}}, \bibinfo {author} {\bibfnamefont {K.}~\bibnamefont {Senkalla}},
  \bibinfo {author} {\bibfnamefont {B.}~\bibnamefont {Tratzmiller}}, \bibinfo
  {author} {\bibfnamefont {J.}~\bibnamefont {Scheuer}}, \bibinfo {author}
  {\bibfnamefont {M.}~\bibnamefont {Kern}}, \bibinfo {author} {\bibfnamefont
  {J.}~\bibnamefont {Achard}}, \bibinfo {author} {\bibfnamefont
  {A.}~\bibnamefont {Tallaire}}, \bibinfo {author} {\bibfnamefont {M.~B.}\
  \bibnamefont {Plenio}}, \bibinfo {author} {\bibfnamefont {P.}~\bibnamefont
  {Siyushev}}, \ and\ \bibinfo {author} {\bibfnamefont {F.}~\bibnamefont
  {Jelezko}},\ }\bibfield  {title} {\enquote {\bibinfo {title} {{Initialization
  and Readout of Nuclear Spins via a Negatively Charged Silicon-Vacancy Center
  in Diamond}},}\ }\href {\doibase 10.1103/PhysRevLett.122.190503} {\bibfield
  {journal} {\bibinfo  {journal} {Phys. Rev. Lett.}\ }\textbf {\bibinfo
  {volume} {122}},\ \bibinfo {pages} {190503} (\bibinfo {year}
  {2019})}\BibitemShut {NoStop}%
\bibitem [{\citenamefont {Sohn}\ \emph {et~al.}(2018)\citenamefont {Sohn},
  \citenamefont {Meesala}, \citenamefont {Pingault}, \citenamefont {Atikian},
  \citenamefont {Holzgrafe}, \citenamefont {G{\"u}ndo{\u g}an}, \citenamefont
  {Stavrakas}, \citenamefont {Stanley}, \citenamefont {Sipahigil},
  \citenamefont {Choi}, \citenamefont {Zhang}, \citenamefont {Pacheco},
  \citenamefont {Abraham}, \citenamefont {Bielejec}, \citenamefont {Lukin},
  \citenamefont {Atat{\"u}re},\ and\ \citenamefont {Lon{\v
  c}ar}}]{sohn2018cont}%
  \BibitemOpen
  \bibfield  {author} {\bibinfo {author} {\bibfnamefont {Y.-I.}\ \bibnamefont
  {Sohn}}, \bibinfo {author} {\bibfnamefont {S.}~\bibnamefont {Meesala}},
  \bibinfo {author} {\bibfnamefont {B.}~\bibnamefont {Pingault}}, \bibinfo
  {author} {\bibfnamefont {H.~A.}\ \bibnamefont {Atikian}}, \bibinfo {author}
  {\bibfnamefont {J.}~\bibnamefont {Holzgrafe}}, \bibinfo {author}
  {\bibfnamefont {M.}~\bibnamefont {G{\"u}ndo{\u g}an}}, \bibinfo {author}
  {\bibfnamefont {C.}~\bibnamefont {Stavrakas}}, \bibinfo {author}
  {\bibfnamefont {M.~J.}\ \bibnamefont {Stanley}}, \bibinfo {author}
  {\bibfnamefont {A.}~\bibnamefont {Sipahigil}}, \bibinfo {author}
  {\bibfnamefont {J.}~\bibnamefont {Choi}}, \bibinfo {author} {\bibfnamefont
  {M.}~\bibnamefont {Zhang}}, \bibinfo {author} {\bibfnamefont {J.~L.}\
  \bibnamefont {Pacheco}}, \bibinfo {author} {\bibfnamefont {J.}~\bibnamefont
  {Abraham}}, \bibinfo {author} {\bibfnamefont {E.}~\bibnamefont {Bielejec}},
  \bibinfo {author} {\bibfnamefont {M.~D.}\ \bibnamefont {Lukin}}, \bibinfo
  {author} {\bibfnamefont {M.}~\bibnamefont {Atat{\"u}re}}, \ and\ \bibinfo
  {author} {\bibfnamefont {M.}~\bibnamefont {Lon{\v c}ar}},\ }\bibfield
  {title} {\enquote {\bibinfo {title} {{Controlling the coherence of a diamond
  spin qubit through its strain environment}},}\ }\href {\doibase
  10.1038/s41467-018-04340-3} {\bibfield  {journal} {\bibinfo  {journal} {Nat.
  Commun.}\ }\textbf {\bibinfo {volume} {9}},\ \bibinfo {pages} {2012}
  (\bibinfo {year} {2018})}\BibitemShut {NoStop}%
\bibitem [{\citenamefont {Iwasaki}\ \emph {et~al.}(2017)\citenamefont
  {Iwasaki}, \citenamefont {Miyamoto}, \citenamefont {Taniguchi}, \citenamefont
  {Siyushev}, \citenamefont {Metsch}, \citenamefont {Jelezko},\ and\
  \citenamefont {Hatano}}]{Iwasaki2}%
  \BibitemOpen
  \bibfield  {author} {\bibinfo {author} {\bibfnamefont {T.}~\bibnamefont
  {Iwasaki}}, \bibinfo {author} {\bibfnamefont {Y.}~\bibnamefont {Miyamoto}},
  \bibinfo {author} {\bibfnamefont {T.}~\bibnamefont {Taniguchi}}, \bibinfo
  {author} {\bibfnamefont {P.}~\bibnamefont {Siyushev}}, \bibinfo {author}
  {\bibfnamefont {M.~H.}\ \bibnamefont {Metsch}}, \bibinfo {author}
  {\bibfnamefont {F.}~\bibnamefont {Jelezko}}, \ and\ \bibinfo {author}
  {\bibfnamefont {M.}~\bibnamefont {Hatano}},\ }\bibfield  {title} {\enquote
  {\bibinfo {title} {Tin-vacancy quantum emitters in diamond},}\ }\href
  {\doibase 10.1103/PhysRevLett.119.253601} {\bibfield  {journal} {\bibinfo
  {journal} {Phys. Rev. Lett.}\ }\textbf {\bibinfo {volume} {119}},\ \bibinfo
  {pages} {253601} (\bibinfo {year} {2017})}\BibitemShut {NoStop}%
\bibitem [{\citenamefont {Iwasaki}\ \emph {et~al.}(2015)\citenamefont
  {Iwasaki}, \citenamefont {Ishibashi}, \citenamefont {Miyamoto}, \citenamefont
  {Doi}, \citenamefont {Kobayashi}, \citenamefont {Miyazaki}, \citenamefont
  {Tahara}, \citenamefont {Jahnke}, \citenamefont {Rogers}, \citenamefont
  {Naydenov}, \citenamefont {Jelezko}, \citenamefont {Yamasaki}, \citenamefont
  {Nagamachi}, \citenamefont {Inubushi}, \citenamefont {Mizuochi},\ and\
  \citenamefont {Hatano}}]{Iwasaki}%
  \BibitemOpen
  \bibfield  {author} {\bibinfo {author} {\bibfnamefont {T.}~\bibnamefont
  {Iwasaki}}, \bibinfo {author} {\bibfnamefont {F.}~\bibnamefont {Ishibashi}},
  \bibinfo {author} {\bibfnamefont {Y.}~\bibnamefont {Miyamoto}}, \bibinfo
  {author} {\bibfnamefont {Y.}~\bibnamefont {Doi}}, \bibinfo {author}
  {\bibfnamefont {S.}~\bibnamefont {Kobayashi}}, \bibinfo {author}
  {\bibfnamefont {T.}~\bibnamefont {Miyazaki}}, \bibinfo {author}
  {\bibfnamefont {K.}~\bibnamefont {Tahara}}, \bibinfo {author} {\bibfnamefont
  {K.~D.}\ \bibnamefont {Jahnke}}, \bibinfo {author} {\bibfnamefont {L.~J.}\
  \bibnamefont {Rogers}}, \bibinfo {author} {\bibfnamefont {B.}~\bibnamefont
  {Naydenov}}, \bibinfo {author} {\bibfnamefont {F.}~\bibnamefont {Jelezko}},
  \bibinfo {author} {\bibfnamefont {S.}~\bibnamefont {Yamasaki}}, \bibinfo
  {author} {\bibfnamefont {S.}~\bibnamefont {Nagamachi}}, \bibinfo {author}
  {\bibfnamefont {T.}~\bibnamefont {Inubushi}}, \bibinfo {author}
  {\bibfnamefont {N.}~\bibnamefont {Mizuochi}}, \ and\ \bibinfo {author}
  {\bibfnamefont {M.}~\bibnamefont {Hatano}},\ }\bibfield  {title} {\enquote
  {\bibinfo {title} {Germanium-vacancy single color centers in diamond},}\
  }\href {https://doi.org/10.1038/srep12882} {\bibfield  {journal} {\bibinfo
  {journal} {Sci. Rep.}\ }\textbf {\bibinfo {volume} {5}},\ \bibinfo {pages}
  {12882} (\bibinfo {year} {2015})}\BibitemShut {NoStop}%
\bibitem [{\citenamefont {Ralchenko}\ \emph {et~al.}(2015)\citenamefont
  {Ralchenko}, \citenamefont {Sedov}, \citenamefont {Khomich}, \citenamefont
  {Krivobok}, \citenamefont {Nikolaev}, \citenamefont {Savin}, \citenamefont
  {Vlasov},\ and\ \citenamefont {Konov}}]{Ralchenko}%
  \BibitemOpen
  \bibfield  {author} {\bibinfo {author} {\bibfnamefont {V.~G.}\ \bibnamefont
  {Ralchenko}}, \bibinfo {author} {\bibfnamefont {V.~S.}\ \bibnamefont
  {Sedov}}, \bibinfo {author} {\bibfnamefont {A.~A.}\ \bibnamefont {Khomich}},
  \bibinfo {author} {\bibfnamefont {V.~S.}\ \bibnamefont {Krivobok}}, \bibinfo
  {author} {\bibfnamefont {S.~N.}\ \bibnamefont {Nikolaev}}, \bibinfo {author}
  {\bibfnamefont {S.}~\bibnamefont {Savin}}, \bibinfo {author} {\bibfnamefont
  {I.~I.}\ \bibnamefont {Vlasov}}, \ and\ \bibinfo {author} {\bibfnamefont
  {V.~I.}\ \bibnamefont {Konov}},\ }\bibfield  {title} {\enquote {\bibinfo
  {title} {{Observation of the Ge-vacancy color center in microcrystalline
  diamond films}},}\ }\href {\doibase 10.3103/S1068335615060020} {\bibfield
  {journal} {\bibinfo  {journal} {Bull. Leb. Phys. Inst.}\ }\textbf {\bibinfo
  {volume} {42}},\ \bibinfo {pages} {165} (\bibinfo {year} {2015})}\BibitemShut
  {NoStop}%
\bibitem [{\citenamefont {Ekimov}\ \emph {et~al.}(2015)\citenamefont {Ekimov},
  \citenamefont {Lyapin}, \citenamefont {Boldyrev}, \citenamefont {Kondrin},
  \citenamefont {Khmelnitskiy}, \citenamefont {Gavva}, \citenamefont
  {Kotereva},\ and\ \citenamefont {Popova}}]{Ekimov}%
  \BibitemOpen
  \bibfield  {author} {\bibinfo {author} {\bibfnamefont {E.~A.}\ \bibnamefont
  {Ekimov}}, \bibinfo {author} {\bibfnamefont {S.~G.}\ \bibnamefont {Lyapin}},
  \bibinfo {author} {\bibfnamefont {K.~N.}\ \bibnamefont {Boldyrev}}, \bibinfo
  {author} {\bibfnamefont {M.~V.}\ \bibnamefont {Kondrin}}, \bibinfo {author}
  {\bibfnamefont {R.}~\bibnamefont {Khmelnitskiy}}, \bibinfo {author}
  {\bibfnamefont {V.~A.}\ \bibnamefont {Gavva}}, \bibinfo {author}
  {\bibfnamefont {T.~V.}\ \bibnamefont {Kotereva}}, \ and\ \bibinfo {author}
  {\bibfnamefont {M.~N.}\ \bibnamefont {Popova}},\ }\bibfield  {title}
  {\enquote {\bibinfo {title} {Germanium--vacancy color center in isotopically
  enriched diamonds synthesized at high pressures},}\ }\href {\doibase
  10.1134/S0021364015230034} {\bibfield  {journal} {\bibinfo  {journal} {JETP
  Lett.}\ }\textbf {\bibinfo {volume} {102}},\ \bibinfo {pages} {701} (\bibinfo
  {year} {2015})}\BibitemShut {NoStop}%
\bibitem [{\citenamefont {Palyanov}\ \emph {et~al.}(2015)\citenamefont
  {Palyanov}, \citenamefont {Kupriyanov}, \citenamefont {Borzdov},\ and\
  \citenamefont {Surovtsev}}]{Palyanov2015germanium}%
  \BibitemOpen
  \bibfield  {author} {\bibinfo {author} {\bibfnamefont {Y.~N.}\ \bibnamefont
  {Palyanov}}, \bibinfo {author} {\bibfnamefont {I.~N.}\ \bibnamefont
  {Kupriyanov}}, \bibinfo {author} {\bibfnamefont {Y.~M.}\ \bibnamefont
  {Borzdov}}, \ and\ \bibinfo {author} {\bibfnamefont {N.~V.}\ \bibnamefont
  {Surovtsev}},\ }\bibfield  {title} {\enquote {\bibinfo {title} {Germanium: a
  new catalyst for diamond synthesis and a new optically active impurity in
  diamond},}\ }\href {https://doi.org/10.1038/srep14789} {\bibfield  {journal}
  {\bibinfo  {journal} {Sci. Rep.}\ }\textbf {\bibinfo {volume} {5}},\ \bibinfo
  {pages} {14789} (\bibinfo {year} {2015})}\BibitemShut {NoStop}%
\bibitem [{\citenamefont {Palyanov}\ \emph {et~al.}(2016)\citenamefont
  {Palyanov}, \citenamefont {Kupriyanov}, \citenamefont {Borzdov},
  \citenamefont {Khokhryakov},\ and\ \citenamefont {Surovtsev}}]{Palyanov}%
  \BibitemOpen
  \bibfield  {author} {\bibinfo {author} {\bibfnamefont {Y.~N.}\ \bibnamefont
  {Palyanov}}, \bibinfo {author} {\bibfnamefont {I.~N.}\ \bibnamefont
  {Kupriyanov}}, \bibinfo {author} {\bibfnamefont {Y.~M.}\ \bibnamefont
  {Borzdov}}, \bibinfo {author} {\bibfnamefont {A.~F.}\ \bibnamefont
  {Khokhryakov}}, \ and\ \bibinfo {author} {\bibfnamefont {N.~V.}\ \bibnamefont
  {Surovtsev}},\ }\bibfield  {title} {\enquote {\bibinfo {title}
  {{High-Pressure Synthesis and Characterization of Ge-Doped Single Crystal
  Diamond}},}\ }\href {\doibase 10.1021/acs.cgd.6b00481} {\bibfield  {journal}
  {\bibinfo  {journal} {Cryst. Growth Des.}\ }\textbf {\bibinfo {volume}
  {16}},\ \bibinfo {pages} {3510} (\bibinfo {year} {2016})}\BibitemShut
  {NoStop}%
\bibitem [{\citenamefont {Ekimov}\ \emph {et~al.}(2017)\citenamefont {Ekimov},
  \citenamefont {Krivobok}, \citenamefont {Lyapin}, \citenamefont {Sherin},
  \citenamefont {Gavva},\ and\ \citenamefont {Kondrin}}]{Ekimov2}%
  \BibitemOpen
  \bibfield  {author} {\bibinfo {author} {\bibfnamefont {E.~A.}\ \bibnamefont
  {Ekimov}}, \bibinfo {author} {\bibfnamefont {V.~S.}\ \bibnamefont
  {Krivobok}}, \bibinfo {author} {\bibfnamefont {S.~G.}\ \bibnamefont
  {Lyapin}}, \bibinfo {author} {\bibfnamefont {P.~S.}\ \bibnamefont {Sherin}},
  \bibinfo {author} {\bibfnamefont {V.~A.}\ \bibnamefont {Gavva}}, \ and\
  \bibinfo {author} {\bibfnamefont {M.~V.}\ \bibnamefont {Kondrin}},\
  }\bibfield  {title} {\enquote {\bibinfo {title} {Anharmonicity effects in
  impurity-vacancy centers in diamond revealed by isotopic shifts and optical
  measurements},}\ }\href {\doibase 10.1103/PhysRevB.95.094113} {\bibfield
  {journal} {\bibinfo  {journal} {Phys. Rev. B}\ }\textbf {\bibinfo {volume}
  {95}},\ \bibinfo {pages} {094113} (\bibinfo {year} {2017})}\BibitemShut
  {NoStop}%
\bibitem [{\citenamefont {Bhaskar}\ \emph {et~al.}(2017)\citenamefont
  {Bhaskar}, \citenamefont {Sukachev}, \citenamefont {Sipahigil}, \citenamefont
  {Evans}, \citenamefont {Burek}, \citenamefont {Nguyen}, \citenamefont
  {Rogers}, \citenamefont {Siyushev}, \citenamefont {Metsch}, \citenamefont
  {Park}, \citenamefont {Jelezko}, \citenamefont {Lon\ifmmode~\check{c}\else
  \v{c}\fi{}ar},\ and\ \citenamefont {Lukin}}]{Bhaskar}%
  \BibitemOpen
  \bibfield  {author} {\bibinfo {author} {\bibfnamefont {M.~K.}\ \bibnamefont
  {Bhaskar}}, \bibinfo {author} {\bibfnamefont {D.~D.}\ \bibnamefont
  {Sukachev}}, \bibinfo {author} {\bibfnamefont {A.}~\bibnamefont {Sipahigil}},
  \bibinfo {author} {\bibfnamefont {R.~E.}\ \bibnamefont {Evans}}, \bibinfo
  {author} {\bibfnamefont {M.~J.}\ \bibnamefont {Burek}}, \bibinfo {author}
  {\bibfnamefont {C.~T.}\ \bibnamefont {Nguyen}}, \bibinfo {author}
  {\bibfnamefont {L.~J.}\ \bibnamefont {Rogers}}, \bibinfo {author}
  {\bibfnamefont {P.}~\bibnamefont {Siyushev}}, \bibinfo {author}
  {\bibfnamefont {M.~H.}\ \bibnamefont {Metsch}}, \bibinfo {author}
  {\bibfnamefont {H.}~\bibnamefont {Park}}, \bibinfo {author} {\bibfnamefont
  {F.}~\bibnamefont {Jelezko}}, \bibinfo {author} {\bibfnamefont
  {M.}~\bibnamefont {Lon\ifmmode~\check{c}\else \v{c}\fi{}ar}}, \ and\ \bibinfo
  {author} {\bibfnamefont {M.~D.}\ \bibnamefont {Lukin}},\ }\bibfield  {title}
  {\enquote {\bibinfo {title} {Quantum nonlinear optics with a
  germanium-vacancy color center in a nanoscale diamond waveguide},}\ }\href
  {\doibase 10.1103/PhysRevLett.118.223603} {\bibfield  {journal} {\bibinfo
  {journal} {Phys. Rev. Lett.}\ }\textbf {\bibinfo {volume} {118}},\ \bibinfo
  {pages} {223603} (\bibinfo {year} {2017})}\BibitemShut {NoStop}%
\bibitem [{\citenamefont {Bray}\ \emph {et~al.}(2018)\citenamefont {Bray},
  \citenamefont {Regan}, \citenamefont {Trycz}, \citenamefont {Previdi},
  \citenamefont {Seniutinas}, \citenamefont {Ganesan}, \citenamefont
  {Kianinia}, \citenamefont {Kim},\ and\ \citenamefont
  {Aharonovich}}]{bray2018single}%
  \BibitemOpen
  \bibfield  {author} {\bibinfo {author} {\bibfnamefont {K.}~\bibnamefont
  {Bray}}, \bibinfo {author} {\bibfnamefont {B.}~\bibnamefont {Regan}},
  \bibinfo {author} {\bibfnamefont {A.}~\bibnamefont {Trycz}}, \bibinfo
  {author} {\bibfnamefont {R.}~\bibnamefont {Previdi}}, \bibinfo {author}
  {\bibfnamefont {G.}~\bibnamefont {Seniutinas}}, \bibinfo {author}
  {\bibfnamefont {K.}~\bibnamefont {Ganesan}}, \bibinfo {author} {\bibfnamefont
  {M.}~\bibnamefont {Kianinia}}, \bibinfo {author} {\bibfnamefont
  {S.}~\bibnamefont {Kim}}, \ and\ \bibinfo {author} {\bibfnamefont
  {I.}~\bibnamefont {Aharonovich}},\ }\bibfield  {title} {\enquote {\bibinfo
  {title} {{Single Crystal Diamond Membranes and Photonic Resonators Containing
  Germanium Vacancy Color Centers}},}\ }\href {\doibase
  10.1021/acsphotonics.8b00930} {\bibfield  {journal} {\bibinfo  {journal} {ACS
  Photonics}\ }\textbf {\bibinfo {volume} {5}},\ \bibinfo {pages} {4817}
  (\bibinfo {year} {2018})}\BibitemShut {NoStop}%
\bibitem [{\citenamefont {Ekimov}\ \emph {et~al.}(2019)\citenamefont {Ekimov},
  \citenamefont {Kondrin}, \citenamefont {Krivobok}, \citenamefont {Khomich},
  \citenamefont {Vlasov}, \citenamefont {Khmelnitskiy}, \citenamefont
  {Iwasaki},\ and\ \citenamefont {Hatano}}]{ekimov2019effect}%
  \BibitemOpen
  \bibfield  {author} {\bibinfo {author} {\bibfnamefont {E.}~\bibnamefont
  {Ekimov}}, \bibinfo {author} {\bibfnamefont {M.}~\bibnamefont {Kondrin}},
  \bibinfo {author} {\bibfnamefont {V.}~\bibnamefont {Krivobok}}, \bibinfo
  {author} {\bibfnamefont {A.}~\bibnamefont {Khomich}}, \bibinfo {author}
  {\bibfnamefont {I.}~\bibnamefont {Vlasov}}, \bibinfo {author} {\bibfnamefont
  {R.}~\bibnamefont {Khmelnitskiy}}, \bibinfo {author} {\bibfnamefont
  {T.}~\bibnamefont {Iwasaki}}, \ and\ \bibinfo {author} {\bibfnamefont
  {M.}~\bibnamefont {Hatano}},\ }\bibfield  {title} {\enquote {\bibinfo {title}
  {{Effect of Si, Ge and Sn dopant elements on structure and photoluminescence
  of nano- and microdiamonds synthesized from organic compounds}},}\ }\href
  {\doibase https://doi.org/10.1016/j.diamond.2019.01.029} {\bibfield
  {journal} {\bibinfo  {journal} {Diam. Relat. Mater.}\ }\textbf {\bibinfo
  {volume} {93}},\ \bibinfo {pages} {75} (\bibinfo {year} {2019})}\BibitemShut
  {NoStop}%
\bibitem [{\citenamefont {Ditalia~Tchernij}\ \emph {et~al.}(2017)\citenamefont
  {Ditalia~Tchernij}, \citenamefont {Herzig}, \citenamefont {Forneris},
  \citenamefont {K\"upper}, \citenamefont {Pezzagna}, \citenamefont {Traina},
  \citenamefont {Moreva}, \citenamefont {Degiovanni}, \citenamefont {Brida},
  \citenamefont {Skukan}, \citenamefont {Genovese}, \citenamefont
  {Jak\ifmmode~\breve{s}\else \u{s}\fi{}i\'c}, \citenamefont {Meijer},\ and\
  \citenamefont {Olivero}}]{tchernij2017single}%
  \BibitemOpen
  \bibfield  {author} {\bibinfo {author} {\bibfnamefont {S.}~\bibnamefont
  {Ditalia~Tchernij}}, \bibinfo {author} {\bibfnamefont {T.}~\bibnamefont
  {Herzig}}, \bibinfo {author} {\bibfnamefont {J.}~\bibnamefont {Forneris}},
  \bibinfo {author} {\bibfnamefont {J.}~\bibnamefont {K\"upper}}, \bibinfo
  {author} {\bibfnamefont {S.}~\bibnamefont {Pezzagna}}, \bibinfo {author}
  {\bibfnamefont {P.}~\bibnamefont {Traina}}, \bibinfo {author} {\bibfnamefont
  {E.}~\bibnamefont {Moreva}}, \bibinfo {author} {\bibfnamefont {I.~P.}\
  \bibnamefont {Degiovanni}}, \bibinfo {author} {\bibfnamefont
  {G.}~\bibnamefont {Brida}}, \bibinfo {author} {\bibfnamefont
  {N.}~\bibnamefont {Skukan}}, \bibinfo {author} {\bibfnamefont
  {M.}~\bibnamefont {Genovese}}, \bibinfo {author} {\bibfnamefont
  {M.}~\bibnamefont {Jak\ifmmode~\breve{s}\else \u{s}\fi{}i\'c}}, \bibinfo
  {author} {\bibfnamefont {J.}~\bibnamefont {Meijer}}, \ and\ \bibinfo {author}
  {\bibfnamefont {P.}~\bibnamefont {Olivero}},\ }\bibfield  {title} {\enquote
  {\bibinfo {title} {{Single-Photon-Emitting Optical Centers in Diamond
  Fabricated upon Sn Implantation}},}\ }\href {\doibase
  10.1021/acsphotonics.7b00904} {\bibfield  {journal} {\bibinfo  {journal} {ACS
  Photonics}\ }\textbf {\bibinfo {volume} {4}},\ \bibinfo {pages} {2580}
  (\bibinfo {year} {2017})}\BibitemShut {NoStop}%
\bibitem [{\citenamefont {Ekimov}, \citenamefont {Lyapin},\ and\ \citenamefont
  {Kondrin}(2018)}]{Ekimov2018tin}%
  \BibitemOpen
  \bibfield  {author} {\bibinfo {author} {\bibfnamefont {E.}~\bibnamefont
  {Ekimov}}, \bibinfo {author} {\bibfnamefont {S.}~\bibnamefont {Lyapin}}, \
  and\ \bibinfo {author} {\bibfnamefont {M.}~\bibnamefont {Kondrin}},\
  }\bibfield  {title} {\enquote {\bibinfo {title} {Tin-vacancy color centers in
  micro- and polycrystalline diamonds synthesized at high pressures},}\ }\href
  {\doibase https://doi.org/10.1016/j.diamond.2018.06.014} {\bibfield
  {journal} {\bibinfo  {journal} {Diam. Relat. Mater.}\ }\textbf {\bibinfo
  {volume} {87}},\ \bibinfo {pages} {223} (\bibinfo {year} {2018})}\BibitemShut
  {NoStop}%
\bibitem [{\citenamefont {Palyanov}, \citenamefont {Kupriyanov},\ and\
  \citenamefont {Borzdov}(2019)}]{palyanov2019high}%
  \BibitemOpen
  \bibfield  {author} {\bibinfo {author} {\bibfnamefont {Y.~N.}\ \bibnamefont
  {Palyanov}}, \bibinfo {author} {\bibfnamefont {I.~N.}\ \bibnamefont
  {Kupriyanov}}, \ and\ \bibinfo {author} {\bibfnamefont {Y.~M.}\ \bibnamefont
  {Borzdov}},\ }\bibfield  {title} {\enquote {\bibinfo {title} {{High-pressure
  synthesis and characterization of Sn-doped single crystal diamond}},}\ }\href
  {\doibase https://doi.org/10.1016/j.carbon.2018.11.084} {\bibfield  {journal}
  {\bibinfo  {journal} {Carbon}\ }\textbf {\bibinfo {volume} {143}},\ \bibinfo
  {pages} {769} (\bibinfo {year} {2019})}\BibitemShut {NoStop}%
\bibitem [{\citenamefont {Alkahtani}\ \emph {et~al.}(2018)\citenamefont
  {Alkahtani}, \citenamefont {Cojocaru}, \citenamefont {Liu}, \citenamefont
  {Herzig}, \citenamefont {Meijer}, \citenamefont {K{\"u}pper}, \citenamefont
  {L{\"u}hmann}, \citenamefont {Akimov},\ and\ \citenamefont
  {Hemmer}}]{Alkahtani2018tin}%
  \BibitemOpen
  \bibfield  {author} {\bibinfo {author} {\bibfnamefont {M.}~\bibnamefont
  {Alkahtani}}, \bibinfo {author} {\bibfnamefont {I.}~\bibnamefont {Cojocaru}},
  \bibinfo {author} {\bibfnamefont {X.}~\bibnamefont {Liu}}, \bibinfo {author}
  {\bibfnamefont {T.}~\bibnamefont {Herzig}}, \bibinfo {author} {\bibfnamefont
  {J.}~\bibnamefont {Meijer}}, \bibinfo {author} {\bibfnamefont
  {J.}~\bibnamefont {K{\"u}pper}}, \bibinfo {author} {\bibfnamefont
  {T.}~\bibnamefont {L{\"u}hmann}}, \bibinfo {author} {\bibfnamefont {A.~V.}\
  \bibnamefont {Akimov}}, \ and\ \bibinfo {author} {\bibfnamefont {P.~R.}\
  \bibnamefont {Hemmer}},\ }\bibfield  {title} {\enquote {\bibinfo {title}
  {Tin-vacancy in diamonds for luminescent thermometry},}\ }\href {\doibase
  10.1063/1.5037053} {\bibfield  {journal} {\bibinfo  {journal} {Appl. Phys.
  Lett.}\ }\textbf {\bibinfo {volume} {112}},\ \bibinfo {pages} {241902}
  (\bibinfo {year} {2018})}\BibitemShut {NoStop}%
\bibitem [{\citenamefont {Rugar}\ \emph {et~al.}(2019)\citenamefont {Rugar},
  \citenamefont {Dory}, \citenamefont {Sun},\ and\ \citenamefont {Vu\ifmmode
  \check{c}\else \v{c}\fi{}kovi\ifmmode~\acute{c}\else
  \'{c}\fi{}}}]{rugar2019char}%
  \BibitemOpen
  \bibfield  {author} {\bibinfo {author} {\bibfnamefont {A.~E.}\ \bibnamefont
  {Rugar}}, \bibinfo {author} {\bibfnamefont {C.}~\bibnamefont {Dory}},
  \bibinfo {author} {\bibfnamefont {S.}~\bibnamefont {Sun}}, \ and\ \bibinfo
  {author} {\bibfnamefont {J.}~\bibnamefont {Vu\ifmmode \check{c}\else
  \v{c}\fi{}kovi\ifmmode~\acute{c}\else \'{c}\fi{}}},\ }\bibfield  {title}
  {\enquote {\bibinfo {title} {{Characterization of optical and spin properties
  of single tin-vacancy centers in diamond nanopillars}},}\ }\href {\doibase
  10.1103/PhysRevB.99.205417} {\bibfield  {journal} {\bibinfo  {journal} {Phys.
  Rev. B}\ }\textbf {\bibinfo {volume} {99}},\ \bibinfo {pages} {205417}
  (\bibinfo {year} {2019})}\BibitemShut {NoStop}%
\bibitem [{\citenamefont {Trusheim}\ \emph {et~al.}(2019)\citenamefont
  {Trusheim}, \citenamefont {Wan}, \citenamefont {Chen}, \citenamefont
  {Ciccarino}, \citenamefont {Flick}, \citenamefont {Sundararaman},
  \citenamefont {Malladi}, \citenamefont {Bersin}, \citenamefont {Walsh},
  \citenamefont {Lienhard}, \citenamefont {Bakhru}, \citenamefont {Narang},\
  and\ \citenamefont {Englund}}]{Trusheim}%
  \BibitemOpen
  \bibfield  {author} {\bibinfo {author} {\bibfnamefont {M.~E.}\ \bibnamefont
  {Trusheim}}, \bibinfo {author} {\bibfnamefont {N.~H.}\ \bibnamefont {Wan}},
  \bibinfo {author} {\bibfnamefont {K.~C.}\ \bibnamefont {Chen}}, \bibinfo
  {author} {\bibfnamefont {C.~J.}\ \bibnamefont {Ciccarino}}, \bibinfo {author}
  {\bibfnamefont {J.}~\bibnamefont {Flick}}, \bibinfo {author} {\bibfnamefont
  {R.}~\bibnamefont {Sundararaman}}, \bibinfo {author} {\bibfnamefont
  {G.}~\bibnamefont {Malladi}}, \bibinfo {author} {\bibfnamefont
  {E.}~\bibnamefont {Bersin}}, \bibinfo {author} {\bibfnamefont
  {M.}~\bibnamefont {Walsh}}, \bibinfo {author} {\bibfnamefont
  {B.}~\bibnamefont {Lienhard}}, \bibinfo {author} {\bibfnamefont
  {H.}~\bibnamefont {Bakhru}}, \bibinfo {author} {\bibfnamefont
  {P.}~\bibnamefont {Narang}}, \ and\ \bibinfo {author} {\bibfnamefont
  {D.}~\bibnamefont {Englund}},\ }\bibfield  {title} {\enquote {\bibinfo
  {title} {Lead-related quantum emitters in diamond},}\ }\href {\doibase
  10.1103/PhysRevB.99.075430} {\bibfield  {journal} {\bibinfo  {journal} {Phys.
  Rev. B}\ }\textbf {\bibinfo {volume} {99}},\ \bibinfo {pages} {075430}
  (\bibinfo {year} {2019})}\BibitemShut {NoStop}%
\bibitem [{\citenamefont {Ditalia~Tchernij}\ \emph {et~al.}(2018)\citenamefont
  {Ditalia~Tchernij}, \citenamefont {L\"{u}hmann}, \citenamefont {Herzig},
  \citenamefont {K\"{u}pper}, \citenamefont {Damin}, \citenamefont
  {Santonocito}, \citenamefont {Signorile}, \citenamefont {Traina},
  \citenamefont {Moreva}, \citenamefont {Celegato}, \citenamefont {Pezzagna},
  \citenamefont {Degiovanni}, \citenamefont {Olivero}, \citenamefont
  {Jak\ifmmode~\breve{s}\else \u{s}\fi{}i\'c}, \citenamefont {Meijer},
  \citenamefont {Genovese},\ and\ \citenamefont
  {Forneris}}]{tchernij2018single}%
  \BibitemOpen
  \bibfield  {author} {\bibinfo {author} {\bibfnamefont {S.}~\bibnamefont
  {Ditalia~Tchernij}}, \bibinfo {author} {\bibfnamefont {T.}~\bibnamefont
  {L\"{u}hmann}}, \bibinfo {author} {\bibfnamefont {T.}~\bibnamefont {Herzig}},
  \bibinfo {author} {\bibfnamefont {J.}~\bibnamefont {K\"{u}pper}}, \bibinfo
  {author} {\bibfnamefont {A.}~\bibnamefont {Damin}}, \bibinfo {author}
  {\bibfnamefont {S.}~\bibnamefont {Santonocito}}, \bibinfo {author}
  {\bibfnamefont {M.}~\bibnamefont {Signorile}}, \bibinfo {author}
  {\bibfnamefont {P.}~\bibnamefont {Traina}}, \bibinfo {author} {\bibfnamefont
  {E.}~\bibnamefont {Moreva}}, \bibinfo {author} {\bibfnamefont
  {F.}~\bibnamefont {Celegato}}, \bibinfo {author} {\bibfnamefont
  {S.}~\bibnamefont {Pezzagna}}, \bibinfo {author} {\bibfnamefont {I.~P.}\
  \bibnamefont {Degiovanni}}, \bibinfo {author} {\bibfnamefont
  {P.}~\bibnamefont {Olivero}}, \bibinfo {author} {\bibfnamefont
  {M.}~\bibnamefont {Jak\ifmmode~\breve{s}\else \u{s}\fi{}i\'c}}, \bibinfo
  {author} {\bibfnamefont {J.}~\bibnamefont {Meijer}}, \bibinfo {author}
  {\bibfnamefont {P.~M.}\ \bibnamefont {Genovese}}, \ and\ \bibinfo {author}
  {\bibfnamefont {J.}~\bibnamefont {Forneris}},\ }\bibfield  {title} {\enquote
  {\bibinfo {title} {{Single-Photon Emitters in Lead-Implanted Single-Crystal
  Diamond}},}\ }\href {\doibase 10.1021/acsphotonics.8b01013} {\bibfield
  {journal} {\bibinfo  {journal} {ACS Photonics}\ }\textbf {\bibinfo {volume}
  {5}},\ \bibinfo {pages} {4864} (\bibinfo {year} {2018})}\BibitemShut
  {NoStop}%
\bibitem [{\citenamefont {Thiering}\ and\ \citenamefont {Gali}(2018)}]{Gali3}%
  \BibitemOpen
  \bibfield  {author} {\bibinfo {author} {\bibfnamefont {G.}~\bibnamefont
  {Thiering}}\ and\ \bibinfo {author} {\bibfnamefont {A.}~\bibnamefont
  {Gali}},\ }\bibfield  {title} {\enquote {\bibinfo {title} {{Ab Initio
  Magneto-Optical Spectrum of Group-IV Vacancy Color Centers in Diamond}},}\
  }\href {\doibase 10.1103/PhysRevX.8.021063} {\bibfield  {journal} {\bibinfo
  {journal} {Phys. Rev. X}\ }\textbf {\bibinfo {volume} {8}},\ \bibinfo {pages}
  {021063} (\bibinfo {year} {2018})}\BibitemShut {NoStop}%
\bibitem [{\citenamefont {Meijer}\ \emph {et~al.}(2005)\citenamefont {Meijer},
  \citenamefont {Burchard}, \citenamefont {Domhan}, \citenamefont {Wittmann},
  \citenamefont {Gaebel}, \citenamefont {Popa}, \citenamefont {Jelezko},\ and\
  \citenamefont {Wrachtrup}}]{meijer2005generation}%
  \BibitemOpen
  \bibfield  {author} {\bibinfo {author} {\bibfnamefont {J.}~\bibnamefont
  {Meijer}}, \bibinfo {author} {\bibfnamefont {B.}~\bibnamefont {Burchard}},
  \bibinfo {author} {\bibfnamefont {M.}~\bibnamefont {Domhan}}, \bibinfo
  {author} {\bibfnamefont {C.}~\bibnamefont {Wittmann}}, \bibinfo {author}
  {\bibfnamefont {T.}~\bibnamefont {Gaebel}}, \bibinfo {author} {\bibfnamefont
  {I.}~\bibnamefont {Popa}}, \bibinfo {author} {\bibfnamefont {F.}~\bibnamefont
  {Jelezko}}, \ and\ \bibinfo {author} {\bibfnamefont {J.}~\bibnamefont
  {Wrachtrup}},\ }\bibfield  {title} {\enquote {\bibinfo {title} {Generation of
  single color centers by focused nitrogen implantation},}\ }\href
  {https://doi.org/10.1063/1.2103389} {\bibfield  {journal} {\bibinfo
  {journal} {Appl. Phys. Lett.}\ }\textbf {\bibinfo {volume} {87}},\ \bibinfo
  {pages} {261909} (\bibinfo {year} {2005})}\BibitemShut {NoStop}%
\bibitem [{\citenamefont {Pezzagna}\ \emph {et~al.}(2010)\citenamefont
  {Pezzagna}, \citenamefont {Naydenov}, \citenamefont {Jelezko}, \citenamefont
  {Wrachtrup},\ and\ \citenamefont {Meijer}}]{pezzagna2010creat}%
  \BibitemOpen
  \bibfield  {author} {\bibinfo {author} {\bibfnamefont {S.}~\bibnamefont
  {Pezzagna}}, \bibinfo {author} {\bibfnamefont {B.}~\bibnamefont {Naydenov}},
  \bibinfo {author} {\bibfnamefont {F.}~\bibnamefont {Jelezko}}, \bibinfo
  {author} {\bibfnamefont {J.}~\bibnamefont {Wrachtrup}}, \ and\ \bibinfo
  {author} {\bibfnamefont {J.}~\bibnamefont {Meijer}},\ }\bibfield  {title}
  {\enquote {\bibinfo {title} {{Creation efficiency of nitrogen-vacancy centres
  in diamond}},}\ }\href {\doibase 10.1088/1367-2630/12/6/065017} {\bibfield
  {journal} {\bibinfo  {journal} {New J. Phys.}\ }\textbf {\bibinfo {volume}
  {12}},\ \bibinfo {pages} {065017} (\bibinfo {year} {2010})}\BibitemShut
  {NoStop}%
\bibitem [{\citenamefont {Rabeau}\ \emph {et~al.}(2006)\citenamefont {Rabeau},
  \citenamefont {Reichart}, \citenamefont {Tamanyan}, \citenamefont {Jamieson},
  \citenamefont {Prawer}, \citenamefont {Jelezko}, \citenamefont {Gaebel},
  \citenamefont {Popa}, \citenamefont {Domhan},\ and\ \citenamefont
  {Wrachtrup}}]{rabeau2006implantation}%
  \BibitemOpen
  \bibfield  {author} {\bibinfo {author} {\bibfnamefont {J.}~\bibnamefont
  {Rabeau}}, \bibinfo {author} {\bibfnamefont {P.}~\bibnamefont {Reichart}},
  \bibinfo {author} {\bibfnamefont {G.}~\bibnamefont {Tamanyan}}, \bibinfo
  {author} {\bibfnamefont {D.}~\bibnamefont {Jamieson}}, \bibinfo {author}
  {\bibfnamefont {S.}~\bibnamefont {Prawer}}, \bibinfo {author} {\bibfnamefont
  {F.}~\bibnamefont {Jelezko}}, \bibinfo {author} {\bibfnamefont
  {T.}~\bibnamefont {Gaebel}}, \bibinfo {author} {\bibfnamefont
  {I.}~\bibnamefont {Popa}}, \bibinfo {author} {\bibfnamefont {M.}~\bibnamefont
  {Domhan}}, \ and\ \bibinfo {author} {\bibfnamefont {J.}~\bibnamefont
  {Wrachtrup}},\ }\bibfield  {title} {\enquote {\bibinfo {title} {{Implantation
  of labelled single nitrogen vacancy centers in diamond using $^{15}$N}},}\
  }\href {https://doi.org/10.1063/1.2158700} {\bibfield  {journal} {\bibinfo
  {journal} {Appl. Phys. Lett.}\ }\textbf {\bibinfo {volume} {88}},\ \bibinfo
  {pages} {023113} (\bibinfo {year} {2006})}\BibitemShut {NoStop}%
\bibitem [{\citenamefont {Weis}\ \emph {et~al.}(2008)\citenamefont {Weis},
  \citenamefont {Schuh}, \citenamefont {Batra}, \citenamefont {Persaud},
  \citenamefont {Rangelow}, \citenamefont {Bokor}, \citenamefont {Lo},
  \citenamefont {Cabrini}, \citenamefont {Sideras-Haddad}, \citenamefont
  {Fuchs} \emph {et~al.}}]{weis2008single}%
  \BibitemOpen
  \bibfield  {author} {\bibinfo {author} {\bibfnamefont {C.~D.}\ \bibnamefont
  {Weis}}, \bibinfo {author} {\bibfnamefont {A.}~\bibnamefont {Schuh}},
  \bibinfo {author} {\bibfnamefont {A.}~\bibnamefont {Batra}}, \bibinfo
  {author} {\bibfnamefont {A.}~\bibnamefont {Persaud}}, \bibinfo {author}
  {\bibfnamefont {I.~W.}\ \bibnamefont {Rangelow}}, \bibinfo {author}
  {\bibfnamefont {J.}~\bibnamefont {Bokor}}, \bibinfo {author} {\bibfnamefont
  {C.~C.}\ \bibnamefont {Lo}}, \bibinfo {author} {\bibfnamefont
  {S.}~\bibnamefont {Cabrini}}, \bibinfo {author} {\bibfnamefont
  {E.}~\bibnamefont {Sideras-Haddad}}, \bibinfo {author} {\bibfnamefont
  {G.}~\bibnamefont {Fuchs}},  \emph {et~al.},\ }\bibfield  {title} {\enquote
  {\bibinfo {title} {Single atom doping for quantum device development in
  diamond and silicon},}\ }\href
  {https://avs.scitation.org/doi/abs/10.1116/1.2968614} {\bibfield  {journal}
  {\bibinfo  {journal} {J. Vac. Sci. Technol. B}\ }\textbf {\bibinfo {volume}
  {26}},\ \bibinfo {pages} {2596} (\bibinfo {year} {2008})}\BibitemShut
  {NoStop}%
\bibitem [{\citenamefont {L\"{u}hmann}\ \emph {et~al.}(2018)\citenamefont
  {L\"{u}hmann}, \citenamefont {Raatz}, \citenamefont {John}, \citenamefont
  {Lesik}, \citenamefont {R\"{o}diger}, \citenamefont {Portail}, \citenamefont
  {Wildanger}, \citenamefont {Klei{\ss}ler}, \citenamefont {Nordlund},
  \citenamefont {Zaitsev}, \citenamefont {Roch}, \citenamefont {Tallaire},
  \citenamefont {Meijer},\ and\ \citenamefont {Pezzagna}}]{Luhmann2018screen}%
  \BibitemOpen
  \bibfield  {author} {\bibinfo {author} {\bibfnamefont {T.}~\bibnamefont
  {L\"{u}hmann}}, \bibinfo {author} {\bibfnamefont {N.}~\bibnamefont {Raatz}},
  \bibinfo {author} {\bibfnamefont {R.}~\bibnamefont {John}}, \bibinfo {author}
  {\bibfnamefont {M.}~\bibnamefont {Lesik}}, \bibinfo {author} {\bibfnamefont
  {J.}~\bibnamefont {R\"{o}diger}}, \bibinfo {author} {\bibfnamefont
  {M.}~\bibnamefont {Portail}}, \bibinfo {author} {\bibfnamefont
  {D.}~\bibnamefont {Wildanger}}, \bibinfo {author} {\bibfnamefont
  {F.}~\bibnamefont {Klei{\ss}ler}}, \bibinfo {author} {\bibfnamefont
  {K.}~\bibnamefont {Nordlund}}, \bibinfo {author} {\bibfnamefont
  {A.}~\bibnamefont {Zaitsev}}, \bibinfo {author} {\bibfnamefont {J.-F.}\
  \bibnamefont {Roch}}, \bibinfo {author} {\bibfnamefont {A.}~\bibnamefont
  {Tallaire}}, \bibinfo {author} {\bibfnamefont {J.}~\bibnamefont {Meijer}}, \
  and\ \bibinfo {author} {\bibfnamefont {S.}~\bibnamefont {Pezzagna}},\
  }\bibfield  {title} {\enquote {\bibinfo {title} {Screening and engineering of
  colour centres in diamond},}\ }\href {\doibase 10.1088/1361-6463/aadfab}
  {\bibfield  {journal} {\bibinfo  {journal} {J. Phys. D: Appl. Phys.}\
  }\textbf {\bibinfo {volume} {51}},\ \bibinfo {pages} {483002} (\bibinfo
  {year} {2018})}\BibitemShut {NoStop}%
\bibitem [{\citenamefont {Schwartz}\ \emph {et~al.}(2011)\citenamefont
  {Schwartz}, \citenamefont {Michaelides}, \citenamefont {Weis},\ and\
  \citenamefont {Schenkel}}]{schwartz2011situ}%
  \BibitemOpen
  \bibfield  {author} {\bibinfo {author} {\bibfnamefont {J.}~\bibnamefont
  {Schwartz}}, \bibinfo {author} {\bibfnamefont {P.}~\bibnamefont
  {Michaelides}}, \bibinfo {author} {\bibfnamefont {C.}~\bibnamefont {Weis}}, \
  and\ \bibinfo {author} {\bibfnamefont {T.}~\bibnamefont {Schenkel}},\
  }\bibfield  {title} {\enquote {\bibinfo {title} {In situ optimization of
  co-implantation and substrate temperature conditions for nitrogen-vacancy
  center formation in single-crystal diamonds},}\ }\href
  {https://doi.org/10.1088\%2F1367-2630\%2F13\%2F3\%2F035022} {\bibfield
  {journal} {\bibinfo  {journal} {New J. Phys.}\ }\textbf {\bibinfo {volume}
  {13}},\ \bibinfo {pages} {035022} (\bibinfo {year} {2011})}\BibitemShut
  {NoStop}%
\bibitem [{\citenamefont {Naydenov}\ \emph {et~al.}(2010)\citenamefont
  {Naydenov}, \citenamefont {Richter}, \citenamefont {Beck}, \citenamefont
  {Steiner}, \citenamefont {Neumann}, \citenamefont {Balasubramanian},
  \citenamefont {Achard}, \citenamefont {Jelezko}, \citenamefont {Wrachtrup},\
  and\ \citenamefont {Kalish}}]{Naydenov}%
  \BibitemOpen
  \bibfield  {author} {\bibinfo {author} {\bibfnamefont {B.}~\bibnamefont
  {Naydenov}}, \bibinfo {author} {\bibfnamefont {V.}~\bibnamefont {Richter}},
  \bibinfo {author} {\bibfnamefont {J.}~\bibnamefont {Beck}}, \bibinfo {author}
  {\bibfnamefont {M.}~\bibnamefont {Steiner}}, \bibinfo {author} {\bibfnamefont
  {P.}~\bibnamefont {Neumann}}, \bibinfo {author} {\bibfnamefont
  {G.}~\bibnamefont {Balasubramanian}}, \bibinfo {author} {\bibfnamefont
  {J.}~\bibnamefont {Achard}}, \bibinfo {author} {\bibfnamefont
  {F.}~\bibnamefont {Jelezko}}, \bibinfo {author} {\bibfnamefont
  {J.}~\bibnamefont {Wrachtrup}}, \ and\ \bibinfo {author} {\bibfnamefont
  {R.}~\bibnamefont {Kalish}},\ }\bibfield  {title} {\enquote {\bibinfo {title}
  {Enhanced generation of single optically active spins in diamond by ion
  implantation},}\ }\href {\doibase 10.1063/1.3409221} {\bibfield  {journal}
  {\bibinfo  {journal} {Appl. Phys. Lett.}\ }\textbf {\bibinfo {volume} {96}},\
  \bibinfo {pages} {163108} (\bibinfo {year} {2010})}\BibitemShut {NoStop}%
\bibitem [{\citenamefont {Yamamoto}\ \emph {et~al.}(2013)\citenamefont
  {Yamamoto}, \citenamefont {Umeda}, \citenamefont {Watanabe}, \citenamefont
  {Onoda}, \citenamefont {Markham}, \citenamefont {Twitchen}, \citenamefont
  {Naydenov}, \citenamefont {McGuinness}, \citenamefont {Teraji}, \citenamefont
  {Koizumi}, \citenamefont {Dolde}, \citenamefont {Fedder}, \citenamefont
  {Honert}, \citenamefont {Wrachtrup}, \citenamefont {Ohshima}, \citenamefont
  {Jelezko},\ and\ \citenamefont {Isoya}}]{Yamamoto}%
  \BibitemOpen
  \bibfield  {author} {\bibinfo {author} {\bibfnamefont {T.}~\bibnamefont
  {Yamamoto}}, \bibinfo {author} {\bibfnamefont {T.}~\bibnamefont {Umeda}},
  \bibinfo {author} {\bibfnamefont {K.}~\bibnamefont {Watanabe}}, \bibinfo
  {author} {\bibfnamefont {S.}~\bibnamefont {Onoda}}, \bibinfo {author}
  {\bibfnamefont {M.~L.}\ \bibnamefont {Markham}}, \bibinfo {author}
  {\bibfnamefont {D.~J.}\ \bibnamefont {Twitchen}}, \bibinfo {author}
  {\bibfnamefont {B.}~\bibnamefont {Naydenov}}, \bibinfo {author}
  {\bibfnamefont {L.~P.}\ \bibnamefont {McGuinness}}, \bibinfo {author}
  {\bibfnamefont {T.}~\bibnamefont {Teraji}}, \bibinfo {author} {\bibfnamefont
  {S.}~\bibnamefont {Koizumi}}, \bibinfo {author} {\bibfnamefont
  {F.}~\bibnamefont {Dolde}}, \bibinfo {author} {\bibfnamefont
  {H.}~\bibnamefont {Fedder}}, \bibinfo {author} {\bibfnamefont
  {J.}~\bibnamefont {Honert}}, \bibinfo {author} {\bibfnamefont
  {J.}~\bibnamefont {Wrachtrup}}, \bibinfo {author} {\bibfnamefont
  {T.}~\bibnamefont {Ohshima}}, \bibinfo {author} {\bibfnamefont
  {F.}~\bibnamefont {Jelezko}}, \ and\ \bibinfo {author} {\bibfnamefont
  {J.}~\bibnamefont {Isoya}},\ }\bibfield  {title} {\enquote {\bibinfo {title}
  {Extending spin coherence times of diamond qubits by high-temperature
  annealing},}\ }\href {\doibase 10.1103/PhysRevB.88.075206} {\bibfield
  {journal} {\bibinfo  {journal} {Phys. Rev. B}\ }\textbf {\bibinfo {volume}
  {88}},\ \bibinfo {pages} {075206} (\bibinfo {year} {2013})}\BibitemShut
  {NoStop}%
\bibitem [{\citenamefont {Hounsome}\ \emph {et~al.}(2005)\citenamefont
  {Hounsome}, \citenamefont {Jones}, \citenamefont {Martineau}, \citenamefont
  {Shaw}, \citenamefont {Briddon}, \citenamefont {{\"O}berg}, \citenamefont
  {Blumenau},\ and\ \citenamefont {Fujita}}]{hounsome2005optical}%
  \BibitemOpen
  \bibfield  {author} {\bibinfo {author} {\bibfnamefont {L.}~\bibnamefont
  {Hounsome}}, \bibinfo {author} {\bibfnamefont {R.}~\bibnamefont {Jones}},
  \bibinfo {author} {\bibfnamefont {P.}~\bibnamefont {Martineau}}, \bibinfo
  {author} {\bibfnamefont {M.}~\bibnamefont {Shaw}}, \bibinfo {author}
  {\bibfnamefont {P.}~\bibnamefont {Briddon}}, \bibinfo {author} {\bibfnamefont
  {S.}~\bibnamefont {{\"O}berg}}, \bibinfo {author} {\bibfnamefont
  {A.}~\bibnamefont {Blumenau}}, \ and\ \bibinfo {author} {\bibfnamefont
  {N.}~\bibnamefont {Fujita}},\ }\bibfield  {title} {\enquote {\bibinfo {title}
  {Optical properties of vacancy related defects in diamond},}\ }\href@noop {}
  {\bibfield  {journal} {\bibinfo  {journal} {physica status solidi (a)}\
  }\textbf {\bibinfo {volume} {202}},\ \bibinfo {pages} {2182--2187} (\bibinfo
  {year} {2005})}\BibitemShut {NoStop}%
\bibitem [{\citenamefont {Iakoubovskii}\ and\ \citenamefont
  {Stesmans}(2004)}]{iakoubovskii2004vacancy}%
  \BibitemOpen
  \bibfield  {author} {\bibinfo {author} {\bibfnamefont {K.}~\bibnamefont
  {Iakoubovskii}}\ and\ \bibinfo {author} {\bibfnamefont {A.}~\bibnamefont
  {Stesmans}},\ }\bibfield  {title} {\enquote {\bibinfo {title} {Vacancy
  clusters in diamond studied by electron spin resonance},}\ }\href@noop {}
  {\bibfield  {journal} {\bibinfo  {journal} {physica status solidi (a)}\
  }\textbf {\bibinfo {volume} {201}},\ \bibinfo {pages} {2509--2515} (\bibinfo
  {year} {2004})}\BibitemShut {NoStop}%
\bibitem [{\citenamefont {F{\'a}varo~de Oliveira}\ \emph
  {et~al.}(2017)\citenamefont {F{\'a}varo~de Oliveira}, \citenamefont
  {Antonov}, \citenamefont {Wang}, \citenamefont {Neumann}, \citenamefont
  {Momenzadeh}, \citenamefont {H{\"a}u{\ss}ermann}, \citenamefont
  {Pasquarelli}, \citenamefont {Denisenko},\ and\ \citenamefont
  {Wrachtrup}}]{Oliveira}%
  \BibitemOpen
  \bibfield  {author} {\bibinfo {author} {\bibfnamefont {F.}~\bibnamefont
  {F{\'a}varo~de Oliveira}}, \bibinfo {author} {\bibfnamefont {D.}~\bibnamefont
  {Antonov}}, \bibinfo {author} {\bibfnamefont {Y.}~\bibnamefont {Wang}},
  \bibinfo {author} {\bibfnamefont {P.}~\bibnamefont {Neumann}}, \bibinfo
  {author} {\bibfnamefont {S.~A.}\ \bibnamefont {Momenzadeh}}, \bibinfo
  {author} {\bibfnamefont {T.}~\bibnamefont {H{\"a}u{\ss}ermann}}, \bibinfo
  {author} {\bibfnamefont {A.}~\bibnamefont {Pasquarelli}}, \bibinfo {author}
  {\bibfnamefont {A.}~\bibnamefont {Denisenko}}, \ and\ \bibinfo {author}
  {\bibfnamefont {J.}~\bibnamefont {Wrachtrup}},\ }\bibfield  {title} {\enquote
  {\bibinfo {title} {Tailoring spin defects in diamond by lattice charging},}\
  }\href {https://doi.org/10.1038/ncomms15409} {\bibfield  {journal} {\bibinfo
  {journal} {Nat. Commun.}\ }\textbf {\bibinfo {volume} {8}},\ \bibinfo {pages}
  {15409} (\bibinfo {year} {2017})}\BibitemShut {NoStop}%
\bibitem [{\citenamefont {Kresse}\ and\ \citenamefont
  {Hafner}(1993)}]{Kresse1}%
  \BibitemOpen
  \bibfield  {author} {\bibinfo {author} {\bibfnamefont {G.}~\bibnamefont
  {Kresse}}\ and\ \bibinfo {author} {\bibfnamefont {J.}~\bibnamefont
  {Hafner}},\ }\bibfield  {title} {\enquote {\bibinfo {title} {\textit{Ab
  initio} molecular dynamics for liquid metals},}\ }\href {\doibase
  10.1103/PhysRevB.47.558} {\bibfield  {journal} {\bibinfo  {journal} {Phys.
  Rev. B}\ }\textbf {\bibinfo {volume} {47}},\ \bibinfo {pages} {558} (\bibinfo
  {year} {1993})}\BibitemShut {NoStop}%
\bibitem [{\citenamefont {Kresse}\ and\ \citenamefont
  {Furthm\"uller}(1996)}]{Kresse2}%
  \BibitemOpen
  \bibfield  {author} {\bibinfo {author} {\bibfnamefont {G.}~\bibnamefont
  {Kresse}}\ and\ \bibinfo {author} {\bibfnamefont {J.}~\bibnamefont
  {Furthm\"uller}},\ }\bibfield  {title} {\enquote {\bibinfo {title} {Efficient
  iterative schemes for \textit{ab initio} total-energy calculations using a
  plane-wave basis set},}\ }\href {\doibase 10.1103/PhysRevB.54.11169}
  {\bibfield  {journal} {\bibinfo  {journal} {Phys. Rev. B}\ }\textbf {\bibinfo
  {volume} {54}},\ \bibinfo {pages} {11169} (\bibinfo {year}
  {1996})}\BibitemShut {NoStop}%
\bibitem [{\citenamefont {Kresse}\ and\ \citenamefont
  {Joubert}(1999)}]{Kresse3}%
  \BibitemOpen
  \bibfield  {author} {\bibinfo {author} {\bibfnamefont {G.}~\bibnamefont
  {Kresse}}\ and\ \bibinfo {author} {\bibfnamefont {D.}~\bibnamefont
  {Joubert}},\ }\bibfield  {title} {\enquote {\bibinfo {title} {From ultrasoft
  pseudopotentials to the projector augmented-wave method},}\ }\href {\doibase
  10.1103/PhysRevB.59.1758} {\bibfield  {journal} {\bibinfo  {journal} {Phys.
  Rev. B}\ }\textbf {\bibinfo {volume} {59}},\ \bibinfo {pages} {1758}
  (\bibinfo {year} {1999})}\BibitemShut {NoStop}%
\bibitem [{\citenamefont {Perdew}, \citenamefont {Burke},\ and\ \citenamefont
  {Ernzerhof}(1996)}]{Perdew}%
  \BibitemOpen
  \bibfield  {author} {\bibinfo {author} {\bibfnamefont {J.~P.}\ \bibnamefont
  {Perdew}}, \bibinfo {author} {\bibfnamefont {K.}~\bibnamefont {Burke}}, \
  and\ \bibinfo {author} {\bibfnamefont {M.}~\bibnamefont {Ernzerhof}},\
  }\bibfield  {title} {\enquote {\bibinfo {title} {Generalized gradient
  approximation made simple},}\ }\href {\doibase 10.1103/PhysRevLett.77.3865}
  {\bibfield  {journal} {\bibinfo  {journal} {Phys. Rev. Lett.}\ }\textbf
  {\bibinfo {volume} {77}},\ \bibinfo {pages} {3865} (\bibinfo {year}
  {1996})}\BibitemShut {NoStop}%
\bibitem [{\citenamefont {Zhang}\ and\ \citenamefont
  {Northrup}(1991)}]{zhang1991chemical}%
  \BibitemOpen
  \bibfield  {author} {\bibinfo {author} {\bibfnamefont {S.~B.}\ \bibnamefont
  {Zhang}}\ and\ \bibinfo {author} {\bibfnamefont {J.~E.}\ \bibnamefont
  {Northrup}},\ }\bibfield  {title} {\enquote {\bibinfo {title} {{Chemical
  potential dependence of defect formation energies in GaAs: Application to Ga
  self-diffusion}},}\ }\href {\doibase 10.1103/PhysRevLett.67.2339} {\bibfield
  {journal} {\bibinfo  {journal} {Phys. Rev. Lett.}\ }\textbf {\bibinfo
  {volume} {67}},\ \bibinfo {pages} {2339} (\bibinfo {year}
  {1991})}\BibitemShut {NoStop}%
\bibitem [{\citenamefont {Freysoldt}\ \emph {et~al.}(2014)\citenamefont
  {Freysoldt}, \citenamefont {Grabowski}, \citenamefont {Hickel}, \citenamefont
  {Neugebauer}, \citenamefont {Kresse}, \citenamefont {Janotti},\ and\
  \citenamefont {Van~de Walle}}]{RevModPhys.86.253}%
  \BibitemOpen
  \bibfield  {author} {\bibinfo {author} {\bibfnamefont {C.}~\bibnamefont
  {Freysoldt}}, \bibinfo {author} {\bibfnamefont {B.}~\bibnamefont
  {Grabowski}}, \bibinfo {author} {\bibfnamefont {T.}~\bibnamefont {Hickel}},
  \bibinfo {author} {\bibfnamefont {J.}~\bibnamefont {Neugebauer}}, \bibinfo
  {author} {\bibfnamefont {G.}~\bibnamefont {Kresse}}, \bibinfo {author}
  {\bibfnamefont {A.}~\bibnamefont {Janotti}}, \ and\ \bibinfo {author}
  {\bibfnamefont {C.~G.}\ \bibnamefont {Van~de Walle}},\ }\bibfield  {title}
  {\enquote {\bibinfo {title} {First-principles calculations for point defects
  in solids},}\ }\href {\doibase 10.1103/RevModPhys.86.253} {\bibfield
  {journal} {\bibinfo  {journal} {Rev. Mod. Phys.}\ }\textbf {\bibinfo {volume}
  {86}},\ \bibinfo {pages} {253} (\bibinfo {year} {2014})}\BibitemShut
  {NoStop}%
\bibitem [{\citenamefont {Vinichenko}\ \emph {et~al.}(2017)\citenamefont
  {Vinichenko}, \citenamefont {Sensoy}, \citenamefont {Friend},\ and\
  \citenamefont {Kaxiras}}]{Vinichenko}%
  \BibitemOpen
  \bibfield  {author} {\bibinfo {author} {\bibfnamefont {D.}~\bibnamefont
  {Vinichenko}}, \bibinfo {author} {\bibfnamefont {M.~G.}\ \bibnamefont
  {Sensoy}}, \bibinfo {author} {\bibfnamefont {C.~M.}\ \bibnamefont {Friend}},
  \ and\ \bibinfo {author} {\bibfnamefont {E.}~\bibnamefont {Kaxiras}},\
  }\bibfield  {title} {\enquote {\bibinfo {title} {Accurate formation energies
  of charged defects in solids: A systematic approach},}\ }\href {\doibase
  10.1103/PhysRevB.95.235310} {\bibfield  {journal} {\bibinfo  {journal} {Phys.
  Rev. B}\ }\textbf {\bibinfo {volume} {95}},\ \bibinfo {pages} {235310}
  (\bibinfo {year} {2017})}\BibitemShut {NoStop}%
\bibitem [{\citenamefont {Henkelman}, \citenamefont {Uberuaga},\ and\
  \citenamefont {J$\Tilde{\rm a}$nsson}(2000)}]{Henkelman}%
  \BibitemOpen
  \bibfield  {author} {\bibinfo {author} {\bibfnamefont {G.}~\bibnamefont
  {Henkelman}}, \bibinfo {author} {\bibfnamefont {B.~P.}\ \bibnamefont
  {Uberuaga}}, \ and\ \bibinfo {author} {\bibfnamefont {H.}~\bibnamefont
  {J$\Tilde{\rm a}$nsson}},\ }\bibfield  {title} {\enquote {\bibinfo {title} {A
  climbing image nudged elastic band method for finding saddle points and
  minimum energy paths},}\ }\href {\doibase 10.1063/1.1329672} {\bibfield
  {journal} {\bibinfo  {journal} {J. Chem. Phys.}\ }\textbf {\bibinfo {volume}
  {113}},\ \bibinfo {pages} {9901} (\bibinfo {year} {2000})}\BibitemShut
  {NoStop}%
\bibitem [{\citenamefont {J$\Tilde{\rm a}$nsson}, \citenamefont {Mills},\ and\
  \citenamefont {Jacobsen}(1998)}]{Jansson}%
  \BibitemOpen
  \bibfield  {author} {\bibinfo {author} {\bibfnamefont {H.}~\bibnamefont
  {J$\Tilde{\rm a}$nsson}}, \bibinfo {author} {\bibfnamefont {G.}~\bibnamefont
  {Mills}}, \ and\ \bibinfo {author} {\bibfnamefont {K.~W.}\ \bibnamefont
  {Jacobsen}},\ }\enquote {\bibinfo {title} {Nudged elastic band method for
  finding minimum energy paths of transitions},}\ in\ \href {\doibase
  10.1142/9789812839664_0016} {\emph {\bibinfo {booktitle} {Classical and
  Quantum Dynamics in Condensed Phase Simulations}}}\ (\bibinfo  {publisher}
  {World Scientific, Singapore},\ \bibinfo {year} {1998})\ pp.\ \bibinfo
  {pages} {385--404}\BibitemShut {NoStop}%
\bibitem [{\citenamefont {Voter}(2007)}]{Voter}%
  \BibitemOpen
  \bibfield  {author} {\bibinfo {author} {\bibfnamefont {A.~F.}\ \bibnamefont
  {Voter}},\ }\enquote {\bibinfo {title} {Introduction to the kinetic monte
  carlo method},}\ in\ \href {\doibase 10.1007/978-1-4020-5295-8_1} {\emph
  {\bibinfo {booktitle} {Radiation Effects in Solids}}},\ \bibinfo {editor}
  {edited by\ \bibinfo {editor} {\bibfnamefont {K.~E.}\ \bibnamefont
  {Sickafus}}, \bibinfo {editor} {\bibfnamefont {E.~A.}\ \bibnamefont
  {Kotomin}}, \ and\ \bibinfo {editor} {\bibfnamefont {B.~P.}\ \bibnamefont
  {Uberuaga}}}\ (\bibinfo  {publisher} {Springer Netherlands},\ \bibinfo
  {address} {Dordrecht},\ \bibinfo {year} {2007})\ pp.\ \bibinfo {pages}
  {1--23}\BibitemShut {NoStop}%
\bibitem [{\citenamefont {Manz}\ and\ \citenamefont {Limas}(2016)}]{Manz2016}%
  \BibitemOpen
  \bibfield  {author} {\bibinfo {author} {\bibfnamefont {T.~A.}\ \bibnamefont
  {Manz}}\ and\ \bibinfo {author} {\bibfnamefont {N.~G.}\ \bibnamefont
  {Limas}},\ }\bibfield  {title} {\enquote {\bibinfo {title} {{Introducing
  DDEC6 atomic population analysis: part 1. Charge partitioning theory and
  methodology}},}\ }\href {\doibase 10.1039/C6RA04656H} {\bibfield  {journal}
  {\bibinfo  {journal} {RSC Adv.}\ }\textbf {\bibinfo {volume} {6}},\ \bibinfo
  {pages} {47771} (\bibinfo {year} {2016})}\BibitemShut {NoStop}%
\bibitem [{\citenamefont {Manz}\ and\ \citenamefont {Sholl}(2010)}]{Manz2010}%
  \BibitemOpen
  \bibfield  {author} {\bibinfo {author} {\bibfnamefont {T.~A.}\ \bibnamefont
  {Manz}}\ and\ \bibinfo {author} {\bibfnamefont {D.~S.}\ \bibnamefont
  {Sholl}},\ }\bibfield  {title} {\enquote {\bibinfo {title} {Chemically
  meaningful atomic charges that reproduce the electrostatic potential in
  periodic and nonperiodic materials},}\ }\href {\doibase 10.1021/ct100125x}
  {\bibfield  {journal} {\bibinfo  {journal} {J. Chem. Theory Comput.}\
  }\textbf {\bibinfo {volume} {6}},\ \bibinfo {pages} {2455} (\bibinfo {year}
  {2010})}\BibitemShut {NoStop}%
\bibitem [{\citenamefont {Cao}\ \emph {et~al.}(1987)\citenamefont {Cao},
  \citenamefont {Gatti}, \citenamefont {MacDougall},\ and\ \citenamefont
  {Bader}}]{Cao1987}%
  \BibitemOpen
  \bibfield  {author} {\bibinfo {author} {\bibfnamefont {W.}~\bibnamefont
  {Cao}}, \bibinfo {author} {\bibfnamefont {C.}~\bibnamefont {Gatti}}, \bibinfo
  {author} {\bibfnamefont {P.}~\bibnamefont {MacDougall}}, \ and\ \bibinfo
  {author} {\bibfnamefont {R.}~\bibnamefont {Bader}},\ }\bibfield  {title}
  {\enquote {\bibinfo {title} {{On the presence of non-nuclear attractors in
  the charge distributions of Li and Na clusters}},}\ }\href {\doibase
  https://doi.org/10.1016/0009-2614(87)85044-3} {\bibfield  {journal} {\bibinfo
   {journal} {Chem. Phys. Lett.}\ }\textbf {\bibinfo {volume} {141}},\ \bibinfo
  {pages} {380} (\bibinfo {year} {1987})}\BibitemShut {NoStop}%
\bibitem [{\citenamefont {Reed}, \citenamefont {Weinstock},\ and\ \citenamefont
  {Weinhold}(1985)}]{Reed1985}%
  \BibitemOpen
  \bibfield  {author} {\bibinfo {author} {\bibfnamefont {A.~E.}\ \bibnamefont
  {Reed}}, \bibinfo {author} {\bibfnamefont {R.~B.}\ \bibnamefont {Weinstock}},
  \ and\ \bibinfo {author} {\bibfnamefont {F.}~\bibnamefont {Weinhold}},\
  }\bibfield  {title} {\enquote {\bibinfo {title} {Natural population
  analysis},}\ }\href {\doibase 10.1063/1.449486} {\bibfield  {journal}
  {\bibinfo  {journal} {J. Chem. Phys.}\ }\textbf {\bibinfo {volume} {83}},\
  \bibinfo {pages} {735} (\bibinfo {year} {1985})}\BibitemShut {NoStop}%
\bibitem [{\citenamefont {Togo}\ and\ \citenamefont {Tanaka}(2015)}]{Togo}%
  \BibitemOpen
  \bibfield  {author} {\bibinfo {author} {\bibfnamefont {A.}~\bibnamefont
  {Togo}}\ and\ \bibinfo {author} {\bibfnamefont {I.}~\bibnamefont {Tanaka}},\
  }\bibfield  {title} {\enquote {\bibinfo {title} {First principles phonon
  calculations in materials science},}\ }\href {\doibase
  https://doi.org/10.1016/j.scriptamat.2015.07.021} {\bibfield  {journal}
  {\bibinfo  {journal} {Scr. Mater.}\ }\textbf {\bibinfo {volume} {108}},\
  \bibinfo {pages} {1} (\bibinfo {year} {2015})}\BibitemShut {NoStop}%
\bibitem [{\citenamefont {Levinshtein}, \citenamefont {Rumyantsev},\ and\
  \citenamefont {Shur}(2001)}]{Levinshtein}%
  \BibitemOpen
  \bibfield  {author} {\bibinfo {author} {\bibfnamefont {M.~E.}\ \bibnamefont
  {Levinshtein}}, \bibinfo {author} {\bibfnamefont {S.~L.}\ \bibnamefont
  {Rumyantsev}}, \ and\ \bibinfo {author} {\bibfnamefont {M.~S.}\ \bibnamefont
  {Shur}},\ }\href@noop {} {\emph {\bibinfo {title} {Properties of Advanced
  Semiconductor Materials: GaN, AlN, InN, BN, SiC, SiGe}}}\ (\bibinfo
  {publisher} {John Wiley \& Sons, Inc.},\ \bibinfo {address} {New York},\
  \bibinfo {year} {2001})\BibitemShut {NoStop}%
\bibitem [{\citenamefont {Isberg}\ \emph {et~al.}(2002)\citenamefont {Isberg},
  \citenamefont {Hammersberg}, \citenamefont {Johansson}, \citenamefont
  {Wikstr{\"o}m}, \citenamefont {Twitchen}, \citenamefont {Whitehead},
  \citenamefont {Coe},\ and\ \citenamefont {Scarsbrook}}]{Isberg}%
  \BibitemOpen
  \bibfield  {author} {\bibinfo {author} {\bibfnamefont {J.}~\bibnamefont
  {Isberg}}, \bibinfo {author} {\bibfnamefont {J.}~\bibnamefont {Hammersberg}},
  \bibinfo {author} {\bibfnamefont {E.}~\bibnamefont {Johansson}}, \bibinfo
  {author} {\bibfnamefont {T.}~\bibnamefont {Wikstr{\"o}m}}, \bibinfo {author}
  {\bibfnamefont {D.~J.}\ \bibnamefont {Twitchen}}, \bibinfo {author}
  {\bibfnamefont {A.~J.}\ \bibnamefont {Whitehead}}, \bibinfo {author}
  {\bibfnamefont {S.~E.}\ \bibnamefont {Coe}}, \ and\ \bibinfo {author}
  {\bibfnamefont {G.~A.}\ \bibnamefont {Scarsbrook}},\ }\bibfield  {title}
  {\enquote {\bibinfo {title} {High carrier mobility in single-crystal
  plasma-deposited diamond},}\ }\href {\doibase 10.1126/science.1074374}
  {\bibfield  {journal} {\bibinfo  {journal} {Science}\ }\textbf {\bibinfo
  {volume} {297}},\ \bibinfo {pages} {1670} (\bibinfo {year}
  {2002})}\BibitemShut {NoStop}%
\bibitem [{\citenamefont {Salustro}\ \emph {et~al.}(2017)\citenamefont
  {Salustro}, \citenamefont {Ferrari}, \citenamefont {Orlando},\ and\
  \citenamefont {Dovesi}}]{Salustro2017}%
  \BibitemOpen
  \bibfield  {author} {\bibinfo {author} {\bibfnamefont {S.}~\bibnamefont
  {Salustro}}, \bibinfo {author} {\bibfnamefont {A.~M.}\ \bibnamefont
  {Ferrari}}, \bibinfo {author} {\bibfnamefont {R.}~\bibnamefont {Orlando}}, \
  and\ \bibinfo {author} {\bibfnamefont {R.}~\bibnamefont {Dovesi}},\
  }\bibfield  {title} {\enquote {\bibinfo {title} {Comparison between cluster
  and supercell approaches: the case of defects in diamond},}\ }\href {\doibase
  10.1007/s00214-017-2071-5} {\bibfield  {journal} {\bibinfo  {journal} {Theor.
  Chem. Acc.}\ }\textbf {\bibinfo {volume} {136}},\ \bibinfo {pages} {42}
  (\bibinfo {year} {2017})}\BibitemShut {NoStop}%
\bibitem [{\citenamefont {Zelferino}\ \emph {et~al.}(2016)\citenamefont
  {Zelferino}, \citenamefont {Salustro}, \citenamefont {Baima}, \citenamefont
  {Lacivita}, \citenamefont {Orlando},\ and\ \citenamefont
  {Dovesi}}]{Zelferino2016}%
  \BibitemOpen
  \bibfield  {author} {\bibinfo {author} {\bibfnamefont {A.}~\bibnamefont
  {Zelferino}}, \bibinfo {author} {\bibfnamefont {S.}~\bibnamefont {Salustro}},
  \bibinfo {author} {\bibfnamefont {J.}~\bibnamefont {Baima}}, \bibinfo
  {author} {\bibfnamefont {V.}~\bibnamefont {Lacivita}}, \bibinfo {author}
  {\bibfnamefont {R.}~\bibnamefont {Orlando}}, \ and\ \bibinfo {author}
  {\bibfnamefont {R.}~\bibnamefont {Dovesi}},\ }\bibfield  {title} {\enquote
  {\bibinfo {title} {The electronic states of the neutral vacancy in diamond: a
  quantum mechanical approach},}\ }\href {\doibase 10.1007/s00214-016-1813-0}
  {\bibfield  {journal} {\bibinfo  {journal} {Theor. Chem. Acc.}\ }\textbf
  {\bibinfo {volume} {135}},\ \bibinfo {pages} {74} (\bibinfo {year}
  {2016})}\BibitemShut {NoStop}%
\bibitem [{\citenamefont {Miyazaki}(2002)}]{Miyazaki2002}%
  \BibitemOpen
  \bibfield  {author} {\bibinfo {author} {\bibfnamefont {T.}~\bibnamefont
  {Miyazaki}},\ }\bibfield  {title} {\enquote {\bibinfo {title} {Theoretical
  studies of sulfur and sulfur-hydrogen complexes in diamond},}\ }\href
  {\doibase 10.1002/1521-396X(200210)193:3<395::AID-PSSA395>3.0.CO;2-1}
  {\bibfield  {journal} {\bibinfo  {journal} {Phys. Status Solidi A}\ }\textbf
  {\bibinfo {volume} {193}},\ \bibinfo {pages} {395} (\bibinfo {year}
  {2002})}\BibitemShut {NoStop}%
\bibitem [{\citenamefont {Collins}(2002)}]{Collins_2002}%
  \BibitemOpen
  \bibfield  {author} {\bibinfo {author} {\bibfnamefont {A.~T.}\ \bibnamefont
  {Collins}},\ }\bibfield  {title} {\enquote {\bibinfo {title} {{The Fermi
  level in diamond}},}\ }\href {\doibase 10.1088/0953-8984/14/14/307}
  {\bibfield  {journal} {\bibinfo  {journal} {Journal of Physics: Condensed
  Matter}\ }\textbf {\bibinfo {volume} {14}},\ \bibinfo {pages} {3743}
  (\bibinfo {year} {2002})}\BibitemShut {NoStop}%
\bibitem [{\citenamefont {Lany}\ and\ \citenamefont {Zunger}(2008)}]{Lany}%
  \BibitemOpen
  \bibfield  {author} {\bibinfo {author} {\bibfnamefont {S.}~\bibnamefont
  {Lany}}\ and\ \bibinfo {author} {\bibfnamefont {A.}~\bibnamefont {Zunger}},\
  }\bibfield  {title} {\enquote {\bibinfo {title} {{Assessment of correction
  methods for the band-gap problem and for finite-size effects in supercell
  defect calculations: Case studies for ZnO and GaAs}},}\ }\href {\doibase
  10.1103/PhysRevB.78.235104} {\bibfield  {journal} {\bibinfo  {journal} {Phys.
  Rev. B}\ }\textbf {\bibinfo {volume} {78}},\ \bibinfo {pages} {235104}
  (\bibinfo {year} {2008})}\BibitemShut {NoStop}%
\bibitem [{\citenamefont {Jacobs}(1974)}]{Kroger}%
  \BibitemOpen
  \bibfield  {author} {\bibinfo {author} {\bibfnamefont {K.}~\bibnamefont
  {Jacobs}},\ }\bibfield  {title} {\enquote {\bibinfo {title} {{F. A. Kr\"oger.
  The Chemistry of Imperfect Crystals. 2nd Revised Edition, Volume 1:
  Preparation, Purification, Crystal Growth And Phase Theory. North-Holland
  Publishing Company - Amsterdam/London 1973 American Elsevier Publishing
  Company, Inc. - New York 313 Seiten, zahlreiche Abbildungen und Tabellen,
  Kunstleder Preis Dfl. 70.00}},}\ }\href {\doibase 10.1002/crat.19740090719}
  {\bibfield  {journal} {\bibinfo  {journal} {Kristall und Technik}\ }\textbf
  {\bibinfo {volume} {9}},\ \bibinfo {pages} {K67} (\bibinfo {year}
  {1974})}\BibitemShut {NoStop}%
\bibitem [{\citenamefont {Stix}(1992)}]{stix1992waves}%
  \BibitemOpen
  \bibfield  {author} {\bibinfo {author} {\bibfnamefont {T.}~\bibnamefont
  {Stix}},\ }\href {https://books.google.com/books?id=OsOWJ8iHpmMC} {\emph
  {\bibinfo {title} {Waves in Plasmas}}}\ (\bibinfo  {publisher} {American
  Inst. of Physics},\ \bibinfo {year} {1992})\BibitemShut {NoStop}%
\bibitem [{\citenamefont {Gali}, \citenamefont {Fyta},\ and\ \citenamefont
  {Kaxiras}(2008)}]{Gali4}%
  \BibitemOpen
  \bibfield  {author} {\bibinfo {author} {\bibfnamefont {A.}~\bibnamefont
  {Gali}}, \bibinfo {author} {\bibfnamefont {M.}~\bibnamefont {Fyta}}, \ and\
  \bibinfo {author} {\bibfnamefont {E.}~\bibnamefont {Kaxiras}},\ }\bibfield
  {title} {\enquote {\bibinfo {title} {Ab initio supercell calculations on
  nitrogen-vacancy center in diamond: Electronic structure and hyperfine
  tensors},}\ }\href {\doibase 10.1103/PhysRevB.77.155206} {\bibfield
  {journal} {\bibinfo  {journal} {Phys. Rev. B}\ }\textbf {\bibinfo {volume}
  {77}},\ \bibinfo {pages} {155206} (\bibinfo {year} {2008})}\BibitemShut
  {NoStop}%
\bibitem [{\citenamefont {Siyushev}\ \emph {et~al.}(2017)\citenamefont
  {Siyushev}, \citenamefont {Metsch}, \citenamefont {Ijaz}, \citenamefont
  {Binder}, \citenamefont {Bhaskar}, \citenamefont {Sukachev}, \citenamefont
  {Sipahigil}, \citenamefont {Evans}, \citenamefont {Nguyen}, \citenamefont
  {Lukin}, \citenamefont {Hemmer}, \citenamefont {Palyanov}, \citenamefont
  {Kupriyanov}, \citenamefont {Borzdov}, \citenamefont {Rogers},\ and\
  \citenamefont {Jelezko}}]{Siyushev}%
  \BibitemOpen
  \bibfield  {author} {\bibinfo {author} {\bibfnamefont {P.}~\bibnamefont
  {Siyushev}}, \bibinfo {author} {\bibfnamefont {M.~H.}\ \bibnamefont
  {Metsch}}, \bibinfo {author} {\bibfnamefont {A.}~\bibnamefont {Ijaz}},
  \bibinfo {author} {\bibfnamefont {J.~M.}\ \bibnamefont {Binder}}, \bibinfo
  {author} {\bibfnamefont {M.~K.}\ \bibnamefont {Bhaskar}}, \bibinfo {author}
  {\bibfnamefont {D.~D.}\ \bibnamefont {Sukachev}}, \bibinfo {author}
  {\bibfnamefont {A.}~\bibnamefont {Sipahigil}}, \bibinfo {author}
  {\bibfnamefont {R.~E.}\ \bibnamefont {Evans}}, \bibinfo {author}
  {\bibfnamefont {C.~T.}\ \bibnamefont {Nguyen}}, \bibinfo {author}
  {\bibfnamefont {M.~D.}\ \bibnamefont {Lukin}}, \bibinfo {author}
  {\bibfnamefont {P.~R.}\ \bibnamefont {Hemmer}}, \bibinfo {author}
  {\bibfnamefont {Y.~N.}\ \bibnamefont {Palyanov}}, \bibinfo {author}
  {\bibfnamefont {I.~N.}\ \bibnamefont {Kupriyanov}}, \bibinfo {author}
  {\bibfnamefont {Y.~M.}\ \bibnamefont {Borzdov}}, \bibinfo {author}
  {\bibfnamefont {L.~J.}\ \bibnamefont {Rogers}}, \ and\ \bibinfo {author}
  {\bibfnamefont {F.}~\bibnamefont {Jelezko}},\ }\bibfield  {title} {\enquote
  {\bibinfo {title} {Optical and microwave control of germanium-vacancy center
  spins in diamond},}\ }\href {\doibase 10.1103/PhysRevB.96.081201} {\bibfield
  {journal} {\bibinfo  {journal} {Phys. Rev. B}\ }\textbf {\bibinfo {volume}
  {96}},\ \bibinfo {pages} {081201} (\bibinfo {year} {2017})}\BibitemShut
  {NoStop}%
\bibitem [{\citenamefont {Kaxiras}(2003)}]{kaxiras2003atomic}%
  \BibitemOpen
  \bibfield  {author} {\bibinfo {author} {\bibfnamefont {E.}~\bibnamefont
  {Kaxiras}},\ }\href {https://books.google.com/books?id=WTL\_vgbWpHEC} {\emph
  {\bibinfo {title} {Atomic and Electronic Structure of Solids}}}\ (\bibinfo
  {publisher} {Cambridge University Press},\ \bibinfo {year}
  {2003})\BibitemShut {NoStop}%
\bibitem [{\citenamefont {Ashcroft}\ and\ \citenamefont
  {Mermin}(1976)}]{ashcroft1976solid}%
  \BibitemOpen
  \bibfield  {author} {\bibinfo {author} {\bibfnamefont {N.}~\bibnamefont
  {Ashcroft}}\ and\ \bibinfo {author} {\bibfnamefont {N.}~\bibnamefont
  {Mermin}},\ }\href {https://books.google.com/books?id=1C9HAQAAIAAJ} {\emph
  {\bibinfo {title} {Solid State Physics}}},\ HRW international editions\
  (\bibinfo  {publisher} {Holt, Rinehart and Winston},\ \bibinfo {year}
  {1976})\BibitemShut {NoStop}%
\bibitem [{\citenamefont {Pestourie}\ \emph {et~al.}(2018)\citenamefont
  {Pestourie}, \citenamefont {P\'{e}rez-Arancibia}, \citenamefont {Lin},
  \citenamefont {Shin}, \citenamefont {Capasso},\ and\ \citenamefont
  {Johnson}}]{Pestourie}%
  \BibitemOpen
  \bibfield  {author} {\bibinfo {author} {\bibfnamefont {R.}~\bibnamefont
  {Pestourie}}, \bibinfo {author} {\bibfnamefont {C.}~\bibnamefont
  {P\'{e}rez-Arancibia}}, \bibinfo {author} {\bibfnamefont {Z.}~\bibnamefont
  {Lin}}, \bibinfo {author} {\bibfnamefont {W.}~\bibnamefont {Shin}}, \bibinfo
  {author} {\bibfnamefont {F.}~\bibnamefont {Capasso}}, \ and\ \bibinfo
  {author} {\bibfnamefont {S.~G.}\ \bibnamefont {Johnson}},\ }\bibfield
  {title} {\enquote {\bibinfo {title} {Inverse design of large-area
  metasurfaces},}\ }\href {\doibase 10.1364/OE.26.033732} {\bibfield  {journal}
  {\bibinfo  {journal} {Opt. Express}\ }\textbf {\bibinfo {volume} {26}},\
  \bibinfo {pages} {33732} (\bibinfo {year} {2018})}\BibitemShut {NoStop}%
\bibitem [{\citenamefont {Zhou}\ \emph {et~al.}(2018)\citenamefont {Zhou},
  \citenamefont {Mu}, \citenamefont {Adamo}, \citenamefont {Bauerdick},
  \citenamefont {Rudzinski}, \citenamefont {Aharonovich},\ and\ \citenamefont
  {Gao}}]{Zhou_2018}%
  \BibitemOpen
  \bibfield  {author} {\bibinfo {author} {\bibfnamefont {Y.}~\bibnamefont
  {Zhou}}, \bibinfo {author} {\bibfnamefont {Z.}~\bibnamefont {Mu}}, \bibinfo
  {author} {\bibfnamefont {G.}~\bibnamefont {Adamo}}, \bibinfo {author}
  {\bibfnamefont {S.}~\bibnamefont {Bauerdick}}, \bibinfo {author}
  {\bibfnamefont {A.}~\bibnamefont {Rudzinski}}, \bibinfo {author}
  {\bibfnamefont {I.}~\bibnamefont {Aharonovich}}, \ and\ \bibinfo {author}
  {\bibfnamefont {W.}~\bibnamefont {Gao}},\ }\bibfield  {title} {\enquote
  {\bibinfo {title} {Direct writing of single germanium vacancy center arrays
  in diamond},}\ }\href {\doibase 10.1088/1367-2630/aaf2ac} {\bibfield
  {journal} {\bibinfo  {journal} {New Journal of Physics}\ }\textbf {\bibinfo
  {volume} {20}},\ \bibinfo {pages} {125004} (\bibinfo {year}
  {2018})}\BibitemShut {NoStop}%
\bibitem [{\citenamefont {Childress}\ \emph {et~al.}(2006)\citenamefont
  {Childress}, \citenamefont {Gurudev~Dutt}, \citenamefont {Taylor},
  \citenamefont {Zibrov}, \citenamefont {Jelezko}, \citenamefont {Wrachtrup},
  \citenamefont {Hemmer},\ and\ \citenamefont {Lukin}}]{Childress}%
  \BibitemOpen
  \bibfield  {author} {\bibinfo {author} {\bibfnamefont {L.}~\bibnamefont
  {Childress}}, \bibinfo {author} {\bibfnamefont {M.~V.}\ \bibnamefont
  {Gurudev~Dutt}}, \bibinfo {author} {\bibfnamefont {J.~M.}\ \bibnamefont
  {Taylor}}, \bibinfo {author} {\bibfnamefont {A.~S.}\ \bibnamefont {Zibrov}},
  \bibinfo {author} {\bibfnamefont {F.}~\bibnamefont {Jelezko}}, \bibinfo
  {author} {\bibfnamefont {J.}~\bibnamefont {Wrachtrup}}, \bibinfo {author}
  {\bibfnamefont {P.~R.}\ \bibnamefont {Hemmer}}, \ and\ \bibinfo {author}
  {\bibfnamefont {M.~D.}\ \bibnamefont {Lukin}},\ }\bibfield  {title} {\enquote
  {\bibinfo {title} {{Coherent Dynamics of Coupled Electron and Nuclear Spin
  Qubits in Diamond}},}\ }\href {\doibase 10.1126/science.1131871} {\bibfield
  {journal} {\bibinfo  {journal} {Science}\ }\textbf {\bibinfo {volume}
  {314}},\ \bibinfo {pages} {281} (\bibinfo {year} {2006})}\BibitemShut
  {NoStop}%
\bibitem [{\citenamefont {Towns}\ \emph {et~al.}(2014)\citenamefont {Towns},
  \citenamefont {Cockerill}, \citenamefont {Dahan}, \citenamefont {Foster},
  \citenamefont {Gaither}, \citenamefont {Grimshaw}, \citenamefont {Hazlewood},
  \citenamefont {Lathrop}, \citenamefont {Lifka}, \citenamefont {Peterson},
  \citenamefont {Roskies}, \citenamefont {Scott},\ and\ \citenamefont
  {Wilkins-Diehr}}]{Towns}%
  \BibitemOpen
  \bibfield  {author} {\bibinfo {author} {\bibfnamefont {J.}~\bibnamefont
  {Towns}}, \bibinfo {author} {\bibfnamefont {T.}~\bibnamefont {Cockerill}},
  \bibinfo {author} {\bibfnamefont {M.}~\bibnamefont {Dahan}}, \bibinfo
  {author} {\bibfnamefont {I.}~\bibnamefont {Foster}}, \bibinfo {author}
  {\bibfnamefont {K.}~\bibnamefont {Gaither}}, \bibinfo {author} {\bibfnamefont
  {A.}~\bibnamefont {Grimshaw}}, \bibinfo {author} {\bibfnamefont
  {V.}~\bibnamefont {Hazlewood}}, \bibinfo {author} {\bibfnamefont
  {S.}~\bibnamefont {Lathrop}}, \bibinfo {author} {\bibfnamefont
  {D.}~\bibnamefont {Lifka}}, \bibinfo {author} {\bibfnamefont {G.~D.}\
  \bibnamefont {Peterson}}, \bibinfo {author} {\bibfnamefont {R.}~\bibnamefont
  {Roskies}}, \bibinfo {author} {\bibfnamefont {J.~R.}\ \bibnamefont {Scott}},
  \ and\ \bibinfo {author} {\bibfnamefont {N.}~\bibnamefont {Wilkins-Diehr}},\
  }\bibfield  {title} {\enquote {\bibinfo {title} {{XSEDE: Accelerating
  Scientific Discovery}},}\ }\href {\doibase 10.1109/MCSE.2014.80} {\bibfield
  {journal} {\bibinfo  {journal} {Computing in Science \& Engineering}\
  }\textbf {\bibinfo {volume} {16}},\ \bibinfo {pages} {62} (\bibinfo {year}
  {2014})}\BibitemShut {NoStop}%
\end{thebibliography}%

\end{document}